\documentclass[acmlarge,screen]{acmart}
\PassOptionsToPackage{table,xcdraw,prologue}{xcolor}
\usepackage{multirow}
\usepackage{longtable}
\usepackage{graphicx}
\usepackage{tabularx}
\usepackage{subfig}
\usepackage{colortbl}
\usepackage{diagbox}
\usepackage{geometry}
\usepackage[linesnumbered,ruled,vlined]{algorithm2e}
\usepackage{xcolor}
\AtBeginDocument{%
  \providecommand\BibTeX{{%
    \normalfont B\kern-0.5em{\scshape i\kern-0.25em b}\kern-0.8em\TeX}}}

\begin{document}

\title[Investigating Deceptive Patterns in Mandarin Chinese Female Synthetic Speech]{The Manipulative Power of Voice Characteristics: Investigating Deceptive Patterns in Mandarin Chinese Female Synthetic Speech}

\author{Shuning Zhang}
\orcid{0000-0002-4145-117X}
\authornotemark[1]
\email{zsn23@mails.tsinghua.edu.cn}
\affiliation{%
  \institution{Tsinghua University}%
  \city{Beijing}%
  \country{China}%
}

\author{Han Chen}
\orcid{0000-0002-4145-117X}
\authornote{These authors contributed equally to this work.}
\email{hanchen241@gmail.com}
\affiliation{%
  \institution{Wuhan Institute of Technology}
  \city{Wuhan}
  \country{China}
}

\author{Yabo Wang}
\orcid{0009-0005-5782-8807}
\email{yb-wang22@mails.tsinghua.edu.cn}
\affiliation{%
  \institution{Tsinghua University}
  \city{Beijing}
  \country{China}
}

\author{Yiqun Xu}
\orcid{0009-0004-9774-9129}
\email{xuyiqun22@mails.tsinghua.edu.cn}
\affiliation{%
  \institution{Tsinghua University}
  \city{Beijing}
  \country{China}
}

\author{Jiaqi Bai}
\orcid{0009-0003-9792-3556}
\email{bjq21@mails.tsinghua.edu.cn}
\affiliation{%
  \institution{Tsinghua University}
  \city{Beijing}
  \country{China}
}

\author{yuanyuan Wu}
\orcid{0009-0004-2633-9691}
\email{buddy.yuan@sjtu.edu.cn}
\affiliation{%
  \institution{Shanghai Jiaotong University}
  \city{Shanghai}
  \country{China}
}

\author{Shixuan Li}
\orcid{0009-0008-6828-6347}
\email{li-sx24@mails.tsinghua.edu.cn}
\affiliation{
    \institution{Tsinghua University}
    \city{Beijing}
    \country{China}
}

\author{Xin Yi}
\orcid{0000-0001-8041-7962}
\authornotemark[2]
\email{yixin@tsinghua.edu.cn}
\affiliation{%
    \institution{Tsinghua University}
    \city{Beijing}
    \country{China}
}

\author{Chunhui Wang}
\orcid{0000-0001-8041-7962}
\authornote{Co-corresponding.}
\email{chunhui_89@163.com}
\affiliation{
    \institution{National Key Laboratory of Human Factors Engineering, China Astronaut Research and Training Center}
    \city{Beijing}
    \country{China}
}

\author{Hewu Li}
\orcid{0000-0002-6331-6542}
\email{lihewu@cernet.edu.cn}
\affiliation{%
    \institution{Tsinghua University}
    \city{Beijing}
    \country{China}
}

\renewcommand{\shortauthors}{Zhang and Chen et al.}

\begin{abstract}
Pervasive voice interaction enables deceptive patterns through subtle voice characteristics, yet empirical investigation into this manipulation lags behind, especially within major non-English language contexts. Addressing this gap, our study presents the first systematic investigation into voice characteristic-based dark patterns employing female synthetic voices in Mandarin Chinese. This focus is crucial given the prevalence of female personas in commercial assistants and the prosodic significance in the Chinese language. Guided by the conceptual framework identifying key influencing factors, we systematically evaluate effectiveness variations by manipulating \textit{voice characteristics} (five characteristics, three intensities) across different \textit{scenarios} (shopping vs. question-answering) with different \textbf{commercial aims}. A preliminary study (N=24) validated the experimental materials and the main study (N=36) revealed significant behavioral manipulation (up to +2027.6\%). Crucially, the analysis showed that effectiveness varied significantly with voice characteristics and scenario, mediated by user perception (of tone, intonation, timbre) and user demographics (individual preferences, though limited demographic impact). These interconnected findings offer evidence-based insights for ethical design.
\end{abstract}

\ccsdesc[300]{Human-centered computing~Ubiquitous and mobile computing design and evaluation methods}
\ccsdesc[500]{Human-centered computing~Auditory feedback}
\ccsdesc[100]{Human-centered computing~Natural language interfaces}

\begin{teaserfigure}
    \centering
    \includegraphics[width=0.9\textwidth]{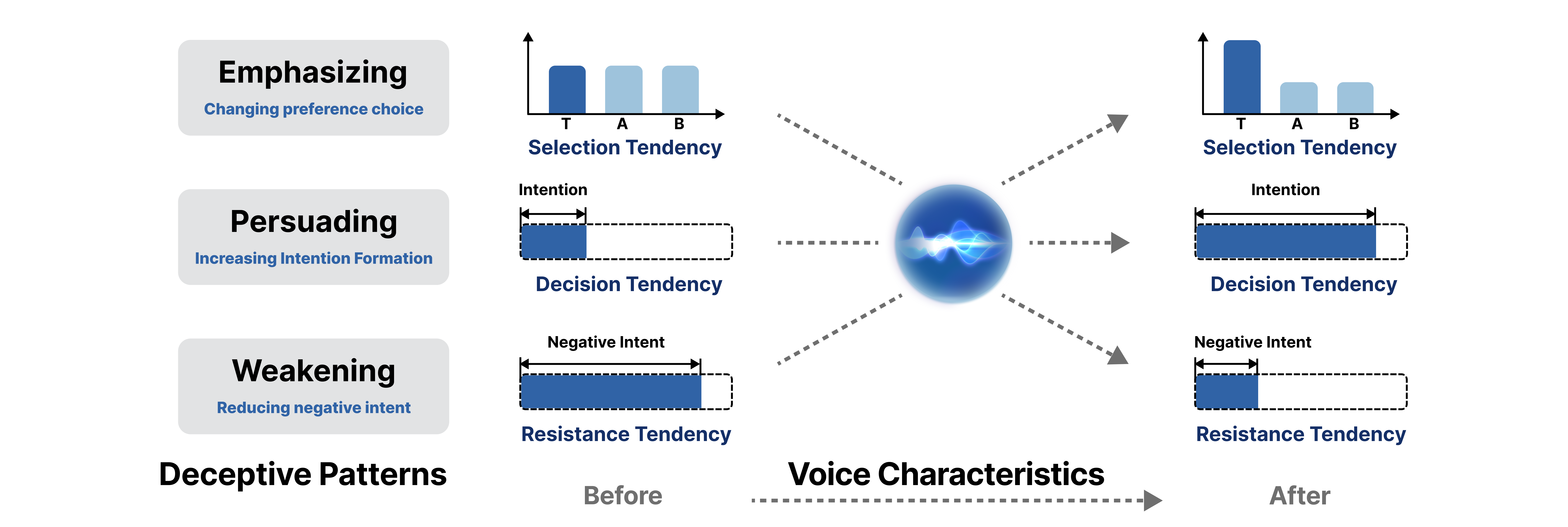}
    \caption{Illustration of the voice characteristic-based dark patterns in this paper. We proposed three different dark patterns with distinct commercial aims. T, A, B denote different information, where T is the targeted false information, and A, B are candidate information.}
\end{teaserfigure}

\keywords{voice characteristics, voice interaction, dark pattern, manipulation}


\maketitle

\section{Introduction}


Deceptive and manipulative design patterns, commonly termed dark patterns, represent interface elements intentionally crafted to mislead or coerce users into actions that primarily serve the designer's interests, often to the detriment of the user~\cite{brignull2018dark,conti2010malicious}. The prevalence and negative consequences of these patterns within traditional Graphical User Interfaces (GUIs) are well-documented (see Table~\ref{tbl:rw_classification}), highlighting their capacity to undermine user autonomy, degrade user experience, and adversely affect user privacy, security, and financial well-being~\cite{luguri2021shining,gray2018dark,mathur2021makes}. Substantial research has focused on identifying and categorizing these patterns in visual interfaces~\cite{gray2018dark,mathur2021makes,conti2010malicious,brignull2018dark}.

With the proliferation of voice interaction techniques~\cite{seaborn2021voice}, manipulative designs have extended beyond GUIs into the domain of Voice User Interfaces (VUIs), which are increasingly integrated into ubiquitous computing environments~\cite{owens2022exploring,hu2025vision}. Voice interaction constitutes a primary mode of human-computer interaction (HCI)~\cite{alberts2024computers}, and the potential threats within this ecosystem have been frequently discussed~\cite{wang2021demystifying,beneteau2020assumptions}. \textbf{VUIs introduce unique vulnerabilities due to the inherent nature of voice communication, creating new attack surfaces along three critical dimensions}:
First, voice characteristics such as tone, pitch, rhythm~\cite{lupianez2022behavioural,belin2011understanding} operate through paralinguistic channels rather than visual elements, creating a covert attack surface that significantly reduces detectability compared to GUI-based patterns. Second, human cognitive systems inherently interpret voice interactions as natural social communications rather than system-mediated exchanges~\cite{fedorenko2024language}, which diminishes users' awareness of potential manipulation patterns. Finally, advancements in voice synthesis technology allow for precise modulation of voice characteristics without sacrificing naturalness~\cite{ju2024naturalspeech3}, thus enabling subtle behavioral manipulation. While prior studies have suggested that voice characteristics might constitute dark patterns~\cite{dubiel2024impact}, a systematic understanding of how these factors combine to enable deceptive patterns that manipulate user behavior remains notably absent, especially in non-English contexts.

This study addresses this gap by systematically investigating voice characteristic-based dark patterns, specifically focusing on female synthetic speech in Mandarin Chinese. We chose Mandarin Chinese as it is one of the major world languages with a vast user base for voice technologies\footnote{\url{https://www.statista.com/statistics/266808/the-most-spoken-languages-worldwide/}}~\cite{liesenfeld2020namespec,liu2024production}, yet significantly underrepresented in VUI dark pattern research compared to English~\cite{dubiel2024impact,owens2022exploring}. Furthermore, as a tonal language, prosody in Mandarin carries significant linguistic and pragmatic weight, making it a pertinent context for exploring the manipulative potential of synthesized voice characteristics~\cite{huang2016improving}. Our focus on female voices is motivated by their widespread use in commercial voice assistants~\cite{cambre2019one}, making this a highly relevant starting point for assessing real-world risks. The central problems we address is the potential for undisclosed, subtle manipulation of user behavior through the deliberate crafting of voice characteristics in widely used languages such as Mandarin Chinese~\cite{liesenfeld2020namespec,liu2024production}, creating significant ethical risks. To dissect this problem, we ask:

\noindent \textbf{RQ1:} How can specific voice characteristics be leveraged to constitute manipulative dark patterns?

\noindent \textbf{RQ2:} How does the effectiveness of voice characteristic-based dark patterns in influencing user behavior vary across different voice characteristics, commercial aims, scenarios, and demographic factors?


For RQ1, we formulated three different voice characteristic-based dark patterns tailored to user vulnerabilities (from user, device, and context perspectives)~\cite{cambre2019one} to achieve representative commercial aims~\cite{mathur2021makes,gray2018dark,gray2023towards}: emphasizing, persuading and weakening. This categorization introduces new voice characteristic-based patterns that extend the reach of traditional dark patterns in visual interfaces~\cite{mathur2019dark,mathur2021makes} to voice characteristics. Different dark patterns were associated with different commercial aims within related scenarios. Through analysis of previous work~\cite{higgins2022sympathy,fahim2021integral,ofuji2018verbal}, we assigned different voice characteristics to enable different deceptive patterns: misleading authority~\cite{herder2023context}, overly friendly~\cite{higgins2022sympathy,kuhne2020human}, emotional manipulation~\cite{fahim2021integral,martelaro2016tell}, false urgency~\cite{ofuji2018verbal,ko2022modeling}, hesitation and uncertainty~\cite{goupil2021listeners}.

We introduced two representative scenarios for examining the effectiveness of voice-characteristic-based dark patterns: \textbf{shopping} and \textbf{question-answering}~\cite{seaborn2021voice}. Shopping was chosen to emphasize commercial profit, while question-answering highlights the objective delivery of factual information. These scenarios were selected because they frequently involve voice assistance, a widely used interface that has gained increasing attention~\cite{mctear2016conversational} and could potentially shape future interaction paradigms. In the \textbf{shopping} scenario, the goal of the dark pattern is to manipulate users' intention towards buying certain products or reducing returns and exchanges. In contrast, in the \textbf{question-answering} scenario, the aim is to manipulate users into accepting misinformation. While we focused on these two scenarios to illustrate the potential of voice-based dark patterns, our approach is not limited to them.

To systematically investigate RQ2, we operationalize the factors known to influence voice perception and interaction outcomes according to our conceptual framework (Section~\ref{sec:framework}, Figure~\ref{fig:factor}). After validating synthesized \textit{voice signal} materials (N=24)~\cite{dubiel2024impact}, our main controlled study (N=36) manipulated core \textit{voice signal} attributes (five characteristics across two intensity levels). We assessed their impact under varying \textit{task factors} by implementing three dark pattern aims within two distinct interaction contexts (Shopping vs. Q\&A). We measured the \textit{manipulation effect} on user behavior (choices, intentions, trust) and gathered data on mediating perceptions and potential \textit{user factors} (demographics, preferences). This approach allows a structured analysis of how these specific framework components interact to determine manipulation success. 


Our findings, analyzed through the lens of our conceptual framework (Section~\ref{sec:framework}, Figure~\ref{fig:factor}), provide a comprehensive answer to RQ2. We confirmed the potent effect of the manipulated \textit{voice characteristic}, demonstrating significant behavioral impact across aims; for example, an authoritative voice increased selection rates from 22.7\% to 77.3\% (+240.5\%) in question-answering and from 2.9\% to 61.7\% (+2027.6\%) in shopping. Effectiveness varied substantially with the specific \textit{voice characteristic} (e.g., authority, emotional, friendly) and its \textit{intensity}. Moderate intensity was often optimal for emotional or friendly tones, whereas exaggerated intensity, while sometimes effective for non-emotional tones such as authority, could provoke resistance in others such as urgency, aligning with cognitive load and appraisal theories~\cite{sweller2011cognitive, mayer2003three, moors2013appraisal}. Crucially, \textit{scenario} strongly modulated these effects; effectiveness differed significantly between the shopping and Q\&A \textit{contexts}, with users preferring authoritative voices for Q\&A but more emotional characteristics for shopping. User \textit{perception} of vocal features (intonation, timbre, tone~\cite{seaborn2021voice}) clearly mediated these outcomes, linking preferences (e.g., for authority in Q\&A) to susceptibility. While \textit{user factors} such as demographics had minimal impact, individual preferences varied (e.g., by gender). Notably, these manipulations often succeeded without users' conscious detection, underscoring their subtlety. To summarize, the contributions of this paper are threefold:

\begin{itemize}
    \item We present a novel categorization of voice characteristic-based dark patterns, offering new insights into how vocal characteristics may potentially be strategically employed to manipulate user behavior.
    \item Our study (N=36) revealed the strong influence of voice characteristic manipulations on user decisions, highlighting the need for careful design in voice interfaces.
    \item We unveil the effect of contexts and causes, with normal intensity of manipulation as most effective, user perceptions as the mediator of manipulations, and provide essential guidelines for minimizing ethical risks in the design of voice characteristics.
\end{itemize}

We are committed to the responsible use and dissemination of the findings presented in this paper. We advocate for developers and regulators to take appropriate measures towards this pervasive dark pattern.

\section{Related Work}\label{sec:related_work}


Research on the potential risks and dark aspects of pervasive voice interactions has long captured the attention of the Ubiquitous computing (Ubicomp) community~\cite{kim2020interruptibility, wang2021demystifying}. We first review studies on voice characteristics, followed by a synthesis of literature on dark patterns and deceptive design.

\subsection{Effect of Voice Characteristics}

Voice interaction, a subject extensively studied across various domains including telephony, mobile applications, and in-vehicle systems~\cite{clark2019state,seaborn2021voice,kim2020interruptibility} is fundamentally shaped by its inherent characteristics beyond mere linguistic content. These characteristics encompass both vocalic (para-linguistic) properties such as tone, loudness, pitch (frequency), and timbre (voice quality)~\cite{poyatos1993paralanguage}, and prosodic (non-verbal) features such as rhythm, intonation (pitch variation), and stress (emphasis)~\cite{seaborn2021voice}. Furthermore, socio-linguistic aspects, including accent, style of speech, and the application of vocal filters, contribute significantly to the overall perception of the voice, a perspective supported by multiple studies~\cite{seaborn2021voice,seaborn2021measuring,kim2021designers}. Collectively, these vocal elements enable the expression of complex social and emotional cues, such as praise~\cite{bracken2004criticism}, empathy~\cite{james2018artificial,niculescu2013making}, and humor~\cite{niculescu2013making}, thereby profoundly influencing the user's experience.

Building upon this foundation, research has demonstrated the significant impact of specific voice attributes on user attitudes, perceptions, and trust. Formality, for instance, has been consistently linked to user trust in voice interactions~\cite{mubarak2020does,jestin2022effects}. Studies indicate that formal versus informal voice assistant styles can affect trust differently, particularly among older adults who may favor informal styles~\cite{zhong2022effects}, while informal styles can foster more positive product attitudes in commercial contexts~\cite{rhee2020effects}. However, preferences can vary, with middle-aged adults showing similar acceptance for both styles~\cite{chin2024like}. Empathy conveyed through voice significantly enhances users' feelings of intimacy, perceived similarity, and interaction ease~\cite{kim2020can}, and positively influences the evaluation of a robot's receptiveness~\cite{niculescu2013making}. Moreover, the expression of emotions, such as happiness, impacts perceived trustworthiness~\cite{knight2021influence,ceha2022expressive}, with positive speaker valence identified as a stronger predictor of trust than arousal~\cite{schirmer2019angry}.

Further investigations have explored the effects of diverse vocal strategies and content manipulations. For example, agent-delivered praise enhanced perceived niceness, although text-only praise yielded higher intelligence ratings~\cite{bracken2004criticism}. Humor has been shown to improve perceptions of task enjoyment, robot personality, and speaking style~\cite{niculescu2013making}. Interestingly, studies involving children found no significant effect of personalization (using names)~\cite{yarosh2018children}, but demonstrated that intent expressed through non-semantic ``gibberish'' was highly recognizable and sometimes more effective than other non-linguistic sounds~\cite{yilmazyildiz2015gibberish,zaga2016help}. Research by Dubiel et al.~\cite{dubiel2020persuasive} confirmed that vocal elements could impact perception, for example, debate-style voices could be perceived as more truthful. Pias et al. also explored features such as tone and age regarding persuasiveness and ethics~\cite{pias2024impact}, while Feldman et al.~\cite{feldman2024voice} noted the potential for manipulation based on voice characteristics without detailing specific deceptive patterns.

The significance of these vocal characteristics is particularly pronounced in tonal languages, where prosodic features are intrinsically linked to lexical meaning. In Mandarin Chinese, for example, extensive research has examined the acoustic correlates of its lexical tones, such as fundamental frequency ($F_0$)~\cite{jongman2006perception}. Building on this, studies have investigated how these tones interact with pragmatic intent. Ouyang et al.~\cite{ouyang2015prosody} demonstrated that corrective words are marked by longer durations and greater ranges in both $F_0$ and intensity. The practical application of voice interfaces in this context reveals further complexities. Researchers have noted that older adults encounter barriers due to limited cultural and linguistic customization, such as difficulty understanding regional dialects~\cite{qian2025exploring}. Addressing such challenges requires targeted research, from investigating the Lombard effect to improve speech recognition~\cite{liu2024production} to designing specialized dialogue systems for culturally specific data such as personal names~\cite{liesenfeld2020namespec}. Our study further investigates the correlation between specific voice styles and potential manipulative patterns by leveraging the links between vocal cues, trust and language-specific prosodic features.

\subsection{Dark Patterns}

Dark patterns were originally described as ``tricks used in websites and apps that make users do things they didn't intend to, like buying or signing up for something.'' ~\cite{brignull2018dark}. Brignull's~\cite{brignull2018dark} initial taxonomy introduced 12 types of dark patterns, including Bait-and-Switch and Confirmshaming. Conti and Sobiesk ~\cite{conti2010malicious} documented 11 categories of problematic user interfaces. Past literature has also explored various dimensions and aspects of dark patterns (Table~\ref{tbl:rw_classification}), providing the foundation for classifying voice characteristic-based dark patterns. However, their work largely focused on UI-based dark patterns.
\begin{table}[htbp]
\centering
\caption{Categories of dark patterns in previous literature.}
\label{tbl:rw_classification}
\resizebox{\textwidth}{!}{
\begin{tabular}{p{3.5cm}|ccccc ccccc ccccc ccccc c}
\toprule
\rowcolor{gray!20}
\textbf{Category} & \raisebox{-0.5ex}{\cite{brignull2018dark}} & \raisebox{-0.5ex}{\cite{gray2018dark}} & \raisebox{-0.5ex}{\cite{bosch2016tales}} & \raisebox{-0.5ex}{\cite{conti2010malicious}} & \raisebox{-0.5ex}{\cite{greenberg2014dark}} & \raisebox{-0.5ex}{\cite{10.1145/3357236.3395486}} & \raisebox{-0.5ex}{\cite{mathur2021makes}} & \raisebox{-0.5ex}{\cite{mathur2019dark}} & \raisebox{-0.5ex}{\cite{10.1145/3411764.3445467}} & \raisebox{-0.5ex}{\cite{burr2018analysis}} & \raisebox{-0.5ex}{\cite{widdicks2020backfiring}} & \raisebox{-0.5ex}{\cite{bongard2021definitely}} & \raisebox{-0.5ex}{\cite{10.1145/3532106.3533562}} & \raisebox{-0.5ex}{\cite{zagal2013dark}} & \raisebox{-0.5ex}{\cite{maier2019dark}} & \raisebox{-0.5ex}{\cite{10.1145/3479600}} & \raisebox{-0.5ex}{\cite{luguri2021shining}} & \raisebox{-0.5ex}{\cite{chatellier2019shaping}} & \raisebox{-0.5ex}{\cite{di2020ui}} & \raisebox{-0.5ex}{\cite{10.1145/3313831.3376672}} & \raisebox{-0.5ex}{\cite{10.1145/3491102.3501899}} \\
\midrule

\textbf{Emphasizing} & $\checkmark$ & $\checkmark$ &  & $\checkmark$ &  & $\checkmark$ & $\checkmark$ & $\checkmark$ & $\checkmark$ &  &  & $\checkmark$ & $\checkmark$ &  & $\checkmark$ & $\checkmark$ & $\checkmark$ & $\checkmark$ & $\checkmark$ & $\checkmark$ & $\checkmark$ \\
\hline

\textbf{Persuading} & $\checkmark$ & $\checkmark$ & $\checkmark$ & $\checkmark$ & $\checkmark$ & $\checkmark$ & $\checkmark$ & $\checkmark$ & $\checkmark$ & $\checkmark$ & $\checkmark$ & $\checkmark$ &  & $\checkmark$ & $\checkmark$ & $\checkmark$ & $\checkmark$ & $\checkmark$ & $\checkmark$ & $\checkmark$ & $\checkmark$ \\
\hline

\textbf{Weakening} & $\checkmark$ & $\checkmark$ & $\checkmark$ & $\checkmark$ & $\checkmark$ & $\checkmark$ & $\checkmark$ & $\checkmark$ &  & $\checkmark$ & $\checkmark$ & $\checkmark$ & $\checkmark$ & $\checkmark$ & $\checkmark$ & $\checkmark$ & $\checkmark$ & $\checkmark$ & $\checkmark$ & $\checkmark$ & $\checkmark$ \\
\hline

\textbf{Malicious} & $\checkmark$ & $\checkmark$ & $\checkmark$ & $\checkmark$ & $\checkmark$ & $\checkmark$ & $\checkmark$ & $\checkmark$ &  & $\checkmark$ & $\checkmark$ & $\checkmark$ & $\checkmark$ & $\checkmark$ & $\checkmark$ & $\checkmark$ & $\checkmark$ & $\checkmark$ & $\checkmark$ & $\checkmark$ & $\checkmark$ \\
\hline

\textbf{Abuse} & $\checkmark$ & $\checkmark$ & $\checkmark$ & $\checkmark$ & $\checkmark$ & $\checkmark$ & $\checkmark$ & $\checkmark$ & $\checkmark$ & $\checkmark$ &  & $\checkmark$ & $\checkmark$ & $\checkmark$ & $\checkmark$ & $\checkmark$ & $\checkmark$ & $\checkmark$ & $\checkmark$ &  &  \\
\bottomrule
\end{tabular}}
\end{table}

A few studies have explored dark patterns beyond GUIs. Kowalczyk et al.~\cite{kowalczyk2023understanding} studied dark patterns in Internet of Things (IoT) devices, finding a higher number of already identified dark patterns and new types specific to IoT devices, such as ``pay for the long term use''. Wang et al. ~\cite{wang2023dark} delved into manipulative designs in Augmented Reality (AR), finding that lighting and object interference impact participants' responses. Lacey and Caudwell ~\cite{lacey2019cuteness} explored ``cuteness'' as a dark pattern in home robot design, evoking emotional responses to collect data. Focusing on voice-based interfaces, Dula et al. ~\cite{dula2023identifying} discussed the parameters of voice as a part of ``dishonest anthropomorphism'', which could be viewed as a deceptive design feature, whereby the human-likeness of the agent is used to influence the users. Owens et al. ~\cite{owens2022exploring} sought expert opinions on potential dark patterns in voice-based interfaces, focusing on a range of problematic scenarios, which included interaction parameters of voice assistant technology and speech properties. However, they did not touch the concept of voice characteristics and mostly focused on the content delivered through the voice channel. Dubiel et al.~\cite{dubiel2024impact} suggested that voice fidelity could act as a dark pattern. However, their study did not explore and validate the broad range of voice characteristics as potential dark patterns, a gap our work addresses.

\subsection{Deceptions Around Voice Interaction}

Voice interfaces are designed for diverse interactions~\cite{kim2020interruptibility,beneteau2020assumptions}, but they can also pose threats~\cite{kim2020interruptibility}, including manipulation for deceptive purposes. Deception takes various forms and serves multiple objectives~\cite{pisanski2021efficacy}. Pisanski et al.~\cite{pisanski2021efficacy} demonstrated that cheaters could manipulate users into overestimating their height by altering vocal frequency and size. Similarly, Zhan et al.~\cite{zhan2023deceptive} highlighted deceptive design potential in ChatGPT's outputs and structures, emphasizing the need for careful interaction design. Owens et al.~\cite{owens2022exploring} identified key vocal characteristics in voice interfaces that could facilitate deceptive design patterns and explored how participants perceived these dark patterns. Dubiel et al.~\cite{dubiel2024impact} suggested that pace and pitch variations might serve as effective manipulative tactics. Feldman et al.~\cite{feldman2024voice} noted that voice can convey credibility, expertise, and charisma, influencing public opinion. However, none of these studies specifically investigated the design of dark patterns through voice characteristics, nor did they formally assess the impact of varying voice styles on deception.

Parallel research has explored how acoustic-prosodic and linguistic features can detect deception~\cite{chen2020acoustic,levitan2022believe}. Chen et al.~\cite{chen2020acoustic} found that some prosodic and lexical cues were perceived as trustworthy but were unreliable indicators of truthfulness. In contrast, Levitan et al.~\cite{levitan2022believe} identified twelve acoustic features that influenced perceptions of trust and mistrust. However, their work did not explore how these voice characteristics might be synthesized into more convincing or deceptive lies, nor did it categorize these features as manipulative patterns.

\section{Formulation of Dark Patterns through Voice Characteristics}\label{sec:design}

Investigating how vocal characteristics enable manipulative dark patterns requires a structured approach. This section presents our methodology, beginning with a conceptual framework (Section~\ref{sec:framework}) that integrates key dimensions: scenario, commercial aim, and voice characteristics. Directly operationalizing this framework, we first scoped the scenarios to carry out the deception (Section~\ref{sec:scenario}), then categorized and designed the commercial aim for the deceptive patterns (Section~\ref{sec:manipulative}). We finally designed how voice characteristics could be leveraged and selected to manipulate users.

\subsection{Conceptual Framework}~\label{sec:framework}

\begin{figure}[!htbp]
    \centering
    \includegraphics[width=1.0\linewidth]{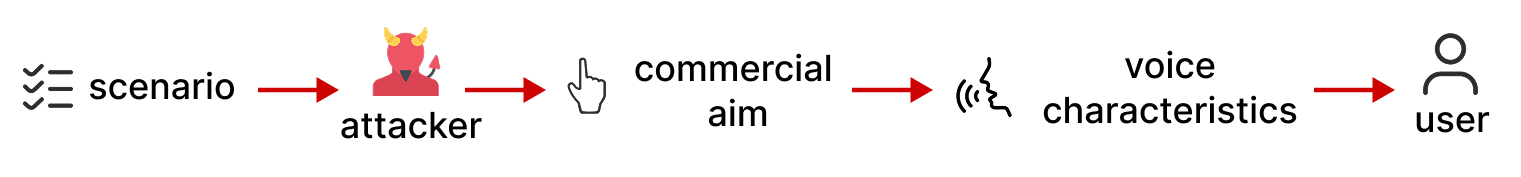}
    \caption{Conceptual threat model illustrating the manipulation process based on scenario, voice characteristics, commercial aims and influenced by user-related factors.}
    \label{fig:factor}
\end{figure}

To structurally understand voice-based manipulation, we synthesize insights from prior established frameworks on voice interaction~\cite{kreiman1993perceptual} into an integrated conceptual framework (Figure~\ref{fig:factor}). This framework outlines the core process of voice-based manipulation, illustrating how specific voice characteristics can be employed as manipulative dark patterns through a sequence of key elements and their interactions.

The manipulative process begins within a specific scenario~\cite{rhee2020effects, huh2023building}, which provides the context, shapes user expectations, and presents opportunities for an attacker. Driven by a commercial aim~\cite{gray2021end}, the attacker designs and deploys particular voice characteristics as the manipulative stimulus. These voice characteristics encompass key vocalic and prosodic properties of speech that are intentionally modulated, such as pitch, speed, and tone~\cite{seaborn2021voice, mullennix2003social}. Subsequently, these characteristics are perceived by the user, whose interpretation and behavioral response are modulated by individual differences. These differences stem from a range of user factors, including demographics (age, gender, and background) and individual preferences~\cite{zhong2022effects, cambre2019one, chin2024like}. The interplay of all these preceding elements in the chain culminates in a manipulation effect on user behavior.

Guided by this framework, we examine how manipulated voice characteristics, within specific scenarios and driven by particular commercial aims, influence user behavior (the manipulation effect). This structured approach allows us to systematically test hypotheses regarding the mechanisms of such voice-based manipulation and, in parallel, how its effectiveness is shaped by context, intensity, and variations stemming from user factors.


\subsection{Manipulation Scenarios}\label{sec:scenario}

The effectiveness of voice-based manipulation is highly dependent on the scenario. We focus our investigation on two prevalent VUI scenarios due to their commonality and distinct characteristics.  

\textbf{Shopping Scenarios:} These interactions are often characterized by subjective user preferences, potentially complex multi-step decision processes (e.g., product comparison, selection, purchase confirmation, post-purchase feedback), and direct financial implications. The goal is often commercially driven.

\textbf{Question-Answering Scenarios:} These interactions involve users seeking factual information or assessing the credibility of presented content. Decision-making here leans more towards objective evaluation and trust assessment, making it relevant for understanding manipulation related to misinformation or belief shaping.

While voice-based dark patterns are not limited to these two contexts, shopping and question-answering represent common VUI applications where manipulation can have significant consequences~\cite{seaborn2021voice}.

\subsection{Commercial Aims and User Vulnerabilities}\label{sec:manipulative}

Within the defined scenarios, the attacker pursues specific \textbf{commercial aims} by exploiting inherent user vulnerabilities. Building on dark pattern taxonomies~\cite{mathur2021makes}, we identify three core aims relevant to voice characteristics manipulation, linked to underlying mechanisms:

\textbf{Aim: Emphasizing (Misleading Mechanism)}: To induce the user to select or favor a particular option predetermined by the attacker.

\textbf{Aim: Persuading (Manipulative Mechanism)}: To prompt the user to make decisions more quickly or with less critical evaluation than they might otherwise.

\textbf{Aim: Weakening (Steer Mechanism)}: To guide the user to overlook, disregard, or devalue certain information, options, or subsequent actions (like returns).

The feasibility of exploiting voice characteristics arises from their unique features, distinct from traditional UIs~\cite{mathur2021makes} or voice patterns~\cite{owens2022exploring}. These features create user vulnerabilities and can be framed through a user-device-context model~\cite{cambre2019one}, informing the design of specific dark patterns.

\textbf{Exploiting emotional and cognitive responses:} voice characteristics directly influence user perception through vocal cues (tone, pitch, rhythm) that are often processed subconsciously as social signals, potentially bypassing rational thought~\cite{cambre2019one}. This allows attackers to target emotions or cognitive heuristics, especially in engagement-driven scenarios such as shopping.

\textbf{Leveraging interaction dynamics:} The typical turn-based, often unidirectional nature of VUI interaction limits users' ability for simultaneous comparison or reflection. The perceived social presence can also lower critical scrutiny~\cite{cambre2019one}. This structure facilitates the imposition of manipulative cues with reduced resistance.

\textbf{Utilizing subtlety:} Voice characteristics permit subtle, real-time modulations (pace, pitch, volume) that can evoke specific responses without necessarily triggering conscious awareness or skepticism~\cite{cambre2019one}. This enables covert influence on perceptions such as urgency or confidence.

When constructing voice characteristic-based dark patterns, we first reviewed existing dark pattern classifications  (Table~\ref{tbl:rw_classification}). Inspired by prior work ~\cite{brignull2018dark,gray2023towards,owens2022exploring,mathur2021makes}, we initially abstracted the potential manipulative intents behind these patterns into five high-level categories: Emphasizing, Persuading, Weakening, Malicious, and Abuse. Considering our focus on instantaneous manipulation via voice characteristics, we evaluated these five categories. The Malicious and Abuse categories were excluded as they rely on content deception or long-term interactions, not immediate vocal changes. We therefore concentrate on Emphasizing, Persuading, and Weakening, as these intents align with achieving instantaneous influence through voice modulation. Social psychological theories regarding attention steering~\cite{10.1145/3386867}, emotional contagion~\cite{hatfield1993emotional, efthymiou2024empathy}, and cognitive load~\cite{sweller2011cognitive} provide theoretical support for Emphasizing, Persuading, and Weakening mechanisms separately. Building upon this focused selection and its theoretical basis, Table~\ref{tbl:purpose} exemplifies these three intents in attack scenarios.


\begin{table}[!htbp] 
    \centering
    \caption{Manifestation of commercial aims in different scenarios.}
    \label{tbl:purpose}
    \vspace{-1ex}
    \begin{tabular}{m{2cm}|m{4cm}|m{4cm}|m{4cm}} 
    \toprule
    \textbf{Manipulative pattern} & \textbf{Purpose} &\textbf{Shopping scenario} & \textbf{Question-answering scenario} \\
    \midrule
    Emphasizing & Inducing users to select a particular option (Misleading) & Pre-purchase: guide product choice & Guide choice towards specific (false) options \\
    \hline
    Persuading & Prompting users to make decisions quickly (Manipulative) & Purchase: speed up placing orders & Encourage trust in (false) beliefs \\
    \hline
    Weakening & Guiding users to overlook certain information (Steer) & Post-purchase: discourage bad reviews/returns & Bias opinions about specific information \\
    \bottomrule
    \end{tabular}
    \vspace{-2ex}
\end{table}

\subsection{Voice Characteristics to Instantiate Manipulation}\label{sec:voice_characteristic}

The practical implementation of these commercial aims involves the attacker strategically selecting and synthesizing specific voice characteristics. The selection is guided by social and psychological theories linking vocal attributes to perceptual outcomes such as trust, authority, urgency, or friendliness. For instance, authority has long been recognized for its manipulative effects, influencing users' social judgments and decision-making processes~\cite{kruglanski2005says}. Similarly, urgency has been shown to trigger irrational and rushed behavior~\cite{billieux2010role}, while expressions of hesitation, friendliness, and emotion have been explored for enhancing customer support~\cite{zhang2011effects}. 

Table~\ref{tbl:voice} details potential voice characteristic profiles designed to achieve the specific commercial aims. Each voice characteristic profile denotes a specific voice characteristic targeted for the specific commercial aim.

\begin{table}[!htbp]
    \caption{Voice Characteristic Profiles for Attack Implementation.}
    \vspace{-1ex}
    \label{tbl:voice}
    \centering
    \resizebox{\textwidth}{!}{ 
    \begin{tabular}{p{1.7cm}p{1.9cm}p{2.5cm}p{1.4cm}p{1.4cm}p{1.4cm}p{2.1cm}}
    \toprule
     \textbf{Commercial Aim} & \textbf{Mechanism} & \textbf{Voice Characteristics Profile} & \textbf{Pace} & \textbf{Pitch} & \textbf{Tone} & \textbf{Pauses} \\
    \midrule
    \multirow{3}{*}{Emphasizing} 
    & \multirow{3}{*}{Misleading} 
    & Authoritative 
    & Slow 
    & Low 
    & Low 
    & Few well-timed pauses \\ \cline{3-7}
    & & Friendly
    & Moderate 
    & Higher 
    & Moderate 
    & Relaxed natural pauses \\ \cline{3-7}
    & & Emotional 
    & Moderate 
    & Variable 
    & High 
    & Strategically placed pauses  \\
    \hline
    \multirow{3}{*}{Persuading} 
    & \multirow{3}{*}{Manipulative} 
    & Urgency
    & Fast 
    & Slightly high 
    & Increased 
    & Short pauses  \\ \cline{3-7}
    & & Emotional 
    & Moderate 
    & Variable 
    & High 
    & Strategically placed pauses \\ \cline{3-7}
    & & Friendly
    & Moderate 
    & Higher 
    & Moderate 
    & Relaxed natural pauses \\
    \hline
    \multirow{3}{*}{Weakening} 
    & \multirow{3}{*}{Steer} 
    & Friendly
    & Moderate 
    & Higher 
    & Moderate 
    & Relaxed natural pauses \\ \cline{3-7}
    & & Emotional
    & Moderate 
    & Variable 
    & High 
    & Strategically placed pauses \\ \cline{3-7}
    & & Uncertainty
    & Slow 
    & Medium 
    & Moderate 
    & Frequent pauses  \\
    \bottomrule
    \end{tabular}
    }
    \vspace{-2ex}
\end{table}

Given that each commercial aim could potentially be achieved via multiple voice profiles, we included several viable candidates (e.g., emphasizing via authority, friendliness, or emotion) to examine potential nuances. It is important to note that the study aims to demonstrate the feasibility of using voice characteristics as dark patterns. Consequently, this formulation focuses on testing representative candidates rather than providing an exhaustive catalog of all possible voice-based dark patterns and characteristics. The rationale for selecting specific profiles includes:

\textbf{False Urgency (denoted urgency):} fast pace and high pitch are shown to induce rushed decisions~\cite{kobayashi2022acoustic,landesberger2020urgent}.

\textbf{Overly Friendly (denoted friendly):} high pitch and moderate tone are linked to intimacy and trust~\cite{pal2023affects,higgins2022sympathy}.

\textbf{Misleading Authority (denoted authoritative):} slow pace and deep pitch are associated with credibility~\cite{van2009authority,herder2023context}.

\textbf{Emotional Manipulation (denoted emotional):} variable pitch and high tone are shown to influence purchase intent~\cite{guo2020positive,fahim2021integral}. 

\textbf{Hesitation and Uncertainty (denoted uncertain):} slow pace and frequent pauses are linked to lower perceived competence~\cite{goupil2021listeners}.

\section{Construction of Manipulative Voice Characteristics}\label{sec:construction}
This section outlines the process for selecting, constructing, and validating manipulative voice characteristics, considering established frameworks~\cite{bhuta2004perceptual} and empirical studies~\cite{dubiel2024impact}, but empirically tuned to the Chinese language context to ensure the quality of synthesized voices. 

\subsection{Determining Parameters of Voice Characteristics}

We relied on existing research to develop the manipulative voice characteristics used in this study, ensuring their ability to influence user perception and behavior:

$\bullet$ \textbf{Urgent Voice}: designed to evoke a sense of immediacy, this voice characteristic encourages users to act quickly. \textit{It is marked by a rapid speech rate, an elevated pitch and increased volume.} This design is informed by Edworthy et al.~\cite{edworthy1991improving}, who demonstrated the role of urgency in communication.

$\bullet$ \textbf{Friendly Voice}: aimed at creating a warm and approachable atmosphere through \textit{a moderate speech rate, a slightly elevated pitch, and a soft, welcoming tone}, particularly effective in fostering rapport and increasing user receptivity~\cite{van2020voice}.

$\bullet$ \textbf{Authoritative Voice}: intended to convey trust, \textit{which features a firm, steady speech rate, a low pitch, and clear enunciation}, all of which have been shown to promote trust in the speaker's expertise~\cite{hosbach2009constitutes}.

$\bullet$ \textbf{Emotional Voice}: crafted to elicit strong emotional reactions, \textit{varied in pitch and tone to convey a range of emotions, including joy and sympathy}. Such emotional manipulation has been linked to increased user engagement and action, particularly in contexts involving persuasion and decision-making~\cite{chasaide2001voice}.

$\bullet$  \textbf{Uncertain Voice}: introduced hesitation through \textit{frequent pauses and fluctuated expressions} to convey the impression of uncertainty~\cite{jiang2017sound,goupil2021listeners}. 


\subsection{Synthesizing Voice Characteristics}

For synthesis, we selected a method representing commercial voice assistants. Among multiple voice conversion engines, we finally used the engine provided by Xunfei\footnote{\url{https://peiyin.xunfei.cn/make}} with over 3.8 billion users. When constructing the manipulative voice characteristics, besides the inherent voice characteristics provided by Xunfei, we also manipulated the pitch and speed, enhancing the realism and impact of the voice patterns. We constructed two intensities of the voice characteristics--normal and exaggerated--by changing the voice characteristics, pitch, and speed~\cite{gobl201011}. To provide a quantitative basis for our stimuli, we analyzed the synthesized audio to extract key acoustic and spectral features. Table~\ref{tab:core_audio_features} (and  Appendix~\ref{sec:audio_feature}) detail these metrics and their calculation, illustrating the objective differences in properties such as fundamental frequency ($F_0$), speech rate, and pause ratio across the different voice categories and intensity levels.

\begin{table}[htbp]
\centering
\caption{Core acoustic and spectral features extracted from audio samples.}
\label{tab:core_audio_features}
\resizebox{\textwidth}{!}{%
\begin{tabular}{@{}llrrrrrrrrrr@{}}
\toprule
\textbf{Category} & \textbf{Intensity} & \textbf{Duration} & \textbf{Mean F0 (SD)} & \textbf{Mean Int.} & \textbf{Speech Rate} & \textbf{Pause Ratio} & \textbf{Jitter} & \textbf{Shimmer} & \textbf{Mean HNR} & \textbf{Spec Centroid} & \textbf{Spec Band} \\
 & & (s) & (Hz) & (dB) & (Onsets/s) & (\%) & (\%) & (\%) & (dB) & (Hz) & (Hz) \\
\midrule
Uncertain & Normal & 607.6 & 241.7 (47.6) & 68.7 & 3.9 & 37.9 & 0.8 & 7.4 & 16.7 & 2036.8 & 1872.0 \\
Uncertain & Exaggerated & 672.2 & 225.0 (45.8) & 68.3 & 3.6 & 37.3 & 0.8 & 6.9 & 17.6 & 2009.1 & 1819.4 \\
Urgency & Normal & 566.8 & 246.2 (45.6) & 71.0 & 4.4 & 33.3 & 0.9 & 8.3 & 15.5 & 1889.3 & 1642.1 \\
Urgency & Exaggerated & 501.7 & 295.2 (43.8) & 61.4 & 4.2 & 36.8 & 0.7 & 10.2 & 14.3 & 1711.4 & 1449.9 \\
Authoritative & Normal & 520.8 & 238.0 (42.8) & 69.2 & 4.0 & 32.4 & 0.9 & 10.6 & 11.7 & 2378.4 & 2041.1 \\
Authoritative & Exaggerated & 583.2 & 222.0 (40.1) & 68.9 & 3.9 & 32.8 & 0.9 & 10.0 & 12.6 & 2213.5 & 1932.9 \\
Emotional & Normal & 477.6 & 323.2 (51.0) & 69.5 & 4.0 & 31.9 & 0.6 & 9.0 & 12.1 & 2405.0 & 2067.2 \\
Emotional & Exaggerated & 577.6 & 293.3 (46.7) & 69.4 & 3.8 & 31.2 & 0.6 & 10.0 & 13.3 & 2424.1 & 2048.8 \\
Friendly & Normal & 605.1 & 225.3 (44.7) & 68.3 & 3.9 & 34.9 & 0.8 & 7.4 & 17.2 & 2041.0 & 1875.2 \\
Friendly & Exaggerated & 539.9 & 258.5 (50.2) & 68.5 & 4.3 & 34.8 & 0.8 & 7.5 & 16.7 & 2136.1 & 1937.9 \\
Neutral & Neutral & 572.8 & 248.4 (38.6) & 67.9 & 4.6 & 35.8 & 0.8 & 7.6 & 16.9 & 2194.5 & 1899.1 \\
\bottomrule
\end{tabular}%
}
\end{table}

\subsection{Validating Manipulative Voice Characteristics}\label{sec:vali}

We conducted a study to validate the voice characteristics' construction following previous practices~\cite{dubiel2024impact}. The IRB-approved study\footnote{approval number: THU-03-2025-0002.} recruited 24 participants (13 males, 11 females, with a mean age of 23.3, SD=3.2) through snowball sampling~\cite{goodman1961snowball}. All participants were familiar with voice interactions and each participant was compensated 75 RMB according to the local wage standard. We selected materials from Section~\ref{sec:experiment_material} and used the same materials as in Section~\ref{sec:experiment_design}. As this study involved 2 intensities $\times$ 5 manipulative characteristics and 1 neutral voice characteristic, each participant needed to hear 2 intensities $\times$ 6 voice characteristics, with the dual aim of first distinguishing the 6 voice characteristics and then determining the 2 intensities. The experiment involved 3 aims $\times$ 2 intensities $\times$ 6 voice characteristics for each participant to balance experiment time (about 20 minutes) and avoid random effects caused by textual material. Participants listened to the audio materials one by one for assessment.

Participants listened to each audio recording without accompanying text and assessed the voice based on the following aspects: the characteristics (selecting from neutral, friendly, emotional, uncertain, authoritative, and urgent), the intensity (selecting from normal and exaggerated), and Mean Opinion Score (MOS) ~\cite{lewis2018investigating}. The MOS (on a 0-10 scale) for voice assessment focused on four aspects: intelligibility, naturalness, prosody, and social impression.

\begin{figure}[!htbp]
    \centering
    \subfloat[Between Characteristics.]{
        \includegraphics[width=0.40\textwidth]{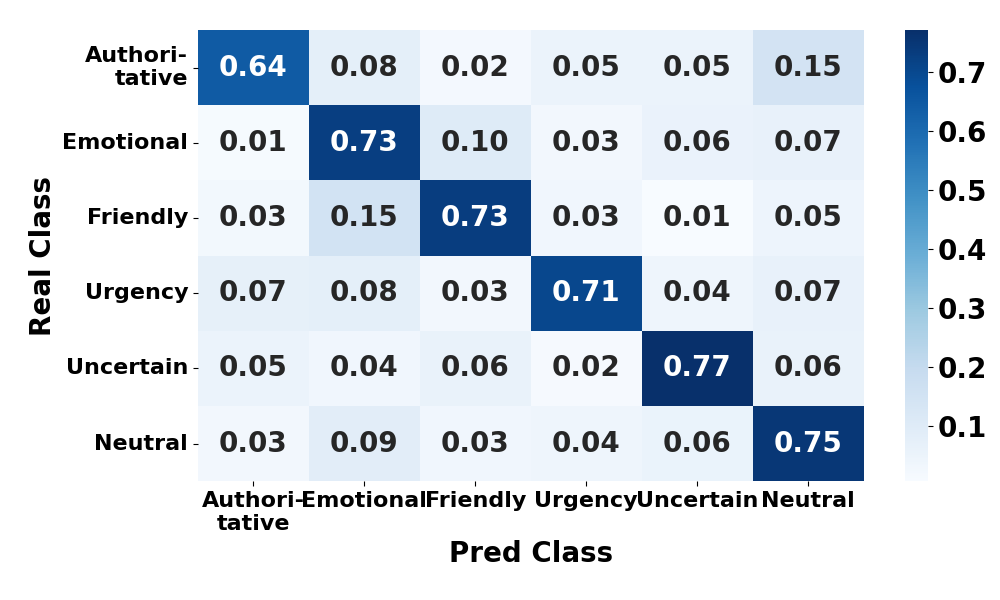}
        \label{fig:characteristic}
    }
    \subfloat[Between Levels.]{
        \includegraphics[width=0.40\textwidth]{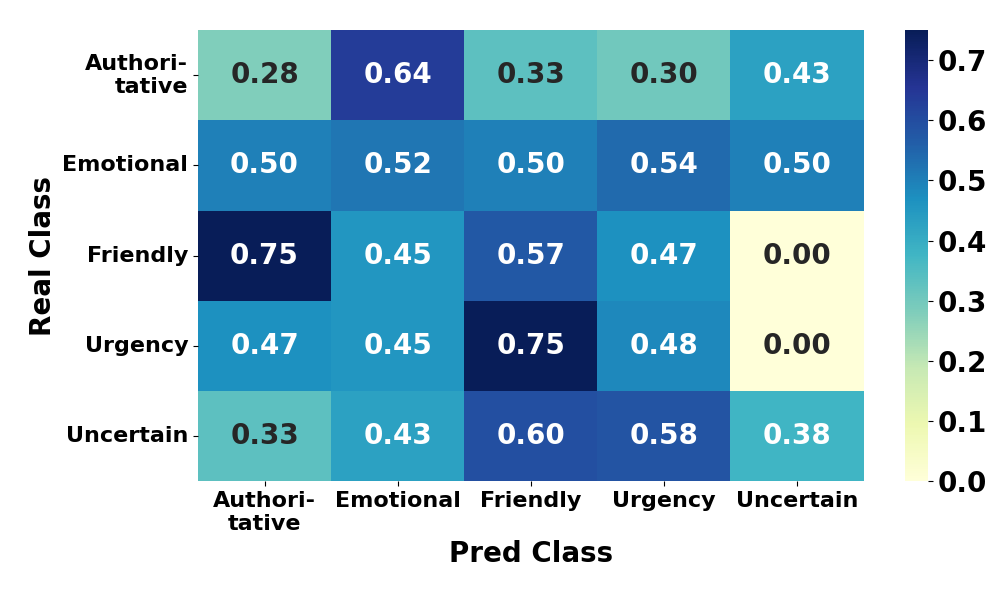}
        \label{fig:level}
    }
    \vspace{-1ex}
    \subfloat[Subjective Ratings.]{
        \includegraphics[width=0.80\textwidth]{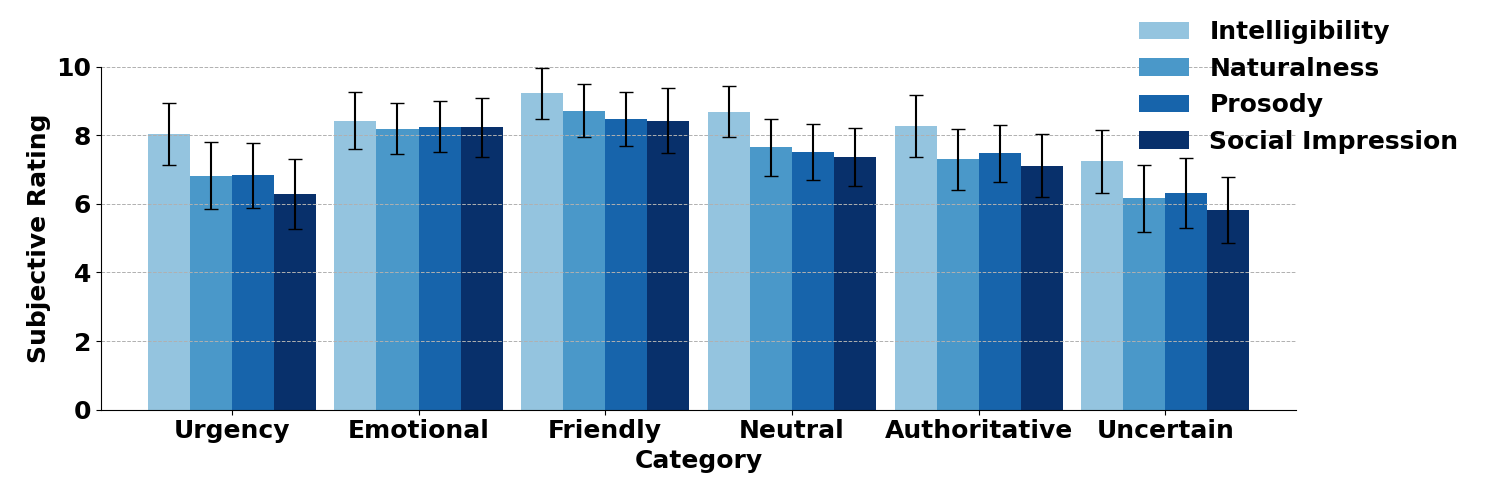}
        \label{fig:subjective_mos}
    }
    \vspace{-2ex}
    \caption{(a) Heatmap showing the ratio of correctly classified samples for each real class (real class) and predicted class (pred class), (b) Heatmap of accuracy for distinguishing different levels, where each value represents the probability of correct classification for the specific real (real class) and predicted class pair (pred class). (c) Subjective MOS ratings, with error bars denoting one standard deviation.}
    \label{fig:study1}
\end{figure}

Results showed taht participants could distinguish different voice characteristics with high accuracy (Figure~\ref{fig:characteristic}), achieving 72.2\% overall accuracy. Uncertain and Neutral characteristics were recognized with 77.2\% and 75.0\% accuracy. Distinguishing levels was harder, with an average accuracy of only 55.7\% (Figure~\ref{fig:level}), indicating that participants were largely insensitive to voice level variations. This supports previous findings~\cite{dubiel2024impact} and suggests that participants may still be influenced even without perceiving these levels (see Section~\ref{sec:manipulative results}). Additionally, participants overall praised the voice materials for their high quality, resulting in MOS scores of over 5 across different dimensions. Friendly and emotional voice characteristics were mostly preferred for their intelligibility and naturalness, with uncertain and urgency voice characteristics scoring relatively lower from the perspective of social influence and prosody. The rating dimensions were all higher for intelligibility and naturalness than for prosody and social influence, probably because all materials were generated using a voice synthesis engine, where AI-powered voice synthesis would enhance the anthropomorphism of the voice and its naturalness. Overall, users' responses proved the success of the voice materials for deployment and Section~\ref{sec:study_setup}.

\section{Experiment Setup}\label{sec:study_setup}

This section analyzes effectiveness variations by examining four key aspects derived from our conceptual framework (Figure~\ref{fig:factor}): voice characteristics (Sections~\ref{sec:manipulative results} and~\ref{sec:analysis}), scenarios (Section~\ref{sec:context}), manipulative aims (commercial aims) (Sections~\ref{sec:manipulative results} and~\ref{sec:analysis}), and users' demographics (Section~\ref{sec:variance}). Investigating these framework components allows a structured analysis of how they interact to determine manipulation success, addressing the core question of effectiveness variability (RQ2).

\subsection{Participants and Apparatus}

The IRB-approved study recruited 36 Chinese participants (14 males, 22 females) with a mean age of 22.6 (SD=1.8) through snowball sampling~\cite{goodman1961snowball}. The sample size was pre-determined based on a power analysis, which indicated that the resulting effect sizes would be $d = 0.3 \sim 0.35$ for main effects and $d = 0.25 \sim 0.3$ for interaction effects. All participants were familiar with voice interactions. One participant had a background in security and privacy, while four had backgrounds in computer science. Each participant was compensated 100 RMB according to the local wage standard. We developed an online shopping website to facilitate the study, and a pilot test with 10 users confirmed its functionality. Furthermore, feedback gathered during the pilot test indicated that the website's visual elements were clear, straightforward, and did not interfere with participants' focus on evaluating the auditory stimuli.

\subsection{Experiment Material Selection}\label{sec:experiment_material}

To ensure diverse and realistic materials, we leveraged data from established datasets. \textbf{For the shopping scenario}, we sampled dialogues from the E-commerce Dialogue Corpus (ECD)~\cite{zhang2018modeling}, Jing Dong Dialogue Corpus (JDDC)~\cite{chen2019jddc}, and Robocall Audio~\cite{robocallDatasetTechReport}. These datasets are widely used in voice shopping research~\cite{liu2023conversational} and align with the phase-based classification of dark patterns: pre-purchase, purchase and post-purchase (Table~\ref{tbl:purpose}). We randomly sampled data from these sources to ensure diversity and quality. \textbf{For the question-answering scenario}, we employed a combination of datasets covering diverse domains: LIAR~\cite{wang2017liar} (political), Climate-fever~\cite{diggelmann2020climate} (environmental) and Medical~\cite{pal2022medmcqa} (health), mapping them to emphasizing, persuading and weakening aims respectively. All datasets are potentially associated with impactful misinformation.

We selected diverse content to minimize material bias. For shopping, we chose varied top products from Amazon\footnote{\url{https://www.amazon.com/shopping/}} in categories suitable for voice interaction (e.g., low substitutability, high frequency), consistent with prior work. For question-answering, we used diverse topics (e.g., environmental issues, health facts). We verified shopping content authenticity and question-answering content factual inaccuracy. We minimally adjusted texts for vocal alignment to maintain consistency and ecological validity~\cite{lee2011can,afshan2022speaker} and developed appropriate contextual descriptions based on prior research~\cite{huh2023building,rzepka2023voice}.

We employed Xunfei\footnote{\url{https://peiyin.xunfei.cn/make}, which supports varied speaker styles, laughter and intonation modulation, and is controllable via code, API, or WebUI.} voice synthesis engine to generate all voice samples~\cite{sun2018lip}. We generated manipulated statements using the target voice characteristics, while the neutral voice mimicked typical commercial assistants (e.g., XiaoAi\footnote{\url{https://xiaoai.mi.com/}}, XiaoDu\footnote{\url{https://global.dumall.com/}}). A pilot study successfully validated both the voice characteristics and experimental materials (Section~\ref{sec:experiment_material}). 

\subsection{Experiment Design}\label{sec:experiment_design}

We adopted a within-subjects design with \textbf{voice characteristics}, \textbf{manipulative patterns}, \textbf{scenarios}, and \textbf{manipulative degree} as within-subject factors (Figure~\ref{fig:experiment_design}). Based on previous research on voice~\cite{hodari2020perception} and facial expressions~\cite{hyde2013perceptual}, we defined three levels for each voice characteristic: no manipulation (neutral, level 1), normal manipulation (level 2), and exaggerated manipulation (level 3). The mapping of manipulative patterns, voice characteristics and scenarios followed Table~\ref{tbl:purpose} and~\ref{tbl:voice}. The experimental tasks mirrored these mappings: in the shopping scenario, tasks involved product recommendations (emphasizing), purchase prompts (persuading), and dissuading returns/negative reviews (weakening); in the question-answering scenario, tasks involved selecting webpage information (emphasizing), assessing information trustworthiness (persuading), and evaluating review credibility (weakening) (Table~\ref{tbl:purpose}, Section~\ref{sec:experiment_material}). Following established practices~\cite{bongard2021definitely}, our measurements included:

\begin{figure}
    \centering
    \includegraphics[width=0.8\textwidth]
    {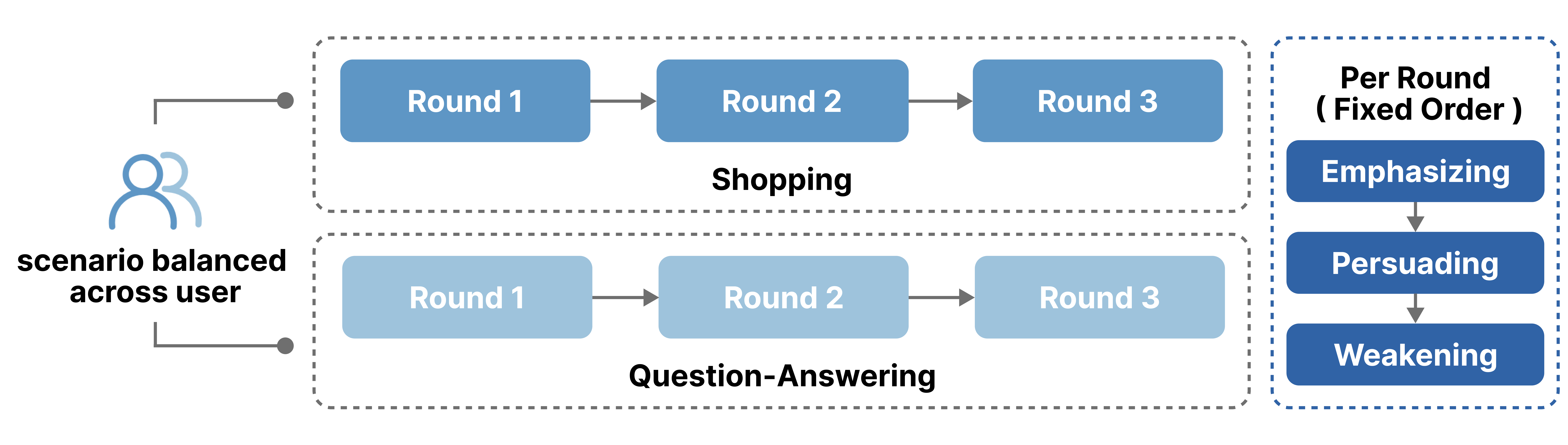}
    \vspace{-2ex}
    \caption{The experimental design, where scenarios and manipulative degrees were counter-balanced while the manipulative patterns' order was fixed.}
    \label{fig:experiment_design}
\end{figure}

\noindent \textbf{1. Demographics}: We asked participants about their gender, age, and professional backgrounds~\cite{bongard2021definitely}.

\noindent \textbf{2. System Usability}: We used System Usability Scale (SUS) to assess the experimental platform's usability.

\noindent \textbf{3. Manipulation Effect}: We assessed the dark patterns' impact by analyzing user behavior. For the ``emphasizing'' aim, we calculated the selection ratio of the targeted option. For ``persuading'' and ``weakening'' aims, we measured the acceptance rate and trust polarization using seven-point Likert scales, comparing manipulated conditions against the baseline.

We conducted retrospective post-experiment interviews following established protocols~\cite{russell2014looking}. Participants reviewed recordings of their sessions and responded to questions regarding their voice perception, system trust, and overall attitudes. We specifically assessed perceived voice naturalness and intelligibility, critical factors in human-voice assistant interaction \cite{viswanathan2005measuring, lewis2018investigating}.

\begin{figure}
    \centering
    \subfloat[Emphasizing.]{
        \includegraphics[page=2,width=0.45\textwidth]{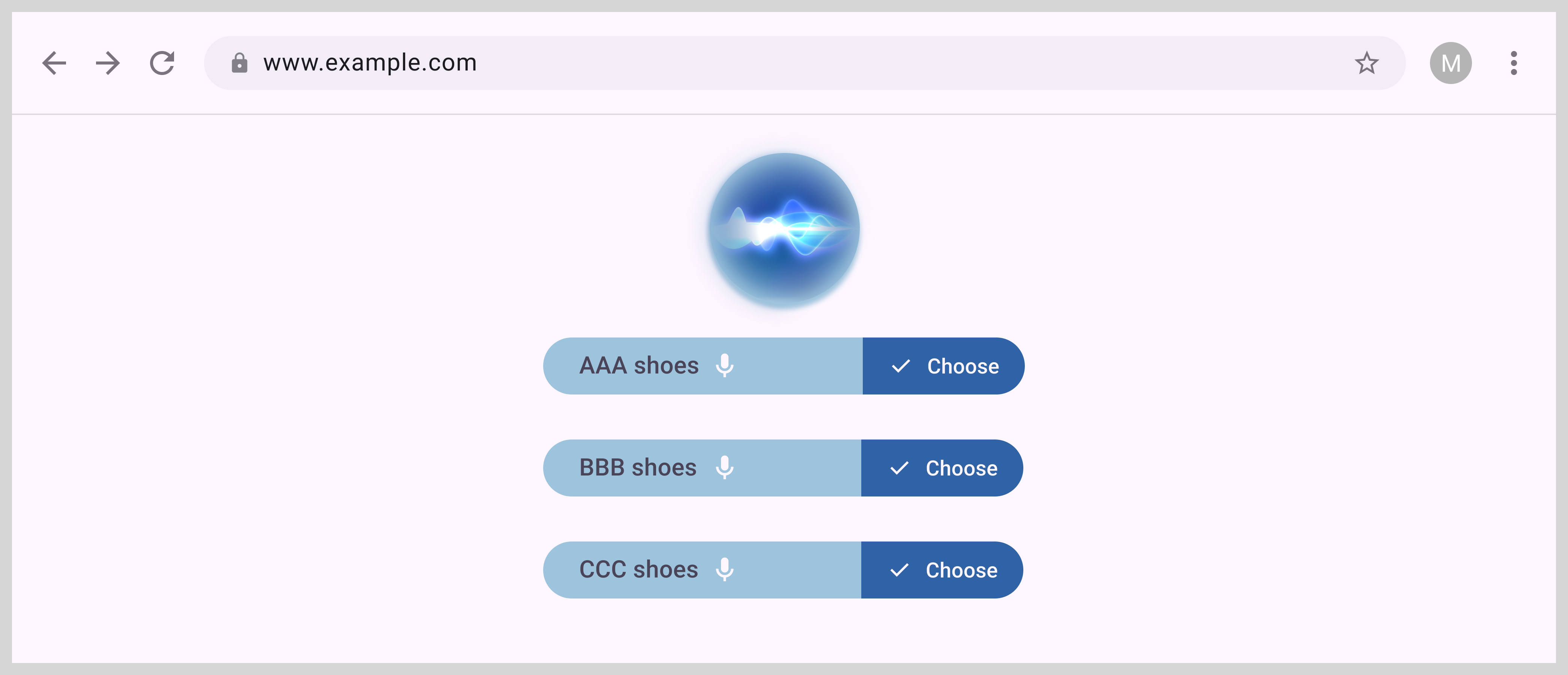}
    }
    \subfloat[Persuading and weakening.]{
        \includegraphics[page=1,width=0.45\textwidth]{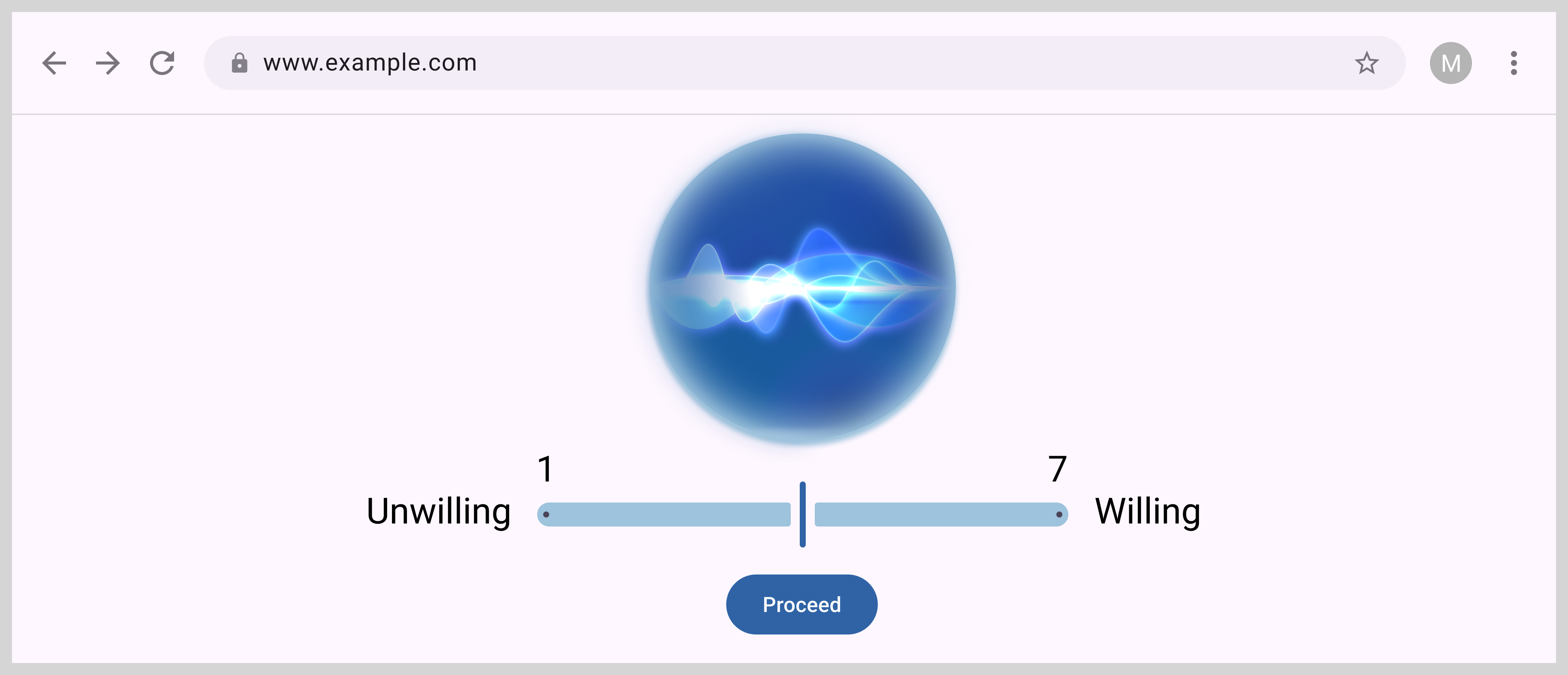}
    }
    \vspace{-2ex}
    \caption{Experimental interfaces: (a) interface for the emphasizing aim; (b) interface for the persuading and weakening aims. The voice would be automatically played and users need to (a) click the choice or (b) drag the slide bar to indicate their preferences.}
    \label{fig:interface}
\end{figure}

\subsection{Procedure}

\begin{figure}
    \vspace{-2ex}\includegraphics[width=0.9\textwidth]{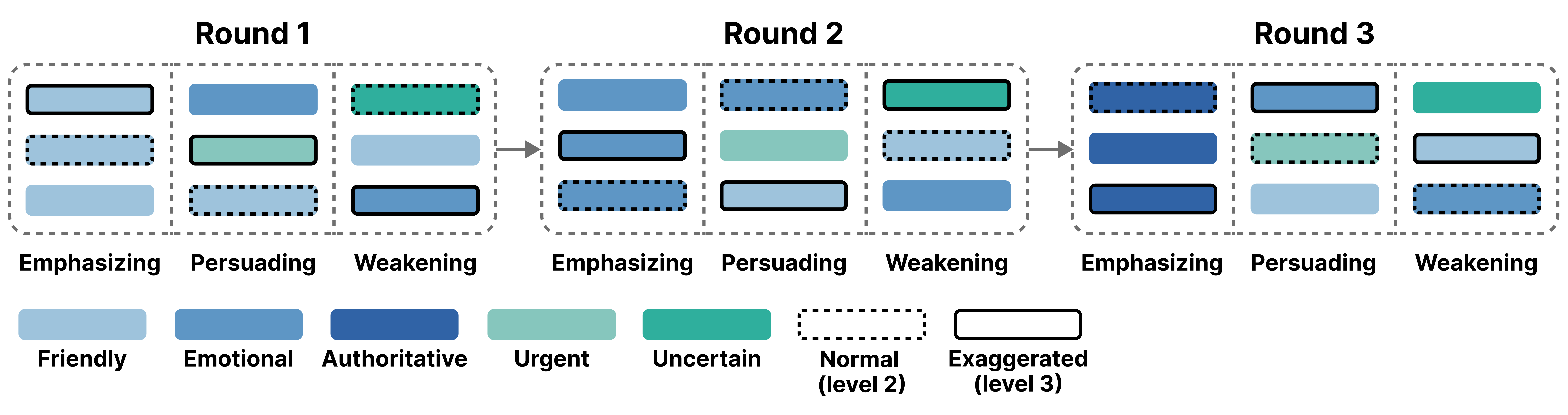}
    \vspace{-2ex}
    \caption{The experiment procedure, where users underwent three rounds of experiment differed by the content, and the experiment materials were counter-balanced within and across users.}
    \label{fig:balance}
    \vspace{-2ex}
\end{figure}

We conducted an online study involving two scenarios: shopping and question-answering. In both scenarios, participants complete the tasks via a deployed website (Figure~\ref{fig:interface}). The website mimicked real voice shopping and question-answering scenarios, except that participants needed to click to proceed with the buying or selecting trusted information in order to clearly indicate the willingness. We briefly introduced the platform to the participants and gave them five minutes to familiarize themselves with the site. Subsequently, they completed the tasks of shopping and question-answering scenarios in a counterbalanced order. Each task simulated real-world processes with embedded manipulative patterns and involved three rounds of interaction. In each round, participants encountered three distinct manipulative patterns—emphasizing, persuading, and weakening. The assignment of voice characteristics and intensities was counter-balanced across rounds to mitigate order effects (see Figure~\ref{fig:balance}). 

Each scenario required approximately 15 minutes to complete. Throughout the tasks, the system recorded all participant interactions (clicks indicating willingness) for analysis. Upon completing both scenarios, participants filled out a post-task questionnaire and participated in a 30-minute audio-recorded retrospective interview. Finally, we conducted a debriefing session, explaining the study's true objectives and alerting participants to the potential existence of manipulative voice patterns. We informed participants beforehand of their right to withdraw at any time, request data deletion, and receive compensation regardless of completion status. No participants withdrew from the study.


\subsection{Analysis Process}


We performed statistical analysis on the quantitative data using a Mixed Linear Regression model to assess the effects of various influencing factors. We used Tukey HSD for post-hoc comparisons with a Bonferroni correction, setting statistical significance at $p < .05$. For the qualitative data, we used thematic coding~\cite{braun2012thematic}. Initially, two experimenters performed axial coding~\cite{williams2019art} to generate high-level codes based on an established voice perception taxonomy (including prosody, speed, pitch~\cite{seaborn2021voice}). Subsequently, four primary authors coded the retrospective interview videos and recorded the frequency of each theme's occurrence.

\section{Results}

We first analyzed the usability of the two scenarios. Participants rated a total of 73.15 out of 100.0 for SUS questionnaire for voice shopping and question-answering on average, which indicated a good quality. All participants at least rated the system quality as OK, proving the successful implementation and its usability. In the next sections, we first examine the effectiveness based on voice characteristics and intensity (Sections~\ref{sec:manipulative results} and~\ref{sec:analysis}). Subsequently, we investigate variations across different contexts (Section~\ref{sec:context}), and consider user factors (Section~\ref{sec:variance}), following the structure outlined in Figure~\ref{fig:factor}. This systematic approach ensures a comprehensive analysis of the variations central to RQ2.

\subsection{Success of Manipulation For Different Voice Signals}\label{sec:manipulative results}

\begin{figure}[!htbp]
    \subfloat[Emphasizing.]{
        \includegraphics[width=0.33\textwidth]{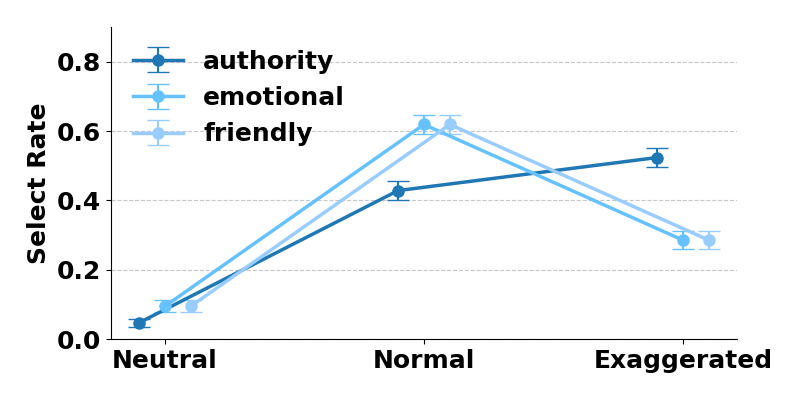}
        \label{fig:emp_shop}
    }
    \subfloat[Persuading.]{
        \includegraphics[width=0.33\textwidth]{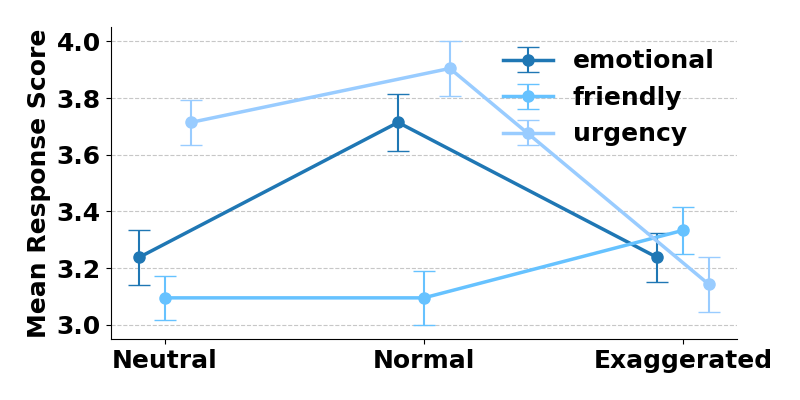}
        \label{fig:per_shop}
    }
    \subfloat[Weakening.]{
        \includegraphics[width=0.33\textwidth]{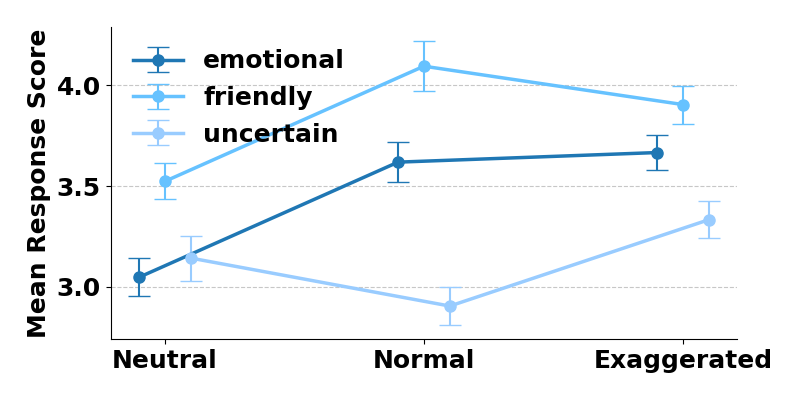}
        \label{fig:weak_shop}
    }

    \subfloat[Emphasizing.]{
        \includegraphics[width=0.33\textwidth]{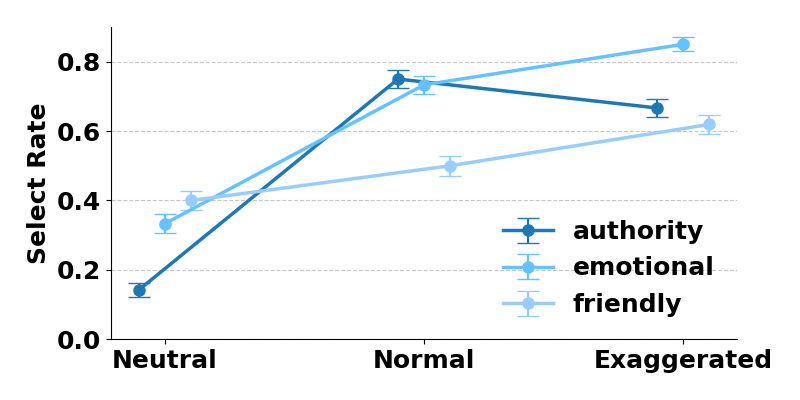}
        \label{fig:emp_qa}
    }
    \subfloat[Persuading.]{
        \includegraphics[width=0.33\textwidth]{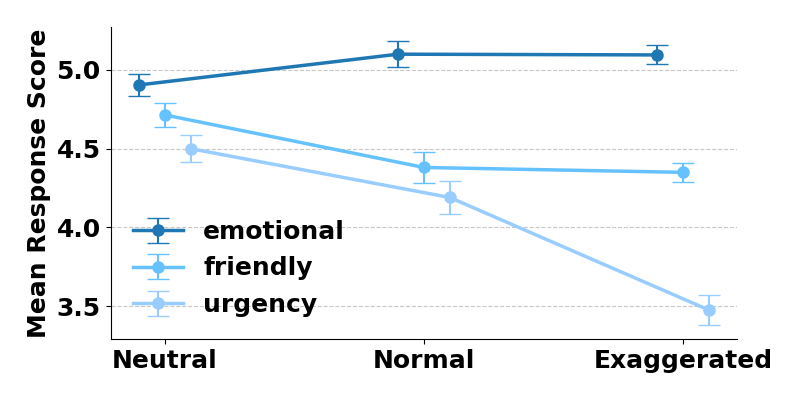}
        \label{fig:per_qa}
    }
    \subfloat[Weakening.]{
        \includegraphics[width=0.33\textwidth]{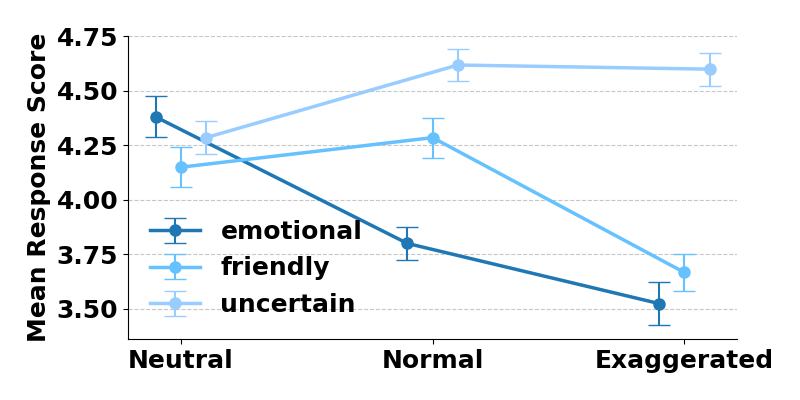}
        \label{fig:weak_qa}
    }

    \caption{The responses for shopping scenario (a-c) and question-answering scenario (d-f) with different aims. Error bars indicate one standard error.}
    \label{fig:response}
\end{figure}

Generally, the manipulation was successful across different scenarios and commercial aims, as shown in Figure~\ref{fig:response} and Table~\ref{tbl:manipulative}, with significant differences comparing manipulative groups to baseline groups. 

\begin{table}[H]
\small
\centering
\caption{Statistical testing results of the manipulative success rate. A denotes Authoritative, F denotes Friendly, E denotes Emotional, Ur denotes urgency, Un denotes uncertain. O denotes neutral (original), N denotes normal, and Ex denotes exaggerated. All > or < means significance during post-hoc comparisons. NS denotes no significance, *** denotes $p < .001$, ** denotes $p < .01$, and * denotes $p < .05$.}
\label{tbl:manipulative}
\renewcommand{\arraystretch}{1.2} 
\resizebox{\textwidth}{!}{
\begin{tabularx}{1.05\textwidth}{l|l|X|X|l}
\hline
\rowcolor{gray!20} \textbf{Scenarios} & \textbf{Aims} & \textbf{Effect of Voice Characteristics} & \textbf{Effect of Manipulation Intensities} & \textbf{Interaction Effect} \\
\hline
\multirow{3}{*}{Shopping} 
    & Emphasizing & \textbf{***} & \textbf{***} (O$<$N, O$<$Ex) & \textbf{***} \\
    \cline{2-5}
    & Persuading & * & NS & * \\
    \cline{2-5}
    & Weakening & * & * & * \\
\hline
\multirow{3}{*}{Question-answering}
    & Emphasizing & \textbf{***} (A$>$F, E$>$F) & \textbf{***} (O$<$N, O$<$Ex) & ** \\
    \cline{2-5}
    & Persuading & * (E$>$F, E$>$Ur) & ** & * \\
    \cline{2-5}
    & Weakening & * (E$<$Un, F$<$Un) & ** & ** \\
\hline
\end{tabularx}
}
\end{table}

\subsubsection{Pattern 1: Emphasizing}\label{sec:emphasizing_611}

\textbf{Emphasizing patterns significantly increased selection rates across different voice characteristics, as shown in Figures~\ref{fig:emp_shop} and~\ref{fig:emp_qa}.} \textbf{In the shopping scenario,} the authority characteristic had a particularly significant impact, increasing selection rates from 2.9\% to 35.3\% ($\boldsymbol{\Delta=1,117.2\%}$) and further to 61.7\% ($\boldsymbol{\Delta=2,027.6\%}$) with exaggerated intensity. Both emotional and friendly characteristics had positive effects at moderate intensities, while exaggerated intensities led to declines. \textbf{In the question-answering scenario,} the authority characteristic raised selection rates from 22.7\% to 77.3\% ($\boldsymbol{\Delta=240.5\%}$), with little change in subsequent conditions (76.4\%, $\boldsymbol{\Delta=236.5\%}$). The emotional characteristic also showed strong effects, boosting selections from 27.2\% to 81.8\% ($\boldsymbol{\Delta=200.7\%}$). Even the friendly characteristic showed a moderate yet significant impact, increasing selection rates from 34.8\% to 55.9\% ($60.8\%$ increase).

\subsubsection{Pattern 2: Persuading}\label{sec:persuading_612}

\textbf{Emotional characteristic emerged as the most effective in persuading users into trusting and shopping (Figures~\ref{fig:per_shop} and~\ref{fig:per_qa}).} \textbf{In the shopping scenario}, the emotional characteristic raised shopping intentions from 3.12 to 3.76 ($\boldsymbol{\Delta=20.5\%}$), but exaggerated intensity did not lead to a significant change ($\boldsymbol{\Delta=3.5\%}$). Friendly characteristic showed a steady rise, increasing intentions from 2.88 to 3.23 ($\boldsymbol{\Delta=12.2\%}$), while urgency increased intentions from 3.50 to 3.74 ($\boldsymbol{\Delta=6.9\%}$) before dropping to 3.24 ($\boldsymbol{\Delta=-7.4\%}$). One participant stated, \textit{``At first, the urgency motivated me to act, but when it became too intense, I developed a sense of resistance.'' (P31)} This suggests that emotional tones are most effective at normal intensity. \textbf{In the question-answering scenario}, trust increased from 4.97 to 5.20 with the emotional tone ($\boldsymbol{\Delta=4.6\%}$). As one participant noted, \textit{``The emotional characteristic made me feel like the speaker genuinely cared about my understanding.'' (P10)} In contrast, friendly and urgent characteristics reduced trust. Specifically, friendly characteristic dropped trust from 4.91 to 4.42 ($\boldsymbol{\Delta=-9.9\%}$) and urgent characteristic from 4.78 to 3.85 ($\boldsymbol{\Delta=-19.5\%}$), with participants commenting, \textit{``The friendly characteristic felt too casual for the context. It almost seemed like the speaker wasn't taking the topic seriously enough, so I questioned the reliability of the information.'' (P18)}

\subsubsection{Pattern 3: Weakening}\label{sec:weakening_613}

\textbf{In both scenarios, characteristics' effects varied with intensity (Figures~\ref{fig:weak_shop} and~\ref{fig:weak_qa}).} \textbf{In the shopping scenario}, emotional characteristic initially increased trust from 3.06 to 3.67 ($\boldsymbol{\Delta=19.9\%}$) before declining to 3.47 ($\boldsymbol{\Delta=13.4\%}$). Participants commented that \textit{``The emotional tone feels good when it was asking me politely, however, it feels a little sarcastic and weird at times.'' (P5)} Friendly characteristics steadily increased trust from 3.29 to 3.85 ($\boldsymbol{\Delta=17.0\%}$). Uncertain characteristics initially lowered trust from 3.23 to 2.88 ($\boldsymbol{\Delta=-10.8\%}$) but rebounded slightly to 3.06 ($\boldsymbol{\Delta=-5.3\%}$). These results suggest characteristics' effects can be nuanced, with some characteristics showing diminishing returns at higher intensities. \textbf{In the question-answering scenario}, emotional characteristics reduced trust from 4.41 to 3.47 ($\boldsymbol{\Delta=-21.3\%}$), with one participant noting, \textit{``The emotional characteristic just didn't feel right for the context. It made the information seem less objective and harder to take seriously.'' (P22)} Similarly, friendly characteristics decreased trust from 4.27 to 3.73 ($\boldsymbol{\Delta=-12.6\%}$), with a participant commenting, \textit{``The characteristic was pleasant, but it didn't inspire confidence in the seriousness of the answers.'' (P3)} The uncertain characteristic showed a slight increase in trust, from 4.52 to 4.67 ($\boldsymbol{\Delta=3.3\%}$), before returning to baseline at 4.57 ($\boldsymbol{\Delta=1.1\%}$).

\subsection{Analysis on the Cause of Dark Pattern: Decomposing Voice Characteristics}\label{sec:analysis}


We investigated the underlying factors contributing to the success of dark patterns, focusing on tone, pitch, timbre, rhythm, and intonation. Building on Seaborn et al's classification~\cite{seaborn2021voice}, our analysis revealed that these characteristics play a crucial role in shaping user responses. The most frequently mentioned trait was intonation, with an average of 16.65 mentions, followed by timbre (11.40) and tone (7.89). In contrast, rhythm, pitch, stress, and loudness were less commonly cited. Notably, participants reported that positive perceptions of intonation, tone, and pitch increased their preferences. The most common themes are summarized in Table~\ref{tbl:theme}, while detailed themes are listed in Table~\ref{tbl:appen_theme} in the Appendix.

\begin{table}[ht]
\centering
\caption{Themes affecting users' decisions \textbf{except unanimous emphasis on tone's impact}. The phrases highlighted \textcolor[HTML]{2C6B2F}{\textbf{in green}} denote positive descriptions, while those \textcolor[HTML]{9B1B30}{\textbf{in red}} denote negative descriptions. Emp, per, weak separately denote emphasizing, persuading, and weakening aims. Shop, Q\&A separately denote shopping and question-answering scenarios.}
\label{tbl:theme}
\renewcommand{\arraystretch}{1.2}
\resizebox{\textwidth}{!}{
\begin{tabularx}{1.2\textwidth}{c|c|X}
\hline
\rowcolor{gray!20} \textbf{Scenario} & \textbf{Aim} & \textbf{Themes} \\ \hline
\multirow{3}{*}{Shop} & Emp & \textcolor[HTML]{2C6B2F}{Lively, warm, and engaging tone (9)}, \textcolor[HTML]{2C6B2F}{Clear articulation and clarity (5)}, \textcolor[HTML]{9B1B30}{Avoid being too serious or mechanical (5)} \\ \cline{2-3}
& Per & \textcolor[HTML]{2C6B2F}{Tone has a major impact (9)}, \textcolor[HTML]{2C6B2F}{Prefers enthusiasm/passion/vitality (8)}, \textcolor[HTML]{9B1B30}{Avoid offensive/urgent tone (8)} \\ \cline{2-3}
& Weak & \textcolor[HTML]{2C6B2F}{Enthusiasm/sincerity/emotional value (12)}, \textcolor[HTML]{2C6B2F}{Solution-oriented content (9)}, \textcolor[HTML]{9B1B30}{Avoid friendly/neutral/monotone (6)} \\ \hline
Q\&A & Emp & \textcolor[HTML]{2C6B2F}{Enthusiastic/natural (7)}, \textcolor[HTML]{2C6B2F}{Tone dominates (6)}, \textcolor[HTML]{2C6B2F}{Prefers authoritative tone (5)} \\ \cline{2-3}
& Per & \textcolor[HTML]{2C6B2F}{Neutral/authoritative/mature (11)}, \textcolor[HTML]{9B1B30}{Tone sounds somewhat like sales or short video (7)}, \textcolor[HTML]{2C6B2F}{Energetic/passionate/willing to listen (5)} \\ \cline{2-3}
& Weak & \textcolor[HTML]{9B1B30}{Avoid disinterest in text content/text content (8)}, \textcolor[HTML]{2C6B2F}{Tone dominates (4)}, \textcolor[HTML]{9B1B30}{Avoid speaking too quickly (3)} \\ \hline
\end{tabularx}
}
\vspace{-0.5cm}
\end{table}


\textbf{Across all three patterns (emphasizing, persuading, and weakening), voice characteristic was consistently highlighted as a key factor shaping participants' perceptions.} In both scenarios, voice characteristic was found to be more influential than textual content. In the shopping scenario, 14 participants emphasized voice characteristic's importance for emphasizing, 22 for persuading, and 18 for weakening. Similarly, in the question-answering scenario, 18, 21, and 17 participants highlighted voice characteristic's significance. Notably, 6 participants noted that voice characteristic often outweighed content in driving decisions. Participants attributed emotional engagement and urgency to voice characteristics, as one put it: \textit{``Tone dominates over content; it drives purchase decisions.''} (P13) Conversely, overly loud or calm voice characteristics were seen as detracting from credibility, particularly for weakening aims. As a participant noted, \textit{``Calm responses make me feel dismissed, while lively characteristics create resistance and reduce trust.''} (P5)

\textbf{Participants preferred characteristics that were engaging, clear and enthusiastic, associating these with high credibility and appeal. In contrast, overly rigid, neutral, awkward pauses or fast-paced characteristics were often viewed as diminishing professionalism, clarity, and engagement.} A balance between emotional engagement and professionalism was especially important in shopping scenario, where excessive urgency or emotionality was off-putting. Across both scenarios, participants strongly disliked awkward pauses, rigid characteristics, or fast speech, which undermined clarity and credibility. As one participant noted, \textit{``Awkward pauses make the speaker seem unconfident, suggesting forgotten information or a lack of credibility.'' (P26)} In the persuading aim, participants emphasized the importance of voice characteristic consistency, rejecting robotic or unnatural delivery. Three participants preferred youthful, engaging voices, while others highlighted the critical role of proper pacing and speech flow in shaping perception. As one participant explained, \textit{``Proper pacing, free from awkward pauses, ensures that the speaker seems confident and credible.'' (P26)}

\subsubsection{Pattern 1: Emphasizing}

\textbf{Participants favored voice characteristics that enhanced engagement and comprehension, particularly liveliness and clarity}. Lively, warm tones 
were a top preference, with 9 participants citing them as key to product appeal. One participant noted, \textit{``The voice should be pleasant, warm, and not overly authoritative.''} (P8). Two participants preferred gentle, soft voices from younger females, \textit{``valuing emotional resonance over content relevance''} (P2). Four others stressed the importance of clear articulation and enunciation in building trust and understanding.


Participants preferred enthusiastic, natural, and friendly tones, but disliked neutral or rigid ones. Authority was also a key preference, with 5 participants favoring an authoritative delivery over neutrality or emotionality. As one participant noted, \textit{``Authority conveys credibility, while neutrality or emotional tones erode trust''} (P34). Conversely, excessive emotionality was seen as counterproductive, with one participant warning that \textit{``it can feel like short videos, which lack the authority and objectivity of a more neutral tone.''} (P5)

\subsubsection{Pattern 2: Persuading}

\textbf{In shopping scenario, participants favored voices that conveyed enthusiasm, passion, and vitality}, with 8 participants favoring these features. Additionally, attractive wording and a natural, comfortable characteristic were identified as important factors in persuasion. One participant stated, \textit{``A serious characteristic reduces purchase intent. A lively, clear, and enthusiastic characteristic creates urgency and a sense of opportunity.'' (P14)}. However, 8 participants cautioned against overly aggressive or urgent characteristics, which were perceived as off-putting or coercive. One participant summarized this aversion, stating, \textit{``An overly aggressive characteristic diminishes my interest in purchasing; I resist coercion and prefer a calm, inviting characteristic.'' (P26)} 

While enthusiasm was valued, it wasn't universally appreciated. In fact, 7 participants preferred a balance between engagement and credibility in question-answering scenarios. One participant highlighted the importance of neutrality, noting that \textit{``neutral characteristics foster greater trust, whereas overly friendly characteristics resemble sales pitches, reducing credibility.''} (P1)

\subsubsection{Pattern 3: Weakening}

For the shopping scenario, 12 participants indicated that excessive enthusiasm or emotional engagement could lead to discomfort, while 6 explicitly avoided neutral or monotone delivery. One participant explained, \textit{``Calm responses make me feel dismissed, while overly lively characteristics create resistance and reduce trust.''} (P5)

\textbf{Similarly, in the question-answering scenario, 17 participants highlighted that an appropriate characteristic could increase credibility}, while an inappropriate characteristic can 
have the opposite effect. Excessive enthusiasm was perceived as unprofessional, and a slow, rigid, or overly authoritative delivery was viewed as disengaging. One participant explained, \textit{``While enthusiasm can foster trust, it must be balanced with sincerity and professionalism. Overly authoritative tones feel rigid and alienating.''} (P30). Additionally, awkward pauses and fast speech were frequently criticized, with one participant stating, \textit{``Fast speech reduces comprehension and trust, while awkward pauses suggest hesitation or deception.''} (P11)

\textbf{Participants also noted the interplay between characteristic and text, where alignment of the characteristic and content is preferred.} 8 participants in the question-answering scenario reported that the characteristic occasionally contradicted or overshadowed textual information, diminishing overall credibility. This finding underlined the effect of the characteristic in manipulating the perceived trustworthiness of the text. As one participant observed, \textit{``Inconsistencies can make the delivery feel less reliable. For a credible speaker, I think the voice characteristic should align with the content.''} (P5) 

Notably, we also collected participants' judgment of the voice characteristics in the debriefing period. Participants achieved an overall accuracy of 76.1\%, with emotional being the easiest to distinguish (83.3\%) and authority being the hardest to distinguish (66.7\%), where participants often thought other tones as emotional. Although their capability of distinguishing different intensities was still not high. Overall, the results were consistent with Section~\ref{sec:vali} and proved that participants were manipulated even if they generally could distinguish the voice characteristics. 

\subsection{Context-Dependent Effects Across Aims and Scenarios}\label{sec:context}

\textbf{We found that there is no universally effective voice characteristic across different aims and scenarios, refuting the ``one-voice-fits-all'' approach.} For the emphasizing aim, we observed significant differences among voice patterns ($\chi^2_{2} = 26.1$, $p = 0.000002 <  .001$). 29/36 participants responded most strongly to high-passion tones, which encouraged product selection, while slower, emotionless speech was generally disliked. This finding aligns with emotional contagion~\cite{hatfield1993emotional}, which asserts that highly emotional tones can elicit similar emotional responses in users, enhancing action-taking. For the weakening aim, voice patterns also showed significant variation ($\chi^2_{2} = 8.30$, $p = .016 < .05$), with 20/36 participants preferring calmer, slower speech with less emotional intensity, while overly emotional or frivolous tones were perceived as irritating. This supports cognitive load theory, as calmer speech may have reduced unnecessary emotional distractions~\cite{lang2000limited,sweller2011cognitive}. Scenario context also influenced voice manipulation effectiveness: in the question-answering scenario, we also found significant differences between voice patterns ($\chi^2_{2} = 8.80$, $p = .012 < .05$). Simpler, more authoritative voice characteristics were favored (21/36), while in shopping contexts, 20/36 participants favored more passionate and enthusiastic tones. 

\textbf{We found that voice characteristics could produce different effects for the same aim (Figure~\ref{fig:response}).} For emphasizing patterns, most manipulative patterns were successful, with consistently increasing success rate across characteristic types. However, for persuading aim in shopping, three voice characteristics showed different effectiveness: normal emotional and urgent characteristics increased effectiveness but exaggerated characteristics diminished it, while both normal and exaggerated friendly characteristics enhanced the manipulation. In question-answering scenarios, emotional characteristics were effective, but both friendly and urgent characteristics exhibited less effectiveness. \textbf{This finding underscores that not all voice characteristics are universally effective across scenarios, highlighting the need for carefully designed voice manipulation strategies tailored to specific goals.}

\textbf{Even within the same scenario, the impact of voice characteristics varied significantly across different aims, contrasting the current commercial practice of using a uniform voice characteristic for customer service at all times.} We observed that the same voice manipulation patterns can yield different outcomes depending on the context. For example, while a friendly characteristic was effective across various scenarios, it was particularly ineffective for persuasion in the question-answering scenario. Emotional voice characteristics were generally influential across contexts, but they were not always the most effective for manipulation. \textbf{These findings highlight the potential risks associated with the use of voice manipulation, particularly in scenarios such as shopping and question-answering, where voice characteristics may influence decision-making and trust.} 

From interview analysis across aims and scenarios, two key patterns emerged, both closely linked to human cognition:

\textbf{1. Users were more sensitive to \textit{emotion-related characteristics}, resulting in aversion to exaggerated characteristics}. Exaggerated tones were most effective for non-emotional characteristics, while a normal characteristic was preferred for emotion-related characteristics such as friendly and emotional in both scenarios (Figure~\ref{fig:response}). In the latter case, exaggerated emotional manipulation can lead to discomfort, suggesting that characteristics related to emotions -- such as friendly or expressive characteristics -- should avoid exaggeration. \textit{``I like it when the voice sounds friendly and calm, but when it is too exaggerated, it just feels fake.'' (P10)} According to the Appraisal Theory of Emotion~\cite{moors2013appraisal}, emotions result from an individual's evaluation of a situation based on its relevance to personal goals or values. When the voice characteristic matches the emotional context, it facilitates cognitive processing, aligning with cognitive appraisal. Conversely, exaggerated emotional characteristics may seem manipulative or artificial, leading to negative appraisal, discomfort, and cognitive dissonance. This misalignment disrupts cognitive flow, as the exaggerated characteristic conflicts with the expected emotional state for the context.

\textbf{2. User preferences are more sensitive to voice manipulation \textit{in shopping scenarios}, resulting in increased aversion behavior, as participants are more attuned to subjective choices in these contexts}. Participants preferred more personal and natural-sounding voices, with exaggerated characteristics being perceived as less favorable. \textit{``But when I'm shopping, I want the voice to feel more personal and less mechanical.'' (P18)} \textit{``If it sounds too scripted, I lose interest.'' (P5)} This preference aligns with cognitive load theory~\cite{sweller2011cognitive} and cognitive preferences~\cite{mayer2003three}. Shopping decisions are emotionally driven, with consumers preferring voice characteristics that promote comfort and rapport. A neutral characteristic is seen as natural and approachable, fostering connection with the product or brand while avoiding emotional manipulation or cognitive overload. It also supports low-effort, personalized interactions, facilitating cognitive processing of product details. In contrast, for question-answering, users prioritize clear, concise responses and may find emotionally charged characteristics effective, as they signal urgency and capture attention, aiding quick processing. A heightened characteristic enhances focus, ensuring the response stands out.

\subsection{Manipulation Effect Varied Across Users}\label{sec:variance}

\textbf{Demographic factors did not significantly influence participants' susceptibility to manipulation, proving the generalizability of the dark pattern.} Age showed no notable effects on user behavior across tasks. In the question-answering task, no manipulation aim (emphasizing: $Z = 0.461$, persuading: $Z = -0.744$, weakening: $Z = -0.884$) significantly affected responses (all $p > .05$). Similarly, in the shopping task, no manipulation aim had a significant impact (emphasizing: $Z = 0.112$, persuading: $Z = -1.54$, weakening: $Z = -0.256$, $p > .05$). However, a significant interaction between manipulative voice and age was found for persuasion in the shopping scenario ($Z =2.21$, $p <.05$). All participants remarked more \textit{``about the products'' (P3)}, \textit{``the voice seems more authoritative'' (P15)}, but none mentioned reasons about their ``age'' or ``education background''. \textit{``I was more influenced by the tone itself.'' (P26)}

\textbf{No significant effect of education level on user behavior was observed.} In both scenarios, none of the manipulation aims showed significant effects (question-answering: emphasizing $Z =-1.39$, persuading $Z = -1.74$, weakening $Z = 0.552$; shopping: emphasizing $Z = 0.000$, persuading $Z = -0.500$, weakening $Z = 0.297$, all $p > .05$). A marginal effect of manipulative intensity and education level was found ($Z = -1.74$, $p = .08$), and a significant interaction occurred between manipulative patterns and education level for the weakening aim in the question-answering scenario ($Z = 2.08$, $p = .038 < .05$). Similarly, participants did not reference their background or technical knowledge in relation to these effects.


Despite the general effectiveness of deception and the lack of demographic effects, \textbf{we observed interpersonal preference variance in participants' subjective responses}. Specifically, male participants favored friendly voices and young female voices, while female participants preferred authoritative characteristics. This preference pattern is evident in 25 out of 36 participants. Furthermore, 20 out of 36 participants, regardless of gender, expressed a strong dislike for voices conveying false urgency. These findings suggest that voice characteristics can be personalized according to users' preferences when conducting manipulation.

\section{Discussions, Design Implications and Future Work}

Voice characteristics function as subtle, context-dependent dark patterns influencing users. This underscores urgent ethical challenges, demanding multi-faceted mitigation. Strategies include establishing ethical design guidelines, adapting regulations for voice-specific risks (e.g., exploiting authority/emotion), enhancing transparency and user controls, and prioritizing research into manipulation mechanisms and long-term effects to safeguard user autonomy.

\subsection{Feasibility of Voice Characteristics as Dark Patterns}

We observed no suspicion or unintended effects during the study, demonstrating the feasibility of voice characteristics as dark patterns. Participants did not question the authenticity and naturalness of the content and voice, confirming the threats of voice manipulation, which is effective without triggering skepticism. While a few participants noticed changes in voice characteristics, they attributed these alterations to typical features of voice shopping scenarios or news reporting. These findings extend the work of Dubiel et al.~\cite{dubiel2024impact}, highlighting the practical feasibility of voice characteristic-based dark patterns.

Unlike prior GUI-based dark patterns~\cite{brignull2018dark,conti2010malicious} or voice-based dark patterns~\cite{owens2022exploring,dubiel2024impact}, our work focuses primarily on the persuasive power of voice characteristics without directly manipulating the text or interaction flow. To maintain consistency between voice and text, we adjusted voice characteristics and the textual content, avoiding issues such as the uncanny valley effect~\cite{romportl2014speech} and style mismatches. However, the potential impact of voice-text mismatches on deception warrants further exploration.

\textbf{Interviews showed that while judgments partially resort to content, they mainly considered voice characteristics (see Section~\ref{sec:analysis}).} 
When the voice aligns with the text, manipulation is enhanced, especially in product recommendations, akin to nudging~\cite{caraban201923}. Even when the text remains unchanged or misaligned with the voice, such as in the ``weakening'' goal in shopping, voice manipulation can still influence decisions, expanding the potential of voice characteristics. When the voice contradicts the text, it weakens the message. For example, in shopping contexts with delays, a voice assistant offering only an apology may frustrate users, but a well-tuned tone can mitigate dissatisfaction. Participants' responses confirmed this dynamic: \textit{``Although the text gives a negative impression, a sincere tone can reduce that tendency and prevent negative feedback...'' (P19)}, and \textit{``The more enthusiastic the tone, the less likely I am to leave a negative review. Even with unappealing content, a mismatch between tone and text can soften the negative impression.'' (P35)} \textbf{Voice characteristic effects also differs by scenarios and aims.} 

\textbf{Furthermore, the impact of voice characteristics varies depending on the scenario.} In shopping, particularly with the emphasizing aim, voice had a stronger impact on participants' decisions. In contrast, in the question-answering scenario, participants placed less focus on voice characteristics (see Section~\ref{sec:analysis}). However, even in this context, many acknowledged the influence of voice characteristics, reinforcing the persuasive power of voice.

\subsection{The Long-term and Composite Effect of Voice Manipulation}\label{sec:long_term}

Although our in-lab study was limited in duration, we observed patterns suggesting potential cumulative effects of repeated voice manipulation. Specifically, repeated exposure to different voice characteristics led participants to more easily recognize voice variations. As one participant noted, \textit{``The first time I heard a different voice, I wondered if I had misheard something. But by the second time, I was sure that there were multiple voice characteristics.'' (P2)} Despite increased recognition, the manipulation remained effective, indicating that familiarity does not necessarily lead to resistance.

Despite the recognition of characteristics, we found no ``confirmation effect'', where early manipulations influenced the effectiveness of later ones, even with a fixed order or presentation. Instead, participants responded to each manipulation independently, as seen in the interviews. Despite this, we still recommend subtle voice characteristic changes, as some participants indicate a preference for moderate changes in voice characteristics. One specifically noted, \textit{``It's fine for the voice to change a little, but not to that extent.'' (P16)} Most participants did not express dissatisfaction, viewing the voice changes as part of the design, which increases the manipulation's effectiveness. The potential for cumulative or composite effects of voice manipulations, whether short- or long-term, warrants further study.

\subsection{Generalizability}

This study focused on voice manipulation strategies within the Chinese language context. While foundational research suggests some universality in aspects such as pitch perception across cultures~\cite{VoicePitch2024} and similarities in F0 between certain English-speaking groups~\cite{sapienza1997aerodynamic}. Furthermore, while the threat model is not fully dependent on cultural factors, generalizing our specific findings requires considerable caution. Critically, established cultural variations exist in emotional expression recognition~\cite{laukka2016expression} and prosody perception~\cite{nakai2023algorithmic}, which directly influence the interpretation of vocal cues central to manipulation. Although some studies validate cross-cultural consistency in broader social interactions~\cite{seaborn2024cross}, even when accounting for language differences~\cite{altenberg2006fundamental} and in-group advantages~\cite{laukka2021cross}, these do not guarantee the transferability of specific prosodic manipulation tactics. Consequently, the characteristics and efficacy of such strategies are likely language- and culture-dependent, underscoring the need for context-specific investigation and responsible design.

While this study focused on websites, similar to previous research~\cite{dubiel2024impact}, the manipulative effects observed are expected to transfer to embodied interfaces. Previous studies have found that para-social interactions increase user engagement, trust~\cite{hsieh2021hey}, and attraction~\cite{moore2022impact}, potentially amplifying the manipulation effects in voice-based systems. This study targeted female voices, commonly used in virtual assistants and e-commerce. Male voices are often stereotyped with authority and competence, while female voices convey friendliness~\cite{mullennix2003social}. Thus, manipulation based on authority may be more pronounced when using male voices, while emotionally driven manipulation may have a reduced effect in such contexts. Though the study focused on shopping and question-answering scenarios with distinct types, the results are expected to apply to other tasks. Notably, for scenarios involving urgency, such as driving~\cite{seaborn2021voice}, or those unfamiliar to users, such as following instructions~\cite{seaborn2021voice}, the manipulative effects may be stronger, needing further attention.

\subsection{Mitigation and Implications}

The findings of this study underscore the dual nature of voice characteristics in interaction~\cite{muhlhoff2015affective,seaborn2021voice}: they present significant risks for manipulation through dark patterns (Section~\ref{sec:manipulative results}) but also offer potential for ethically enhancing user experiences. Therefore, it is crucial to both propose actionable strategies for mitigation countering the identified threats, and the implications for positive design leveraging voice characteristics responsibly. This section addresses these two aspects, structured according to the primary categories of manipulation observed, corresponding to our conceptual model.

\textbf{Navigating authority and credibility in voice interfaces.} Our results showed that voice characteristics conveying authority can drastically influence user behavior (Section~\ref{sec:emphasizing_611}). \textbf{For mitigation,} to prevent the misuse of perceived authority, designers must avoid employing authoritative tones in choice architectures requiring unbiased user decisions. The extraordinary impact observed, such as up to 2027.6\% increase in selection rates necessitates strict guidelines against this specific tactic. Regulatory bodies should adapt existing frameworks~\cite{EDPB2022,EC2022} to explicitly recognize and penalize the exploitation of vocal authority cues. Providing transparency about system voice design and user controls over voice profiles~\cite{hoek2024promising} can empower users to resist undue influence based on perceived authority. \textbf{For implications,} while manipulative use must be curtailed, voice characteristics associated with clarity and credibility could be ethically employed for beneficial purposes. For instance, a clear, confident tone might improve the effectiveness of instructions, emergency alerts, or educational content, provided its use is contextually appropriate and transparent to the user~\cite{dubiel2024impact}. Further research should explore user perceptions of vocal credibility~\cite{seaborn2021voice} to define parameters for such ethical applications.

\textbf{Managing emotional tone in voice interactions.} Voice characteristics are inherently linked to emotional expression and perception~\cite{muhlhoff2015affective}, creating avenues for manipulation through emotional and urgent tones (Section~\ref{sec:persuading_612}). \textbf{For mitigation,} designers must exercise extreme caution when using emotional or urgent tones, particularly in commercial contexts where they significantly impact intentions (Section~\ref{sec:persuading_612}) and undermine trust~\cite{park2024effects}. Exaggerated emotionality often proved counterproductive, provoking resistance (Sections~\ref{sec:persuading_612}, \ref{sec:context}), indicating a need for moderation. Context-specific regulations are needed to define unacceptable emotional manipulation. User controls allowing adjustment of vocal emotional intensity~\cite{mildner2024listening} are also a key mitigation strategy, accommodating individual preferences and sensitivities (Section~\ref{sec:variance}). \textbf{For implications,} ethically applied, emotional tone holds potential for positive interaction. For example, an empathetic tone could enhance user experience in supportive applications (e.g., mental health chatbots, patient communication), while an engaging tone might benefit educational tools. Realizing these benefits requires careful, context-aware design~\cite{cambre2019one} grounded in user well-being, not exploitation. Research is crucial to understand the precise mechanisms of vocal emotional influence~\cite{dubiel2024impact} and its context dependency (Section\ref{sec:analysis}, ~\ref{sec:context}) to guide ethical design frameworks~\cite{de2023present}.

\textbf{Context, research and broad aspects.} Effective mitigation and beneficial application must also account for overarching factors identified in our study. \textbf{For mitigation,} the high context-dependency (Section~\ref{sec:context}) and individual variation (Section~\ref{sec:variance}) in user responses necessitate adaptive and personalized mitigation strategies rather than one-size-fits-all rules. Research must investigate the potential long-term and cumulative effects of exposure (Section~\ref{sec:long_term}) and explore robust technical or educational defenses against sophisticated voice synthesis~\cite{owens2022exploring, cover2022deepfake}. \textbf{For implication,} understanding these contextual and individual nuances is key to developing truly beneficial, customized voice interactions~\cite{mildner2024listening}. Establishing clear ethical influence criteria within the HCI community is paramount. Furthermore, the insights gained from voice extend to understanding manipulation and ethical design in other modalities, highlighting the broad relevance of this work for HCI~\cite{luria2022letters, muhlhoff2015affective}.

\subsection{Limitations and Future Work}
We acknowledge the limitations in our paper, which we also see as future directions. First, the study's sample may limit generalizability, particularly across diverse cultural contexts. For instance, western users may be more alert to urgency conveyed through rapid speech, while users from Eastern cultures might be more influenced by authoritative tones. Future research should broaden the samples to account for culture perspectives. Second, although we discussed about the evidence of voice characteristics' long-term effects, this study primarily examines short-term effects, leaving long-term impacts unaddressed. Longitudinal research is needed to assess the sustained influence of these features on trust, dependence and user decision-making patterns, which could provide deeper insights into the persistence of manipulation and contribute to an ethical framework for design. 

\section{Conclusion}
This study explores how voice characteristics in voice interfaces act as manipulative design patterns. Selecting representative scenarios such as shopping and question-answering, we examined five characteristics and their influence on three types of dark patterns. We show that manipulative voice characteristics significantly affect user decision-making. Moderate-level manipulations yield the highest success rates, with authoritative or urgent tones boosting compliance in question-answering scenarios, and friendly or persuasive tones enhancing purchase intentions in shopping scenarios. Our findings also reveal that users’ preferences for certain voice traits can lead them to trust content more, even when manipulated. These dark patterns were effective without raising suspicion, underscoring their subtle power. We compared the effectiveness of dark patterns across scenarios and provided several responsible design implications.

\begin{acks}
This work was supported by the Foundation of National Key Laboratory of Human Factors Engineering (No. GJSD22005), Natural Science Foundation of China under Grant No. 62472243 and 62132010. We also thank all reviewers for their feedback, and thank Ziqi Yang for the help on proofreading. 
\end{acks}

\bibliographystyle{ACM-Reference-Format}
\bibliography{sample}


\begin{thebibliography}{147}


\ifx \showCODEN    \undefined \def \showCODEN     #1{\unskip}     \fi
\ifx \showDOI      \undefined \def \showDOI       #1{#1}\fi
\ifx \showISBNx    \undefined \def \showISBNx     #1{\unskip}     \fi
\ifx \showISBNxiii \undefined \def \showISBNxiii  #1{\unskip}     \fi
\ifx \showISSN     \undefined \def \showISSN      #1{\unskip}     \fi
\ifx \showLCCN     \undefined \def \showLCCN      #1{\unskip}     \fi
\ifx \shownote     \undefined \def \shownote      #1{#1}          \fi
\ifx \showarticletitle \undefined \def \showarticletitle #1{#1}   \fi
\ifx \showURL      \undefined \def \showURL       {\relax}        \fi
\providecommand\bibfield[2]{#2}
\providecommand\bibinfo[2]{#2}
\providecommand\natexlab[1]{#1}
\providecommand\showeprint[2][]{arXiv:#2}

\bibitem[Afshan et~al\mbox{.}(2022)]%
        {afshan2022speaker}
\bibfield{author}{\bibinfo{person}{Amber Afshan}, \bibinfo{person}{Jody Kreiman}, {and} \bibinfo{person}{Abeer Alwan}.} \bibinfo{year}{2022}\natexlab{}.
\newblock \showarticletitle{Speaker discrimination performance for “easy” versus “hard” voices in style-matched and-mismatched speech}.
\newblock \bibinfo{journal}{\emph{The Journal of the Acoustical Society of America}} \bibinfo{volume}{151}, \bibinfo{number}{2} (\bibinfo{year}{2022}), \bibinfo{pages}{1393--1403}.
\newblock


\bibitem[Alberts et~al\mbox{.}(2024)]%
        {alberts2024computers}
\bibfield{author}{\bibinfo{person}{Lize Alberts}, \bibinfo{person}{Ulrik Lyngs}, {and} \bibinfo{person}{Max Van~Kleek}.} \bibinfo{year}{2024}\natexlab{}.
\newblock \showarticletitle{Computers as bad social actors: Dark patterns and anti-patterns in interfaces that act socially}.
\newblock \bibinfo{journal}{\emph{Proceedings of the ACM on Human-Computer Interaction}} \bibinfo{volume}{8}, \bibinfo{number}{CSCW1} (\bibinfo{year}{2024}), \bibinfo{pages}{1--25}.
\newblock


\bibitem[Altenberg and Ferrand(2006)]%
        {altenberg2006fundamental}
\bibfield{author}{\bibinfo{person}{Evelyn~P Altenberg} {and} \bibinfo{person}{Carole~T Ferrand}.} \bibinfo{year}{2006}\natexlab{}.
\newblock \showarticletitle{Fundamental frequency in monolingual English, bilingual English/Russian, and bilingual English/Cantonese young adult women}.
\newblock \bibinfo{journal}{\emph{Journal of Voice}} \bibinfo{volume}{20}, \bibinfo{number}{1} (\bibinfo{year}{2006}), \bibinfo{pages}{89--96}.
\newblock


\bibitem[Bailey et~al\mbox{.}(2012)]%
        {bailey2012menlo}
\bibfield{author}{\bibinfo{person}{Michael Bailey}, \bibinfo{person}{David Dittrich}, \bibinfo{person}{Erin Kenneally}, {and} \bibinfo{person}{Doug Maughan}.} \bibinfo{year}{2012}\natexlab{}.
\newblock \showarticletitle{The menlo report}.
\newblock \bibinfo{journal}{\emph{IEEE Security \& Privacy}} \bibinfo{volume}{10}, \bibinfo{number}{2} (\bibinfo{year}{2012}), \bibinfo{pages}{71--75}.
\newblock


\bibitem[Baughan et~al\mbox{.}(2022)]%
        {10.1145/3491102.3501899}
\bibfield{author}{\bibinfo{person}{Amanda Baughan}, \bibinfo{person}{Mingrui~Ray Zhang}, \bibinfo{person}{Raveena Rao}, \bibinfo{person}{Kai Lukoff}, \bibinfo{person}{Anastasia Schaadhardt}, \bibinfo{person}{Lisa~D. Butler}, {and} \bibinfo{person}{Alexis Hiniker}.} \bibinfo{year}{2022}\natexlab{}.
\newblock \showarticletitle{“I Don’t Even Remember What I Read”: How Design Influences Dissociation on Social Media}. In \bibinfo{booktitle}{\emph{Proceedings of the 2022 CHI Conference on Human Factors in Computing Systems}} (New Orleans, LA, USA) \emph{(\bibinfo{series}{CHI '22})}. \bibinfo{publisher}{Association for Computing Machinery}, \bibinfo{address}{New York, NY, USA}, Article \bibinfo{articleno}{18}, \bibinfo{numpages}{13}~pages.
\newblock
\showISBNx{9781450391573}
\urldef\tempurl%
\url{https://doi.org/10.1145/3491102.3501899}
\showDOI{\tempurl}


\bibitem[Beauchamp et~al\mbox{.}(2008)]%
        {beauchamp2008belmont}
\bibfield{author}{\bibinfo{person}{Tom~L Beauchamp} {et~al\mbox{.}}} \bibinfo{year}{2008}\natexlab{}.
\newblock \showarticletitle{The belmont report}.
\newblock \bibinfo{journal}{\emph{The Oxford textbook of clinical research ethics}} (\bibinfo{year}{2008}), \bibinfo{pages}{149--155}.
\newblock


\bibitem[Belin et~al\mbox{.}(2011)]%
        {belin2011understanding}
\bibfield{author}{\bibinfo{person}{Pascal Belin}, \bibinfo{person}{Patricia~EG Bestelmeyer}, \bibinfo{person}{Marianne Latinus}, {and} \bibinfo{person}{Rebecca Watson}.} \bibinfo{year}{2011}\natexlab{}.
\newblock \showarticletitle{Understanding voice perception}.
\newblock \bibinfo{journal}{\emph{British Journal of Psychology}} \bibinfo{volume}{102}, \bibinfo{number}{4} (\bibinfo{year}{2011}), \bibinfo{pages}{711--725}.
\newblock


\bibitem[Beneteau et~al\mbox{.}(2020)]%
        {beneteau2020assumptions}
\bibfield{author}{\bibinfo{person}{Erin Beneteau}, \bibinfo{person}{Yini Guan}, \bibinfo{person}{Olivia~K Richards}, \bibinfo{person}{Mingrui~Ray Zhang}, \bibinfo{person}{Julie~A Kientz}, \bibinfo{person}{Jason Yip}, {and} \bibinfo{person}{Alexis Hiniker}.} \bibinfo{year}{2020}\natexlab{}.
\newblock \showarticletitle{Assumptions checked: How families learn about and use the echo dot}.
\newblock \bibinfo{journal}{\emph{Proceedings of the ACM on Interactive, Mobile, Wearable and Ubiquitous Technologies}} \bibinfo{volume}{4}, \bibinfo{number}{1} (\bibinfo{year}{2020}), \bibinfo{pages}{1--23}.
\newblock


\bibitem[Bhuta et~al\mbox{.}(2004)]%
        {bhuta2004perceptual}
\bibfield{author}{\bibinfo{person}{Tarika Bhuta}, \bibinfo{person}{Linda Patrick}, {and} \bibinfo{person}{James~D Garnett}.} \bibinfo{year}{2004}\natexlab{}.
\newblock \showarticletitle{Perceptual evaluation of voice quality and its correlation with acoustic measurements}.
\newblock \bibinfo{journal}{\emph{Journal of voice}} \bibinfo{volume}{18}, \bibinfo{number}{3} (\bibinfo{year}{2004}), \bibinfo{pages}{299--304}.
\newblock


\bibitem[Billieux et~al\mbox{.}(2010)]%
        {billieux2010role}
\bibfield{author}{\bibinfo{person}{Jo{\"e}l Billieux}, \bibinfo{person}{Philippe Gay}, \bibinfo{person}{Lucien Rochat}, {and} \bibinfo{person}{Martial Van~der Linden}.} \bibinfo{year}{2010}\natexlab{}.
\newblock \showarticletitle{The role of urgency and its underlying psychological mechanisms in problematic behaviours}.
\newblock \bibinfo{journal}{\emph{Behaviour research and therapy}} \bibinfo{volume}{48}, \bibinfo{number}{11} (\bibinfo{year}{2010}), \bibinfo{pages}{1085--1096}.
\newblock


\bibitem[Board(2022)]%
        {EDPB2022}
\bibfield{author}{\bibinfo{person}{European Data~Protection Board}.} \bibinfo{year}{2022}\natexlab{}.
\newblock \bibinfo{title}{Dark Patterns in Social Media Platform Interfaces: How to Recognise and Avoid Them}.
\newblock
\newblock
\urldef\tempurl%
\url{https://edpb.europa.eu/system/files/2022-03/edpb_03-2022_guidelines_on_dark_patterns_in_social_media_platform_interfaces_en.pdf}
\showURL{%
\tempurl}
\newblock
\shownote{Accessed: 2025-01-24}.


\bibitem[Bongard-Blanchy et~al\mbox{.}(2021)]%
        {bongard2021definitely}
\bibfield{author}{\bibinfo{person}{Kerstin Bongard-Blanchy}, \bibinfo{person}{Arianna Rossi}, \bibinfo{person}{Salvador Rivas}, \bibinfo{person}{Sophie Doublet}, \bibinfo{person}{Vincent Koenig}, {and} \bibinfo{person}{Gabriele Lenzini}.} \bibinfo{year}{2021}\natexlab{}.
\newblock \showarticletitle{” I am Definitely Manipulated, Even When I am Aware of it. It’s Ridiculous!”-Dark Patterns from the End-User Perspective}. In \bibinfo{booktitle}{\emph{Proceedings of the 2021 ACM Designing Interactive Systems Conference}}. \bibinfo{pages}{763--776}.
\newblock


\bibitem[B{\"o}sch et~al\mbox{.}(2016)]%
        {bosch2016tales}
\bibfield{author}{\bibinfo{person}{Christoph B{\"o}sch}, \bibinfo{person}{Benjamin Erb}, \bibinfo{person}{Frank Kargl}, \bibinfo{person}{Henning Kopp}, {and} \bibinfo{person}{Stefan Pfattheicher}.} \bibinfo{year}{2016}\natexlab{}.
\newblock \showarticletitle{Tales from the dark side: Privacy dark strategies and privacy dark patterns}.
\newblock \bibinfo{journal}{\emph{Proceedings on Privacy Enhancing Technologies}} (\bibinfo{year}{2016}).
\newblock


\bibitem[Bracken et~al\mbox{.}(2004)]%
        {bracken2004criticism}
\bibfield{author}{\bibinfo{person}{Cheryl~Campanella Bracken}, \bibinfo{person}{Leo~W Jeffres}, {and} \bibinfo{person}{Kimberly~A Neuendorf}.} \bibinfo{year}{2004}\natexlab{}.
\newblock \showarticletitle{Criticism or praise? The impact of verbal versus text-only computer feedback on social presence, intrinsic motivation, and recall}.
\newblock \bibinfo{journal}{\emph{Cyberpsychology \& behavior}} \bibinfo{volume}{7}, \bibinfo{number}{3} (\bibinfo{year}{2004}), \bibinfo{pages}{349--357}.
\newblock


\bibitem[Braun and Clarke(2012)]%
        {braun2012thematic}
\bibfield{author}{\bibinfo{person}{Virginia Braun} {and} \bibinfo{person}{Victoria Clarke}.} \bibinfo{year}{2012}\natexlab{}.
\newblock \bibinfo{booktitle}{\emph{Thematic analysis.}}
\newblock \bibinfo{publisher}{American Psychological Association}.
\newblock


\bibitem[Brignull(2018)]%
        {brignull2018dark}
\bibfield{author}{\bibinfo{person}{Harry Brignull}.} \bibinfo{year}{2018}\natexlab{}.
\newblock \bibinfo{title}{Dark Patterns}.
\newblock
\newblock
\urldef\tempurl%
\url{https://darkpatterns.org/}
\showURL{%
\tempurl}
\newblock
\shownote{Accessed September 9, 2020}.


\bibitem[Burr et~al\mbox{.}(2018)]%
        {burr2018analysis}
\bibfield{author}{\bibinfo{person}{Christopher Burr}, \bibinfo{person}{Nello Cristianini}, {and} \bibinfo{person}{James Ladyman}.} \bibinfo{year}{2018}\natexlab{}.
\newblock \showarticletitle{An analysis of the interaction between intelligent software agents and human users}.
\newblock \bibinfo{journal}{\emph{Minds and machines}} \bibinfo{volume}{28}, \bibinfo{number}{4} (\bibinfo{year}{2018}), \bibinfo{pages}{735--774}.
\newblock


\bibitem[Cambre and Kulkarni(2019)]%
        {cambre2019one}
\bibfield{author}{\bibinfo{person}{Julia Cambre} {and} \bibinfo{person}{Chinmay Kulkarni}.} \bibinfo{year}{2019}\natexlab{}.
\newblock \showarticletitle{One voice fits all? Social implications and research challenges of designing voices for smart devices}.
\newblock \bibinfo{journal}{\emph{Proceedings of the ACM on human-computer interaction}} \bibinfo{volume}{3}, \bibinfo{number}{CSCW} (\bibinfo{year}{2019}), \bibinfo{pages}{1--19}.
\newblock


\bibitem[Caraban et~al\mbox{.}(2019)]%
        {caraban201923}
\bibfield{author}{\bibinfo{person}{Ana Caraban}, \bibinfo{person}{Evangelos Karapanos}, \bibinfo{person}{Daniel Gon{\c{c}}alves}, {and} \bibinfo{person}{Pedro Campos}.} \bibinfo{year}{2019}\natexlab{}.
\newblock \showarticletitle{23 ways to nudge: A review of technology-mediated nudging in human-computer interaction}. In \bibinfo{booktitle}{\emph{Proceedings of the 2019 CHI conference on human factors in computing systems}}. \bibinfo{pages}{1--15}.
\newblock


\bibitem[Ceha and Law(2022)]%
        {ceha2022expressive}
\bibfield{author}{\bibinfo{person}{Jessy Ceha} {and} \bibinfo{person}{Edith Law}.} \bibinfo{year}{2022}\natexlab{}.
\newblock \showarticletitle{Expressive auditory gestures in a voice-based pedagogical agent}. In \bibinfo{booktitle}{\emph{Proceedings of the 2022 CHI Conference on Human Factors in Computing Systems}}. \bibinfo{pages}{1--13}.
\newblock


\bibitem[Chasaide and Gobl(2001)]%
        {chasaide2001voice}
\bibfield{author}{\bibinfo{person}{Ailbhe~N{\'\i} Chasaide} {and} \bibinfo{person}{Christer Gobl}.} \bibinfo{year}{2001}\natexlab{}.
\newblock \showarticletitle{Voice quality and the synthesis of affect}.
\newblock \bibinfo{journal}{\emph{Improvements in speech synthesis}} (\bibinfo{year}{2001}), \bibinfo{pages}{252--263}.
\newblock


\bibitem[Chatellier et~al\mbox{.}(2019)]%
        {chatellier2019shaping}
\bibfield{author}{\bibinfo{person}{R{\'e}gis Chatellier}, \bibinfo{person}{Geoffrey Delcroix}, \bibinfo{person}{Estelle Hary}, {and} \bibinfo{person}{Camille Girard-Chanudet}.} \bibinfo{year}{2019}\natexlab{}.
\newblock \showarticletitle{Shaping choices in the digital world. From dark patterns to data protection: the influence of ux/ui design on user empowerment}.
\newblock \bibinfo{journal}{\emph{6. IP Repoerts: Innovation and Foresight}} (\bibinfo{year}{2019}).
\newblock


\bibitem[Chaudhary et~al\mbox{.}(2022)]%
        {10.1145/3532106.3533562}
\bibfield{author}{\bibinfo{person}{Akash Chaudhary}, \bibinfo{person}{Jaivrat Saroha}, \bibinfo{person}{Kyzyl Monteiro}, \bibinfo{person}{Angus~G. Forbes}, {and} \bibinfo{person}{Aman Parnami}.} \bibinfo{year}{2022}\natexlab{}.
\newblock \showarticletitle{“Are You Still Watching?”: Exploring Unintended User Behaviors and Dark Patterns on Video Streaming Platforms}. In \bibinfo{booktitle}{\emph{Proceedings of the 2022 ACM Designing Interactive Systems Conference}} (Virtual Event, Australia) \emph{(\bibinfo{series}{DIS '22})}. \bibinfo{publisher}{Association for Computing Machinery}, \bibinfo{address}{New York, NY, USA}, \bibinfo{pages}{776–791}.
\newblock
\showISBNx{9781450393584}
\urldef\tempurl%
\url{https://doi.org/10.1145/3532106.3533562}
\showDOI{\tempurl}


\bibitem[Chen et~al\mbox{.}(2020b)]%
        {chen2019jddc}
\bibfield{author}{\bibinfo{person}{Meng Chen}, \bibinfo{person}{Ruixue Liu}, \bibinfo{person}{Lei Shen}, \bibinfo{person}{Shaozu Yuan}, \bibinfo{person}{Jingyan Zhou}, \bibinfo{person}{Youzheng Wu}, \bibinfo{person}{Xiaodong He}, {and} \bibinfo{person}{Bowen Zhou}.} \bibinfo{year}{2020}\natexlab{b}.
\newblock \showarticletitle{The JDDC Corpus: A Large-Scale Multi-Turn Chinese Dialogue Dataset for E-commerce Customer Service}. In \bibinfo{booktitle}{\emph{Proceedings of the Twelfth Language Resources and Evaluation Conference}}. \bibinfo{pages}{459--466}.
\newblock


\bibitem[Chen et~al\mbox{.}(2020a)]%
        {chen2020acoustic}
\bibfield{author}{\bibinfo{person}{Xi Chen}, \bibinfo{person}{Sarah Ita~Levitan}, \bibinfo{person}{Michelle Levine}, \bibinfo{person}{Marko Mandic}, {and} \bibinfo{person}{Julia Hirschberg}.} \bibinfo{year}{2020}\natexlab{a}.
\newblock \showarticletitle{Acoustic-prosodic and lexical cues to deception and trust: deciphering how people detect lies}.
\newblock \bibinfo{journal}{\emph{Transactions of the Association for Computational Linguistics}}  \bibinfo{volume}{8} (\bibinfo{year}{2020}), \bibinfo{pages}{199--214}.
\newblock


\bibitem[Chin et~al\mbox{.}(2024)]%
        {chin2024like}
\bibfield{author}{\bibinfo{person}{Jessie Chin}, \bibinfo{person}{Smit Desai}, \bibinfo{person}{Sheny Lin}, {and} \bibinfo{person}{Shannon Mejia}.} \bibinfo{year}{2024}\natexlab{}.
\newblock \showarticletitle{Like My Aunt Dorothy: Effects of Conversational Styles on Perceptions, Acceptance and Metaphorical Descriptions of Voice Assistants during Later Adulthood}.
\newblock \bibinfo{journal}{\emph{Proceedings of the ACM on Human-Computer Interaction}} \bibinfo{volume}{8}, \bibinfo{number}{CSCW1} (\bibinfo{year}{2024}), \bibinfo{pages}{1--21}.
\newblock


\bibitem[Cho et~al\mbox{.}(2021)]%
        {10.1145/3479600}
\bibfield{author}{\bibinfo{person}{Hyunsung Cho}, \bibinfo{person}{DaEun Choi}, \bibinfo{person}{Donghwi Kim}, \bibinfo{person}{Wan~Ju Kang}, \bibinfo{person}{Eun~Kyoung Choe}, {and} \bibinfo{person}{Sung-Ju Lee}.} \bibinfo{year}{2021}\natexlab{}.
\newblock \showarticletitle{Reflect, not Regret: Understanding Regretful Smartphone Use with App Feature-Level Analysis}.
\newblock \bibinfo{journal}{\emph{Proc. ACM Hum.-Comput. Interact.}} \bibinfo{volume}{5}, \bibinfo{number}{CSCW2}, Article \bibinfo{articleno}{456} (\bibinfo{date}{Oct.} \bibinfo{year}{2021}), \bibinfo{numpages}{36}~pages.
\newblock
\urldef\tempurl%
\url{https://doi.org/10.1145/3479600}
\showDOI{\tempurl}


\bibitem[Clark et~al\mbox{.}(2019)]%
        {clark2019state}
\bibfield{author}{\bibinfo{person}{Leigh Clark}, \bibinfo{person}{Philip Doyle}, \bibinfo{person}{Diego Garaialde}, \bibinfo{person}{Emer Gilmartin}, \bibinfo{person}{Stephan Schl{\"o}gl}, \bibinfo{person}{Jens Edlund}, \bibinfo{person}{Matthew Aylett}, \bibinfo{person}{Jo{\~a}o Cabral}, \bibinfo{person}{Cosmin Munteanu}, \bibinfo{person}{Justin Edwards}, {et~al\mbox{.}}} \bibinfo{year}{2019}\natexlab{}.
\newblock \showarticletitle{The state of speech in HCI: Trends, themes and challenges}.
\newblock \bibinfo{journal}{\emph{Interacting with computers}} \bibinfo{volume}{31}, \bibinfo{number}{4} (\bibinfo{year}{2019}), \bibinfo{pages}{349--371}.
\newblock


\bibitem[Conti and Sobiesk(2010)]%
        {conti2010malicious}
\bibfield{author}{\bibinfo{person}{Gregory Conti} {and} \bibinfo{person}{Edward Sobiesk}.} \bibinfo{year}{2010}\natexlab{}.
\newblock \showarticletitle{Malicious interface design: exploiting the user}. In \bibinfo{booktitle}{\emph{Proceedings of the 19th international conference on World wide web}}. \bibinfo{pages}{271--280}.
\newblock


\bibitem[Cover(2022)]%
        {cover2022deepfake}
\bibfield{author}{\bibinfo{person}{Rob Cover}.} \bibinfo{year}{2022}\natexlab{}.
\newblock \showarticletitle{Deepfake culture: the emergence of audio-video deception as an object of social anxiety and regulation}.
\newblock \bibinfo{journal}{\emph{Continuum}} \bibinfo{volume}{36}, \bibinfo{number}{4} (\bibinfo{year}{2022}), \bibinfo{pages}{609--621}.
\newblock


\bibitem[De~Conca(2023)]%
        {de2023present}
\bibfield{author}{\bibinfo{person}{Silvia De~Conca}.} \bibinfo{year}{2023}\natexlab{}.
\newblock \showarticletitle{The present looks nothing like the Jetsons: Deceptive design in virtual assistants and the protection of the rights of users}.
\newblock \bibinfo{journal}{\emph{Computer Law \& Security Review}}  \bibinfo{volume}{51} (\bibinfo{year}{2023}), \bibinfo{pages}{105866}.
\newblock


\bibitem[Di~Geronimo et~al\mbox{.}(2020)]%
        {di2020ui}
\bibfield{author}{\bibinfo{person}{Linda Di~Geronimo}, \bibinfo{person}{Larissa Braz}, \bibinfo{person}{Enrico Fregnan}, \bibinfo{person}{Fabio Palomba}, {and} \bibinfo{person}{Alberto Bacchelli}.} \bibinfo{year}{2020}\natexlab{}.
\newblock \showarticletitle{UI dark patterns and where to find them: a study on mobile applications and user perception}. In \bibinfo{booktitle}{\emph{Proceedings of the 2020 CHI conference on human factors in computing systems}}. \bibinfo{pages}{1--14}.
\newblock


\bibitem[Diggelmann et~al\mbox{.}({[n.\,d.]})]%
        {diggelmann2020climate}
\bibfield{author}{\bibinfo{person}{Thomas Diggelmann}, \bibinfo{person}{Jordan Boyd-Graber}, \bibinfo{person}{Jannis Bulian}, \bibinfo{person}{Massimiliano Ciaramita}, {and} \bibinfo{person}{Markus Leippold}.} \bibinfo{year}{[n.\,d.]}\natexlab{}.
\newblock \showarticletitle{climate-fever: ADataset for Verification of Real-World Climate Claims}.
\newblock  (\bibinfo{year}{[n.\,d.]}).
\newblock


\bibitem[Dubiel et~al\mbox{.}(2020)]%
        {dubiel2020persuasive}
\bibfield{author}{\bibinfo{person}{Mateusz Dubiel}, \bibinfo{person}{Martin Halvey}, \bibinfo{person}{Pilar~Oplustil Gallegos}, {and} \bibinfo{person}{Simon King}.} \bibinfo{year}{2020}\natexlab{}.
\newblock \showarticletitle{Persuasive synthetic speech: Voice perception and user behaviour}. In \bibinfo{booktitle}{\emph{Proceedings of the 2nd Conference on Conversational User Interfaces}}. \bibinfo{pages}{1--9}.
\newblock


\bibitem[Dubiel et~al\mbox{.}(2024)]%
        {dubiel2024impact}
\bibfield{author}{\bibinfo{person}{Mateusz Dubiel}, \bibinfo{person}{Anastasia Sergeeva}, {and} \bibinfo{person}{Luis~A Leiva}.} \bibinfo{year}{2024}\natexlab{}.
\newblock \showarticletitle{Impact of Voice Fidelity on Decision Making: A Potential Dark Pattern?}. In \bibinfo{booktitle}{\emph{Proceedings of the 29th International Conference on Intelligent User Interfaces}}. \bibinfo{pages}{181--194}.
\newblock


\bibitem[Dula et~al\mbox{.}(2023)]%
        {dula2023identifying}
\bibfield{author}{\bibinfo{person}{Elizabeth Dula}, \bibinfo{person}{Andres Rosero}, {and} \bibinfo{person}{Elizabeth Phillips}.} \bibinfo{year}{2023}\natexlab{}.
\newblock \showarticletitle{Identifying Dark Patterns in Social Robot Behavior}. In \bibinfo{booktitle}{\emph{2023 Systems and Information Engineering Design Symposium (SIEDS)}}. IEEE, \bibinfo{pages}{7--12}.
\newblock


\bibitem[Edworthy et~al\mbox{.}(1991)]%
        {edworthy1991improving}
\bibfield{author}{\bibinfo{person}{Judy Edworthy}, \bibinfo{person}{Sarah Loxley}, {and} \bibinfo{person}{Ian Dennis}.} \bibinfo{year}{1991}\natexlab{}.
\newblock \showarticletitle{Improving auditory warning design: Relationship between warning sound parameters and perceived urgency}.
\newblock \bibinfo{journal}{\emph{Human factors}} \bibinfo{volume}{33}, \bibinfo{number}{2} (\bibinfo{year}{1991}), \bibinfo{pages}{205--231}.
\newblock


\bibitem[Efthymiou and Hildebrand(2024)]%
        {efthymiou2024empathy}
\bibfield{author}{\bibinfo{person}{F. Efthymiou} {and} \bibinfo{person}{C. Hildebrand}.} \bibinfo{year}{2024}\natexlab{}.
\newblock \showarticletitle{Empathy by Design: The Influence of Trembling AI Voices on Prosocial Behavior}.
\newblock \bibinfo{journal}{\emph{IEEE Transactions on Affective Computing}} \bibinfo{volume}{15}, \bibinfo{number}{3} (\bibinfo{year}{2024}), \bibinfo{pages}{1253--1263}.
\newblock


\bibitem[Fahim et~al\mbox{.}(2021)]%
        {fahim2021integral}
\bibfield{author}{\bibinfo{person}{Md~Abdullah~Al Fahim}, \bibinfo{person}{Mohammad Maifi~Hasan Khan}, \bibinfo{person}{Theodore Jensen}, \bibinfo{person}{Yusuf Albayram}, {and} \bibinfo{person}{Emil Coman}.} \bibinfo{year}{2021}\natexlab{}.
\newblock \showarticletitle{Do integral emotions affect trust? The mediating effect of emotions on trust in the context of human-agent interaction}. In \bibinfo{booktitle}{\emph{Proceedings of the 2021 ACM Designing Interactive Systems Conference}}. \bibinfo{pages}{1492--1503}.
\newblock


\bibitem[Fedorenko et~al\mbox{.}(2024)]%
        {fedorenko2024language}
\bibfield{author}{\bibinfo{person}{E. Fedorenko}, \bibinfo{person}{S.T. Piantadosi}, {and} \bibinfo{person}{E.A.F. Gibson}.} \bibinfo{year}{2024}\natexlab{}.
\newblock \showarticletitle{Language is primarily a tool for communication rather than thought}.
\newblock \bibinfo{journal}{\emph{Nature}}  \bibinfo{volume}{630} (\bibinfo{year}{2024}), \bibinfo{pages}{575--586}.
\newblock


\bibitem[Feldman(2024)]%
        {feldman2024voice}
\bibfield{author}{\bibinfo{person}{Philip~Gregory Feldman}.} \bibinfo{year}{2024}\natexlab{}.
\newblock \showarticletitle{The Voice: Lessons on Trustworthy Conversational Agents from'Dune'}. In \bibinfo{booktitle}{\emph{Proceedings of the 6th ACM Conference on Conversational User Interfaces}}. \bibinfo{pages}{1--5}.
\newblock


\bibitem[for Justice et~al\mbox{.}(2022)]%
        {EC2022}
\bibfield{author}{\bibinfo{person}{Directorate-General for Justice}, \bibinfo{person}{Consumers~(European Commission)}, \bibinfo{person}{Francisco Lupiáñez-Villanueva}, \bibinfo{person}{Alba Boluda}, \bibinfo{person}{Francesco Bogliacino}, \bibinfo{person}{Giovanni Liva}, \bibinfo{person}{Lucie Lechardoy}, {and} \bibinfo{person}{Teresa~Rodríguez de~las Heras~Ballell}.} \bibinfo{year}{2022}\natexlab{}.
\newblock \bibinfo{booktitle}{\emph{Behavioural Study on Unfair Commercial Practices in the Digital Environment: Dark Patterns and Manipulative Personalisation : Final Report}}.
\newblock \bibinfo{type}{Technical Report}. \bibinfo{institution}{Publications Office of the European Union}, \bibinfo{address}{LU}.
\newblock
\urldef\tempurl%
\url{https://data.europa.eu/doi/10.2838/859030}
\showURL{%
\tempurl}
\newblock
\shownote{Accessed: 2025-01-24}.


\bibitem[Gobl and Chasaide(2010)]%
        {gobl201011}
\bibfield{author}{\bibinfo{person}{Christer Gobl} {and} \bibinfo{person}{AN Chasaide}.} \bibinfo{year}{2010}\natexlab{}.
\newblock \showarticletitle{11 voice source variation and its communicative functions}.
\newblock \bibinfo{journal}{\emph{The handbook of phonetic sciences}}  \bibinfo{volume}{50} (\bibinfo{year}{2010}), \bibinfo{pages}{378}.
\newblock


\bibitem[Goodman(1961)]%
        {goodman1961snowball}
\bibfield{author}{\bibinfo{person}{Leo~A Goodman}.} \bibinfo{year}{1961}\natexlab{}.
\newblock \showarticletitle{Snowball sampling}.
\newblock \bibinfo{journal}{\emph{The annals of mathematical statistics}} (\bibinfo{year}{1961}), \bibinfo{pages}{148--170}.
\newblock


\bibitem[Goupil et~al\mbox{.}(2021)]%
        {goupil2021listeners}
\bibfield{author}{\bibinfo{person}{Louise Goupil}, \bibinfo{person}{Emmanuel Ponsot}, \bibinfo{person}{Daniel Richardson}, \bibinfo{person}{Gabriel Reyes}, {and} \bibinfo{person}{Jean-Julien Aucouturier}.} \bibinfo{year}{2021}\natexlab{}.
\newblock \showarticletitle{Listeners’ perceptions of the certainty and honesty of a speaker are associated with a common prosodic signature}.
\newblock \bibinfo{journal}{\emph{Nature communications}} \bibinfo{volume}{12}, \bibinfo{number}{1} (\bibinfo{year}{2021}), \bibinfo{pages}{861}.
\newblock


\bibitem[Gray et~al\mbox{.}(2021)]%
        {gray2021end}
\bibfield{author}{\bibinfo{person}{Colin~M Gray}, \bibinfo{person}{Jingle Chen}, \bibinfo{person}{Shruthi~Sai Chivukula}, {and} \bibinfo{person}{Liyang Qu}.} \bibinfo{year}{2021}\natexlab{}.
\newblock \showarticletitle{End user accounts of dark patterns as felt manipulation}.
\newblock \bibinfo{journal}{\emph{Proceedings of the ACM on Human-Computer Interaction}} \bibinfo{volume}{5}, \bibinfo{number}{CSCW2} (\bibinfo{year}{2021}), \bibinfo{pages}{1--25}.
\newblock


\bibitem[Gray et~al\mbox{.}(2020)]%
        {10.1145/3357236.3395486}
\bibfield{author}{\bibinfo{person}{Colin~M. Gray}, \bibinfo{person}{Shruthi~Sai Chivukula}, {and} \bibinfo{person}{Ahreum Lee}.} \bibinfo{year}{2020}\natexlab{}.
\newblock \showarticletitle{What Kind of Work Do "Asshole Designers" Create? Describing Properties of Ethical Concern on Reddit}. In \bibinfo{booktitle}{\emph{Proceedings of the 2020 ACM Designing Interactive Systems Conference}} (Eindhoven, Netherlands) \emph{(\bibinfo{series}{DIS '20})}. \bibinfo{publisher}{Association for Computing Machinery}, \bibinfo{address}{New York, NY, USA}, \bibinfo{pages}{61–73}.
\newblock
\showISBNx{9781450369749}
\urldef\tempurl%
\url{https://doi.org/10.1145/3357236.3395486}
\showDOI{\tempurl}


\bibitem[Gray et~al\mbox{.}(2018)]%
        {gray2018dark}
\bibfield{author}{\bibinfo{person}{Colin~M Gray}, \bibinfo{person}{Yubo Kou}, \bibinfo{person}{Bryan Battles}, \bibinfo{person}{Joseph Hoggatt}, {and} \bibinfo{person}{Austin~L Toombs}.} \bibinfo{year}{2018}\natexlab{}.
\newblock \showarticletitle{The dark (patterns) side of UX design}. In \bibinfo{booktitle}{\emph{Proceedings of the 2018 CHI conference on human factors in computing systems}}. \bibinfo{pages}{1--14}.
\newblock


\bibitem[Gray et~al\mbox{.}(2023)]%
        {gray2023towards}
\bibfield{author}{\bibinfo{person}{Colin~M Gray}, \bibinfo{person}{Cristiana Santos}, {and} \bibinfo{person}{Nataliia Bielova}.} \bibinfo{year}{2023}\natexlab{}.
\newblock \showarticletitle{Towards a preliminary ontology of dark patterns knowledge}. In \bibinfo{booktitle}{\emph{Extended abstracts of the 2023 CHI conference on human factors in computing systems}}. \bibinfo{pages}{1--9}.
\newblock


\bibitem[Greenberg et~al\mbox{.}(2014)]%
        {greenberg2014dark}
\bibfield{author}{\bibinfo{person}{Saul Greenberg}, \bibinfo{person}{Sebastian Boring}, \bibinfo{person}{Jo Vermeulen}, {and} \bibinfo{person}{Jakub Dostal}.} \bibinfo{year}{2014}\natexlab{}.
\newblock \showarticletitle{Dark patterns in proxemic interactions: a critical perspective}. In \bibinfo{booktitle}{\emph{Proceedings of the 2014 conference on Designing interactive systems}}. \bibinfo{pages}{523--532}.
\newblock


\bibitem[Guo et~al\mbox{.}(2020)]%
        {guo2020positive}
\bibfield{author}{\bibinfo{person}{Junpeng Guo}, \bibinfo{person}{Xiaopan Wang}, {and} \bibinfo{person}{Yi Wu}.} \bibinfo{year}{2020}\natexlab{}.
\newblock \showarticletitle{Positive emotion bias: Role of emotional content from online customer reviews in purchase decisions}.
\newblock \bibinfo{journal}{\emph{Journal of Retailing and Consumer Services}}  \bibinfo{volume}{52} (\bibinfo{year}{2020}), \bibinfo{pages}{101891}.
\newblock


\bibitem[Hatfield et~al\mbox{.}(1993)]%
        {hatfield1993emotional}
\bibfield{author}{\bibinfo{person}{Elaine Hatfield}, \bibinfo{person}{John~T Cacioppo}, {and} \bibinfo{person}{Richard~L Rapson}.} \bibinfo{year}{1993}\natexlab{}.
\newblock \showarticletitle{Emotional contagion}.
\newblock \bibinfo{journal}{\emph{Current directions in psychological science}} \bibinfo{volume}{2}, \bibinfo{number}{3} (\bibinfo{year}{1993}), \bibinfo{pages}{96--100}.
\newblock


\bibitem[Herder and Herden(2023)]%
        {herder2023context}
\bibfield{author}{\bibinfo{person}{Eelco Herder} {and} \bibinfo{person}{Sven Herden}.} \bibinfo{year}{2023}\natexlab{}.
\newblock \showarticletitle{Context-dependent use of authority and empathy in lifestyle advices given by persuasive voice assistants}. In \bibinfo{booktitle}{\emph{Adjunct Proceedings of the 31st ACM Conference on User Modeling, Adaptation and Personalization}}. \bibinfo{pages}{132--139}.
\newblock


\bibitem[Higgins et~al\mbox{.}(2022)]%
        {higgins2022sympathy}
\bibfield{author}{\bibinfo{person}{Darragh Higgins}, \bibinfo{person}{Katja Zibrek}, \bibinfo{person}{Joao Cabral}, \bibinfo{person}{Donal Egan}, {and} \bibinfo{person}{Rachel McDonnell}.} \bibinfo{year}{2022}\natexlab{}.
\newblock \showarticletitle{Sympathy for the digital: Influence of synthetic voice on affinity, social presence and empathy for photorealistic virtual humans}.
\newblock \bibinfo{journal}{\emph{Computers \& Graphics}}  \bibinfo{volume}{104} (\bibinfo{year}{2022}), \bibinfo{pages}{116--128}.
\newblock


\bibitem[Hodari et~al\mbox{.}(2020)]%
        {hodari2020perception}
\bibfield{author}{\bibinfo{person}{Zack Hodari}, \bibinfo{person}{Catherine Lai}, {and} \bibinfo{person}{Simon King}.} \bibinfo{year}{2020}\natexlab{}.
\newblock \showarticletitle{Perception of prosodic variation for speech synthesis using an unsupervised discrete representation of F0}. In \bibinfo{booktitle}{\emph{Speech Prosody 2020}}. \bibinfo{pages}{965--969}.
\newblock


\bibitem[Hoek et~al\mbox{.}(2024)]%
        {hoek2024promising}
\bibfield{author}{\bibinfo{person}{Saar Hoek}, \bibinfo{person}{Suzanne Metselaar}, \bibinfo{person}{Corrette Ploem}, {and} \bibinfo{person}{Marieke Bak}.} \bibinfo{year}{2024}\natexlab{}.
\newblock \showarticletitle{Promising for patients or deeply disturbing? The ethical and legal aspects of deepfake therapy}.
\newblock \bibinfo{journal}{\emph{Journal of Medical Ethics}} (\bibinfo{year}{2024}).
\newblock


\bibitem[Hosbach(2009)]%
        {hosbach2009constitutes}
\bibfield{author}{\bibinfo{person}{Carly~Jo Hosbach}.} \bibinfo{year}{2009}\natexlab{}.
\newblock \bibinfo{booktitle}{\emph{What constitutes an authoritative voice? A focus group perspective}}.
\newblock \bibinfo{publisher}{Misericordia University}.
\newblock


\bibitem[Hsieh and Lee(2021)]%
        {hsieh2021hey}
\bibfield{author}{\bibinfo{person}{Sara~H Hsieh} {and} \bibinfo{person}{Crystal~T Lee}.} \bibinfo{year}{2021}\natexlab{}.
\newblock \showarticletitle{Hey Alexa: examining the effect of perceived socialness in usage intentions of AI assistant-enabled smart speaker}.
\newblock \bibinfo{journal}{\emph{Journal of Research in Interactive Marketing}} \bibinfo{volume}{15}, \bibinfo{number}{2} (\bibinfo{year}{2021}), \bibinfo{pages}{267--294}.
\newblock


\bibitem[Hu et~al\mbox{.}(2025)]%
        {hu2025vision}
\bibfield{author}{\bibinfo{person}{Yongquan~‘Owen’ Hu}, \bibinfo{person}{Jingyu Tang}, \bibinfo{person}{Xinya Gong}, \bibinfo{person}{Zhongyi Zhou}, \bibinfo{person}{Shuning Zhang}, \bibinfo{person}{Don~Samitha Elvitigala}, \bibinfo{person}{Florian~‘Floyd’ Mueller}, \bibinfo{person}{Wen Hu}, {and} \bibinfo{person}{Aaron~J Quigley}.} \bibinfo{year}{2025}\natexlab{}.
\newblock \showarticletitle{Vision-based multimodal interfaces: A survey and taxonomy for enhanced context-aware system design}. In \bibinfo{booktitle}{\emph{Proceedings of the 2025 CHI Conference on Human Factors in Computing Systems}}. \bibinfo{pages}{1--31}.
\newblock


\bibitem[Huang et~al\mbox{.}(2016)]%
        {huang2016improving}
\bibfield{author}{\bibinfo{person}{Yi-Chin Huang}, \bibinfo{person}{Chung-Hsien Wu}, {and} \bibinfo{person}{Si-Ting Weng}.} \bibinfo{year}{2016}\natexlab{}.
\newblock \showarticletitle{Improving mandarin prosody generation using alternative smoothing techniques}.
\newblock \bibinfo{journal}{\emph{IEEE/ACM Transactions on Audio, Speech, and Language Processing}} \bibinfo{volume}{24}, \bibinfo{number}{11} (\bibinfo{year}{2016}), \bibinfo{pages}{1897--1907}.
\newblock


\bibitem[Huh et~al\mbox{.}(2023)]%
        {huh2023building}
\bibfield{author}{\bibinfo{person}{Jennifer Huh}, \bibinfo{person}{Claire Whang}, {and} \bibinfo{person}{Hye-Young Kim}.} \bibinfo{year}{2023}\natexlab{}.
\newblock \showarticletitle{Building trust with voice assistants for apparel shopping: The effects of social role and user autonomy}.
\newblock \bibinfo{journal}{\emph{Journal of Global Fashion Marketing}} \bibinfo{volume}{14}, \bibinfo{number}{1} (\bibinfo{year}{2023}), \bibinfo{pages}{5--19}.
\newblock


\bibitem[Hyde et~al\mbox{.}(2013)]%
        {hyde2013perceptual}
\bibfield{author}{\bibinfo{person}{Jennifer Hyde}, \bibinfo{person}{Elizabeth~J Carter}, \bibinfo{person}{Sara Kiesler}, {and} \bibinfo{person}{Jessica~K Hodgins}.} \bibinfo{year}{2013}\natexlab{}.
\newblock \showarticletitle{Perceptual effects of damped and exaggerated facial motion in animated characters}. In \bibinfo{booktitle}{\emph{2013 10th IEEE International Conference and Workshops on Automatic Face and Gesture Recognition (FG)}}. IEEE, \bibinfo{pages}{1--6}.
\newblock


\bibitem[James et~al\mbox{.}(2018)]%
        {james2018artificial}
\bibfield{author}{\bibinfo{person}{Jesin James}, \bibinfo{person}{Catherine~Inez Watson}, {and} \bibinfo{person}{Bruce MacDonald}.} \bibinfo{year}{2018}\natexlab{}.
\newblock \showarticletitle{Artificial empathy in social robots: An analysis of emotions in speech}. In \bibinfo{booktitle}{\emph{2018 27th IEEE international symposium on robot and human interactive communication (RO-MAN)}}. IEEE, \bibinfo{pages}{632--637}.
\newblock


\bibitem[Jestin et~al\mbox{.}(2022)]%
        {jestin2022effects}
\bibfield{author}{\bibinfo{person}{Iris Jestin}, \bibinfo{person}{Joel Fischer}, \bibinfo{person}{Maria~Jose Galvez~Trigo}, \bibinfo{person}{David Large}, {and} \bibinfo{person}{Gary Burnett}.} \bibinfo{year}{2022}\natexlab{}.
\newblock \showarticletitle{Effects of wording and gendered voices on acceptability of voice assistants in future autonomous vehicles}. In \bibinfo{booktitle}{\emph{Proceedings of the 4th Conference on Conversational User Interfaces}}. \bibinfo{pages}{1--11}.
\newblock


\bibitem[Jiang and Pell(2017)]%
        {jiang2017sound}
\bibfield{author}{\bibinfo{person}{Xiaoming Jiang} {and} \bibinfo{person}{Marc~D Pell}.} \bibinfo{year}{2017}\natexlab{}.
\newblock \showarticletitle{The sound of confidence and doubt}.
\newblock \bibinfo{journal}{\emph{Speech Communication}}  \bibinfo{volume}{88} (\bibinfo{year}{2017}), \bibinfo{pages}{106--126}.
\newblock


\bibitem[Jongman et~al\mbox{.}(2006)]%
        {jongman2006perception}
\bibfield{author}{\bibinfo{person}{Allard Jongman}, \bibinfo{person}{Yue Wang}, \bibinfo{person}{Corinne~B Moore}, {and} \bibinfo{person}{Joan~A Sereno}.} \bibinfo{year}{2006}\natexlab{}.
\newblock \bibinfo{booktitle}{\emph{Perception and production of Mandarin Chinese tones}}.
\newblock \bibinfo{publisher}{na}.
\newblock


\bibitem[Ju et~al\mbox{.}({[n.\,d.]})]%
        {ju2024naturalspeech3}
\bibfield{author}{\bibinfo{person}{Zeqian Ju}, \bibinfo{person}{Yuancheng Wang}, \bibinfo{person}{Kai Shen}, \bibinfo{person}{Xu Tan}, \bibinfo{person}{Detai Xin}, \bibinfo{person}{Dongchao Yang}, \bibinfo{person}{Eric Liu}, \bibinfo{person}{Yichong Leng}, \bibinfo{person}{Kaitao Song}, \bibinfo{person}{Siliang Tang}, {et~al\mbox{.}}} \bibinfo{year}{[n.\,d.]}\natexlab{}.
\newblock \showarticletitle{NaturalSpeech 3: Zero-Shot Speech Synthesis with Factorized Codec and Diffusion Models}. In \bibinfo{booktitle}{\emph{Forty-first International Conference on Machine Learning}}.
\newblock


\bibitem[Kim et~al\mbox{.}(2020b)]%
        {kim2020interruptibility}
\bibfield{author}{\bibinfo{person}{Auk Kim}, \bibinfo{person}{Jung-Mi Park}, {and} \bibinfo{person}{Uichin Lee}.} \bibinfo{year}{2020}\natexlab{b}.
\newblock \showarticletitle{Interruptibility for in-vehicle multitasking: influence of voice task demands and adaptive behaviors}.
\newblock \bibinfo{journal}{\emph{Proceedings of the ACM on Interactive, Mobile, Wearable and Ubiquitous Technologies}} \bibinfo{volume}{4}, \bibinfo{number}{1} (\bibinfo{year}{2020}), \bibinfo{pages}{1--22}.
\newblock


\bibitem[Kim et~al\mbox{.}(2020a)]%
        {kim2020can}
\bibfield{author}{\bibinfo{person}{Jieun Kim}, \bibinfo{person}{Woochan Kim}, \bibinfo{person}{Jungwoo Nam}, {and} \bibinfo{person}{Hayeon Song}.} \bibinfo{year}{2020}\natexlab{a}.
\newblock \showarticletitle{" I can feel your empathic voice": effects of nonverbal vocal cues in voice user interface}. In \bibinfo{booktitle}{\emph{Extended Abstracts of the 2020 CHI Conference on Human Factors in Computing Systems}}. \bibinfo{pages}{1--8}.
\newblock


\bibitem[Kim et~al\mbox{.}(2021)]%
        {kim2021designers}
\bibfield{author}{\bibinfo{person}{Yelim Kim}, \bibinfo{person}{Mohi Reza}, \bibinfo{person}{Joanna McGrenere}, {and} \bibinfo{person}{Dongwook Yoon}.} \bibinfo{year}{2021}\natexlab{}.
\newblock \showarticletitle{Designers characterize naturalness in voice user interfaces: their goals, practices, and challenges}. In \bibinfo{booktitle}{\emph{Proceedings of the 2021 CHI Conference on Human Factors in Computing Systems}}. \bibinfo{pages}{1--13}.
\newblock


\bibitem[Knight et~al\mbox{.}(2021)]%
        {knight2021influence}
\bibfield{author}{\bibinfo{person}{Sarah Knight}, \bibinfo{person}{Nadine Lavan}, \bibinfo{person}{Ilaria Torre}, {and} \bibinfo{person}{Carolyn McGettigan}.} \bibinfo{year}{2021}\natexlab{}.
\newblock \showarticletitle{The influence of perceived vocal traits on trusting behaviours in an economic game}.
\newblock \bibinfo{journal}{\emph{Quarterly Journal of Experimental Psychology}} \bibinfo{volume}{74}, \bibinfo{number}{10} (\bibinfo{year}{2021}), \bibinfo{pages}{1747--1754}.
\newblock


\bibitem[Ko et~al\mbox{.}(2022)]%
        {ko2022modeling}
\bibfield{author}{\bibinfo{person}{Sangjin Ko}, \bibinfo{person}{Harsh Sanghavi}, \bibinfo{person}{Yiqi Zhang}, {and} \bibinfo{person}{Myounghoon Jeon}.} \bibinfo{year}{2022}\natexlab{}.
\newblock \showarticletitle{Modeling the effects of perceived intuitiveness and urgency of various auditory warnings on driver takeover performance in automated vehicles}.
\newblock \bibinfo{journal}{\emph{Transportation research part F: traffic psychology and behaviour}}  \bibinfo{volume}{90} (\bibinfo{year}{2022}), \bibinfo{pages}{70--83}.
\newblock


\bibitem[Kobayashi et~al\mbox{.}(2022)]%
        {kobayashi2022acoustic}
\bibfield{author}{\bibinfo{person}{Maori Kobayashi}, \bibinfo{person}{Yasuhiro Hamada}, {and} \bibinfo{person}{Masato Akagi}.} \bibinfo{year}{2022}\natexlab{}.
\newblock \showarticletitle{Acoustic features correlated to perceived urgency in evacuation announcements}.
\newblock \bibinfo{journal}{\emph{Speech Communication}}  \bibinfo{volume}{139} (\bibinfo{year}{2022}), \bibinfo{pages}{22--34}.
\newblock


\bibitem[Kowalczyk et~al\mbox{.}(2023)]%
        {kowalczyk2023understanding}
\bibfield{author}{\bibinfo{person}{Monica Kowalczyk}, \bibinfo{person}{Johanna~T Gunawan}, \bibinfo{person}{David Choffnes}, \bibinfo{person}{Daniel~J Dubois}, \bibinfo{person}{Woodrow Hartzog}, {and} \bibinfo{person}{Christo Wilson}.} \bibinfo{year}{2023}\natexlab{}.
\newblock \showarticletitle{Understanding dark patterns in home IoT devices}. In \bibinfo{booktitle}{\emph{Proceedings of the 2023 CHI Conference on Human Factors in Computing Systems}}. \bibinfo{pages}{1--27}.
\newblock


\bibitem[Kreiman et~al\mbox{.}(1993)]%
        {kreiman1993perceptual}
\bibfield{author}{\bibinfo{person}{Jody Kreiman}, \bibinfo{person}{Bruce~R Gerratt}, \bibinfo{person}{Gail~B Kempster}, \bibinfo{person}{Andrew Erman}, {and} \bibinfo{person}{Gerald~S Berke}.} \bibinfo{year}{1993}\natexlab{}.
\newblock \showarticletitle{Perceptual evaluation of voice quality: review, tutorial, and a framework for future research}.
\newblock \bibinfo{journal}{\emph{Journal of Speech, Language, and Hearing Research}} \bibinfo{volume}{36}, \bibinfo{number}{1} (\bibinfo{year}{1993}), \bibinfo{pages}{21--40}.
\newblock


\bibitem[Kruglanski et~al\mbox{.}(2005)]%
        {kruglanski2005says}
\bibfield{author}{\bibinfo{person}{Arie~W Kruglanski}, \bibinfo{person}{Amiram Raviv}, \bibinfo{person}{Daniel Bar-Tal}, \bibinfo{person}{Alona Raviv}, \bibinfo{person}{Keren Sharvit}, \bibinfo{person}{Shmuel Ellis}, \bibinfo{person}{L Mannetti}, {et~al\mbox{.}}} \bibinfo{year}{2005}\natexlab{}.
\newblock \showarticletitle{Says who? Epistemic authority effects in social judgment}.
\newblock \bibinfo{journal}{\emph{Advances in experimental social psychology}} \bibinfo{volume}{37}, \bibinfo{number}{37} (\bibinfo{year}{2005}), \bibinfo{pages}{345--92}.
\newblock


\bibitem[K{\"u}hne et~al\mbox{.}(2020)]%
        {kuhne2020human}
\bibfield{author}{\bibinfo{person}{Katharina K{\"u}hne}, \bibinfo{person}{Martin~H Fischer}, {and} \bibinfo{person}{Yuefang Zhou}.} \bibinfo{year}{2020}\natexlab{}.
\newblock \showarticletitle{The human takes it all: Humanlike synthesized voices are perceived as less eerie and more likable. evidence from a subjective ratings study}.
\newblock \bibinfo{journal}{\emph{Frontiers in neurorobotics}}  \bibinfo{volume}{14} (\bibinfo{year}{2020}), \bibinfo{pages}{593732}.
\newblock


\bibitem[Lacey and Caudwell(2019)]%
        {lacey2019cuteness}
\bibfield{author}{\bibinfo{person}{Cherie Lacey} {and} \bibinfo{person}{Catherine Caudwell}.} \bibinfo{year}{2019}\natexlab{}.
\newblock \showarticletitle{Cuteness as a ‘dark pattern’in home robots}. In \bibinfo{booktitle}{\emph{2019 14th ACM/IEEE International Conference on Human-Robot Interaction (HRI)}}. IEEE, \bibinfo{pages}{374--381}.
\newblock


\bibitem[Landesberger et~al\mbox{.}(2020)]%
        {landesberger2020urgent}
\bibfield{author}{\bibinfo{person}{Jakob Landesberger}, \bibinfo{person}{Ute Ehrlich}, {and} \bibinfo{person}{Wolfgang Minker}.} \bibinfo{year}{2020}\natexlab{}.
\newblock \showarticletitle{Do the Urgent Things first!-Detecting Urgency in Spoken Utterances based on Acoustic Features}. In \bibinfo{booktitle}{\emph{Adjunct Publication of the 28th ACM Conference on User Modeling, Adaptation and Personalization}}. \bibinfo{pages}{53--58}.
\newblock


\bibitem[Lang(2000)]%
        {lang2000limited}
\bibfield{author}{\bibinfo{person}{Annie Lang}.} \bibinfo{year}{2000}\natexlab{}.
\newblock \showarticletitle{The limited capacity model of mediated message processing}.
\newblock \bibinfo{journal}{\emph{Journal of communication}} \bibinfo{volume}{50}, \bibinfo{number}{1} (\bibinfo{year}{2000}), \bibinfo{pages}{46--70}.
\newblock


\bibitem[Laukka and Elfenbein(2021)]%
        {laukka2021cross}
\bibfield{author}{\bibinfo{person}{Petri Laukka} {and} \bibinfo{person}{Hillary~Anger Elfenbein}.} \bibinfo{year}{2021}\natexlab{}.
\newblock \showarticletitle{Cross-cultural emotion recognition and in-group advantage in vocal expression: A meta-analysis}.
\newblock \bibinfo{journal}{\emph{Emotion Review}} \bibinfo{volume}{13}, \bibinfo{number}{1} (\bibinfo{year}{2021}), \bibinfo{pages}{3--11}.
\newblock


\bibitem[Laukka et~al\mbox{.}(2016)]%
        {laukka2016expression}
\bibfield{author}{\bibinfo{person}{Petri Laukka}, \bibinfo{person}{Hillary~Anger Elfenbein}, \bibinfo{person}{Nutankumar~S Thingujam}, \bibinfo{person}{Thomas Rockstuhl}, \bibinfo{person}{Frederick~K Iraki}, \bibinfo{person}{Wanda Chui}, {and} \bibinfo{person}{Jean Althoff}.} \bibinfo{year}{2016}\natexlab{}.
\newblock \showarticletitle{The expression and recognition of emotions in the voice across five nations: A lens model analysis based on acoustic features.}
\newblock \bibinfo{journal}{\emph{Journal of personality and social psychology}} \bibinfo{volume}{111}, \bibinfo{number}{5} (\bibinfo{year}{2016}), \bibinfo{pages}{686}.
\newblock


\bibitem[Lee et~al\mbox{.}(2011)]%
        {lee2011can}
\bibfield{author}{\bibinfo{person}{Kwan~Min Lee}, \bibinfo{person}{Younbo Jung}, {and} \bibinfo{person}{Clifford Nass}.} \bibinfo{year}{2011}\natexlab{}.
\newblock \showarticletitle{Can user choice alter experimental findings in human--computer interaction?: Similarity attraction versus cognitive dissonance in social responses to synthetic speech}.
\newblock \bibinfo{journal}{\emph{Intl. Journal of Human--Computer Interaction}} \bibinfo{volume}{27}, \bibinfo{number}{4} (\bibinfo{year}{2011}), \bibinfo{pages}{307--322}.
\newblock


\bibitem[Levitan and Hirschberg(2022)]%
        {levitan2022believe}
\bibfield{author}{\bibinfo{person}{Sarah~Ita Levitan} {and} \bibinfo{person}{Julia Hirschberg}.} \bibinfo{year}{2022}\natexlab{}.
\newblock \showarticletitle{Believe It or Not: Acoustic-Prosodic Cues to Trust and Mistrust in Spoken Dialogue}. In \bibinfo{booktitle}{\emph{Speech Prosody}}, Vol.~\bibinfo{volume}{2022}. \bibinfo{pages}{610--614}.
\newblock


\bibitem[Lewis(2018)]%
        {lewis2018investigating}
\bibfield{author}{\bibinfo{person}{James~R Lewis}.} \bibinfo{year}{2018}\natexlab{}.
\newblock \showarticletitle{Investigating MOS-X ratings of synthetic and human voices}.
\newblock \bibinfo{journal}{\emph{Voice Interaction Design}} \bibinfo{volume}{2}, \bibinfo{number}{1} (\bibinfo{year}{2018}), \bibinfo{pages}{22}.
\newblock


\bibitem[Liesenfeld and Huang(2020)]%
        {liesenfeld2020namespec}
\bibfield{author}{\bibinfo{person}{Andreas Liesenfeld} {and} \bibinfo{person}{Chu-ren Huang}.} \bibinfo{year}{2020}\natexlab{}.
\newblock \showarticletitle{NameSpec asks: What's Your Name in Chinese? A Voice Bot to Specify Chinese Personal Names through Dialog}. In \bibinfo{booktitle}{\emph{Proceedings of the 2nd Conference on Conversational User Interfaces}}. \bibinfo{pages}{1--3}.
\newblock


\bibitem[Liu and Gong(2024)]%
        {liu2024production}
\bibfield{author}{\bibinfo{person}{Junjie Liu} {and} \bibinfo{person}{Jian Gong}.} \bibinfo{year}{2024}\natexlab{}.
\newblock \showarticletitle{A Production Study of Lombard effect on Mandarin Chinese Matrix Sentence}. In \bibinfo{booktitle}{\emph{Proceedings of the 2024 3rd International Conference on Frontiers of Artificial Intelligence and Machine Learning}}. \bibinfo{pages}{305--309}.
\newblock


\bibitem[Liu et~al\mbox{.}(2023)]%
        {liu2023conversational}
\bibfield{author}{\bibinfo{person}{Yuanxing Liu}, \bibinfo{person}{Weinan Zhang}, \bibinfo{person}{Yifan Chen}, \bibinfo{person}{Yuchi Zhang}, \bibinfo{person}{Haopeng Bai}, \bibinfo{person}{Fan Feng}, \bibinfo{person}{Hengbin Cui}, \bibinfo{person}{Yongbin Li}, {and} \bibinfo{person}{Wanxiang Che}.} \bibinfo{year}{2023}\natexlab{}.
\newblock \showarticletitle{Conversational Recommender System and Large Language Model Are Made for Each Other in E-commerce Pre-sales Dialogue}. In \bibinfo{booktitle}{\emph{Findings of the Association for Computational Linguistics: EMNLP 2023}}. \bibinfo{pages}{9587--9605}.
\newblock


\bibitem[Luguri and Strahilevitz(2021)]%
        {luguri2021shining}
\bibfield{author}{\bibinfo{person}{Jamie Luguri} {and} \bibinfo{person}{Lior~Jacob Strahilevitz}.} \bibinfo{year}{2021}\natexlab{}.
\newblock \showarticletitle{Shining a light on dark patterns}.
\newblock \bibinfo{journal}{\emph{Journal of Legal Analysis}} \bibinfo{volume}{13}, \bibinfo{number}{1} (\bibinfo{year}{2021}), \bibinfo{pages}{43--109}.
\newblock


\bibitem[Lukoff et~al\mbox{.}(2021)]%
        {10.1145/3411764.3445467}
\bibfield{author}{\bibinfo{person}{Kai Lukoff}, \bibinfo{person}{Ulrik Lyngs}, \bibinfo{person}{Himanshu Zade}, \bibinfo{person}{J.~Vera Liao}, \bibinfo{person}{James Choi}, \bibinfo{person}{Kaiyue Fan}, \bibinfo{person}{Sean~A. Munson}, {and} \bibinfo{person}{Alexis Hiniker}.} \bibinfo{year}{2021}\natexlab{}.
\newblock \showarticletitle{How the Design of YouTube Influences User Sense of Agency}. In \bibinfo{booktitle}{\emph{Proceedings of the 2021 CHI Conference on Human Factors in Computing Systems}} (Yokohama, Japan) \emph{(\bibinfo{series}{CHI '21})}. \bibinfo{publisher}{Association for Computing Machinery}, \bibinfo{address}{New York, NY, USA}, Article \bibinfo{articleno}{368}, \bibinfo{numpages}{17}~pages.
\newblock
\showISBNx{9781450380966}
\urldef\tempurl%
\url{https://doi.org/10.1145/3411764.3445467}
\showDOI{\tempurl}


\bibitem[Lupi{\'a}{\~n}ez-Villanueva et~al\mbox{.}(2022)]%
        {lupianez2022behavioural}
\bibfield{author}{\bibinfo{person}{Francisco Lupi{\'a}{\~n}ez-Villanueva}, \bibinfo{person}{Alba Boluda}, \bibinfo{person}{Francesco Bogliacino}, \bibinfo{person}{Giovanni Liva}, \bibinfo{person}{Lucie Lechardoy}, {and} \bibinfo{person}{Teresa~Rodr{\'\i}guez de~las Heras~Ballell}.} \bibinfo{year}{2022}\natexlab{}.
\newblock \bibinfo{booktitle}{\emph{Behavioural study on unfair commercial practices in the digital environment: dark patterns and manipulative personalisation}}.
\newblock \bibinfo{publisher}{Publications Office of the European Union}.
\newblock


\bibitem[Luria and Candy(2022)]%
        {luria2022letters}
\bibfield{author}{\bibinfo{person}{Michal Luria} {and} \bibinfo{person}{Stuart Candy}.} \bibinfo{year}{2022}\natexlab{}.
\newblock \showarticletitle{Letters from the future: Exploring ethical dilemmas in the design of social agents}. In \bibinfo{booktitle}{\emph{Proceedings of the 2022 CHI Conference on Human Factors in Computing Systems}}. \bibinfo{pages}{1--13}.
\newblock


\bibitem[Lyngs et~al\mbox{.}(2020)]%
        {10.1145/3313831.3376672}
\bibfield{author}{\bibinfo{person}{Ulrik Lyngs}, \bibinfo{person}{Kai Lukoff}, \bibinfo{person}{Petr Slovak}, \bibinfo{person}{William Seymour}, \bibinfo{person}{Helena Webb}, \bibinfo{person}{Marina Jirotka}, \bibinfo{person}{Jun Zhao}, \bibinfo{person}{Max Van~Kleek}, {and} \bibinfo{person}{Nigel Shadbolt}.} \bibinfo{year}{2020}\natexlab{}.
\newblock \showarticletitle{'I Just Want to Hack Myself to Not Get Distracted': Evaluating Design Interventions for Self-Control on Facebook}. In \bibinfo{booktitle}{\emph{Proceedings of the 2020 CHI Conference on Human Factors in Computing Systems}} (Honolulu, HI, USA) \emph{(\bibinfo{series}{CHI '20})}. \bibinfo{publisher}{Association for Computing Machinery}, \bibinfo{address}{New York, NY, USA}, \bibinfo{pages}{1–15}.
\newblock
\showISBNx{9781450367080}
\urldef\tempurl%
\url{https://doi.org/10.1145/3313831.3376672}
\showDOI{\tempurl}


\bibitem[Maier(2019)]%
        {maier2019dark}
\bibfield{author}{\bibinfo{person}{Maximilian Maier}.} \bibinfo{year}{2019}\natexlab{}.
\newblock \bibinfo{title}{Dark patterns--An end user perspective}.
\newblock
\newblock


\bibitem[Martelaro et~al\mbox{.}(2016)]%
        {martelaro2016tell}
\bibfield{author}{\bibinfo{person}{Nikolas Martelaro}, \bibinfo{person}{Victoria~C Nneji}, \bibinfo{person}{Wendy Ju}, {and} \bibinfo{person}{Pamela Hinds}.} \bibinfo{year}{2016}\natexlab{}.
\newblock \showarticletitle{Tell me more designing hri to encourage more trust, disclosure, and companionship}. In \bibinfo{booktitle}{\emph{2016 11th ACM/IEEE International Conference on Human-Robot Interaction (HRI)}}. IEEE, \bibinfo{pages}{181--188}.
\newblock


\bibitem[Mathur et~al\mbox{.}(2019)]%
        {mathur2019dark}
\bibfield{author}{\bibinfo{person}{Arunesh Mathur}, \bibinfo{person}{Gunes Acar}, \bibinfo{person}{Michael~J Friedman}, \bibinfo{person}{Eli Lucherini}, \bibinfo{person}{Jonathan Mayer}, \bibinfo{person}{Marshini Chetty}, {and} \bibinfo{person}{Arvind Narayanan}.} \bibinfo{year}{2019}\natexlab{}.
\newblock \showarticletitle{Dark patterns at scale: Findings from a crawl of 11K shopping websites}.
\newblock \bibinfo{journal}{\emph{Proceedings of the ACM on human-computer interaction}} \bibinfo{volume}{3}, \bibinfo{number}{CSCW} (\bibinfo{year}{2019}), \bibinfo{pages}{1--32}.
\newblock


\bibitem[Mathur et~al\mbox{.}(2021)]%
        {mathur2021makes}
\bibfield{author}{\bibinfo{person}{Arunesh Mathur}, \bibinfo{person}{Mihir Kshirsagar}, {and} \bibinfo{person}{Jonathan Mayer}.} \bibinfo{year}{2021}\natexlab{}.
\newblock \showarticletitle{What makes a dark pattern... dark? Design attributes, normative considerations, and measurement methods}. In \bibinfo{booktitle}{\emph{Proceedings of the 2021 CHI conference on human factors in computing systems}}. \bibinfo{pages}{1--18}.
\newblock


\bibitem[Mayer and Massa(2003)]%
        {mayer2003three}
\bibfield{author}{\bibinfo{person}{Richard~E Mayer} {and} \bibinfo{person}{Laura~J Massa}.} \bibinfo{year}{2003}\natexlab{}.
\newblock \showarticletitle{Three facets of visual and verbal learners: Cognitive ability, cognitive style, and learning preference.}
\newblock \bibinfo{journal}{\emph{Journal of educational psychology}} \bibinfo{volume}{95}, \bibinfo{number}{4} (\bibinfo{year}{2003}), \bibinfo{pages}{833}.
\newblock


\bibitem[McTear(2016)]%
        {mctear2016conversational}
\bibfield{author}{\bibinfo{person}{Michael McTear}.} \bibinfo{year}{2016}\natexlab{}.
\newblock \bibinfo{title}{The Conversational Interface: Talking to Smart Devices}.
\newblock
\newblock


\bibitem[Mildner et~al\mbox{.}(2024)]%
        {mildner2024listening}
\bibfield{author}{\bibinfo{person}{Thomas Mildner}, \bibinfo{person}{Orla Cooney}, \bibinfo{person}{Anna-Maria Meck}, \bibinfo{person}{Marion Bartl}, \bibinfo{person}{Gian-Luca Savino}, \bibinfo{person}{Philip~R Doyle}, \bibinfo{person}{Diego Garaialde}, \bibinfo{person}{Leigh Clark}, \bibinfo{person}{John Sloan}, \bibinfo{person}{Nina Wenig}, {et~al\mbox{.}}} \bibinfo{year}{2024}\natexlab{}.
\newblock \showarticletitle{Listening to the Voices: Describing Ethical Caveats of Conversational User Interfaces According to Experts and Frequent Users}. In \bibinfo{booktitle}{\emph{Proceedings of the CHI Conference on Human Factors in Computing Systems}}. \bibinfo{pages}{1--18}.
\newblock


\bibitem[Moore and Urakami(2022)]%
        {moore2022impact}
\bibfield{author}{\bibinfo{person}{Billie~Akwa Moore} {and} \bibinfo{person}{Jacqueline Urakami}.} \bibinfo{year}{2022}\natexlab{}.
\newblock \showarticletitle{The impact of the physical and social embodiment of voice user interfaces on user distraction}.
\newblock \bibinfo{journal}{\emph{International Journal of Human-Computer Studies}}  \bibinfo{volume}{161} (\bibinfo{year}{2022}), \bibinfo{pages}{102784}.
\newblock


\bibitem[Moors et~al\mbox{.}(2013)]%
        {moors2013appraisal}
\bibfield{author}{\bibinfo{person}{Agnes Moors}, \bibinfo{person}{Phoebe~C Ellsworth}, \bibinfo{person}{Klaus~R Scherer}, {and} \bibinfo{person}{Nico~H Frijda}.} \bibinfo{year}{2013}\natexlab{}.
\newblock \showarticletitle{Appraisal theories of emotion: State of the art and future development}.
\newblock \bibinfo{journal}{\emph{Emotion review}} \bibinfo{volume}{5}, \bibinfo{number}{2} (\bibinfo{year}{2013}), \bibinfo{pages}{119--124}.
\newblock


\bibitem[Mubarak et~al\mbox{.}(2020)]%
        {mubarak2020does}
\bibfield{author}{\bibinfo{person}{Eman Mubarak}, \bibinfo{person}{Tooba Shahid}, \bibinfo{person}{Maryam Mustafa}, {and} \bibinfo{person}{Mustafa Naseem}.} \bibinfo{year}{2020}\natexlab{}.
\newblock \showarticletitle{Does gender and accent of voice matter? an interactive voice response (ivr) experiment}. In \bibinfo{booktitle}{\emph{Proceedings of the 2020 International Conference on Information and Communication Technologies and Development}}. \bibinfo{pages}{1--5}.
\newblock


\bibitem[M{\"u}hlhoff(2015)]%
        {muhlhoff2015affective}
\bibfield{author}{\bibinfo{person}{Rainer M{\"u}hlhoff}.} \bibinfo{year}{2015}\natexlab{}.
\newblock \showarticletitle{Affective resonance and social interaction}.
\newblock \bibinfo{journal}{\emph{Phenomenology and the Cognitive Sciences}}  \bibinfo{volume}{14} (\bibinfo{year}{2015}), \bibinfo{pages}{1001--1019}.
\newblock


\bibitem[Mullennix et~al\mbox{.}(2003)]%
        {mullennix2003social}
\bibfield{author}{\bibinfo{person}{John~W Mullennix}, \bibinfo{person}{Steven~E Stern}, \bibinfo{person}{Stephen~J Wilson}, {and} \bibinfo{person}{Corrie-lynn Dyson}.} \bibinfo{year}{2003}\natexlab{}.
\newblock \showarticletitle{Social perception of male and female computer synthesized speech}.
\newblock \bibinfo{journal}{\emph{Computers in Human Behavior}} \bibinfo{volume}{19}, \bibinfo{number}{4} (\bibinfo{year}{2003}), \bibinfo{pages}{407--424}.
\newblock


\bibitem[Nakai et~al\mbox{.}(2023)]%
        {nakai2023algorithmic}
\bibfield{author}{\bibinfo{person}{Tomoya Nakai}, \bibinfo{person}{Laura Rachman}, \bibinfo{person}{Pablo Arias~Sarah}, \bibinfo{person}{Kazuo Okanoya}, {and} \bibinfo{person}{Jean-Julien Aucouturier}.} \bibinfo{year}{2023}\natexlab{}.
\newblock \showarticletitle{Algorithmic voice transformations reveal the phonological basis of language-familiarity effects in cross-cultural emotion judgments}.
\newblock \bibinfo{journal}{\emph{Plos one}} \bibinfo{volume}{18}, \bibinfo{number}{5} (\bibinfo{year}{2023}), \bibinfo{pages}{e0285028}.
\newblock


\bibitem[News(2024)]%
        {VoicePitch2024}
\bibfield{author}{\bibinfo{person}{Neuroscience News}.} \bibinfo{year}{2024}\natexlab{}.
\newblock \showarticletitle{Voice Pitch Influences Social Perceptions Globally}.
\newblock \bibinfo{journal}{\emph{Neuroscience News}} (\bibinfo{date}{9 Feb} \bibinfo{year}{2024}).
\newblock
\urldef\tempurl%
\url{https://www.neurosciencenews.com/voice-pitch-social-perceptions-22227/}
\showURL{%
\tempurl}
\newblock
\shownote{Summary: A new study unveiled the significant role of voice pitch in shaping social perceptions across different cultures. The study involved over 3,100 participants from 22 countries who evaluated voice recordings for attractiveness, formidability, and prestige. Lower voice pitch was universally preferred for long-term relationships and associated with increased formidability and prestige in males. This cross-cultural research underscores the importance of voice pitch in social evaluations.}.


\bibitem[Niculescu et~al\mbox{.}(2013)]%
        {niculescu2013making}
\bibfield{author}{\bibinfo{person}{Andreea Niculescu}, \bibinfo{person}{Betsy Van~Dijk}, \bibinfo{person}{Anton Nijholt}, \bibinfo{person}{Haizhou Li}, {and} \bibinfo{person}{Swee~Lan See}.} \bibinfo{year}{2013}\natexlab{}.
\newblock \showarticletitle{Making social robots more attractive: the effects of voice pitch, humor and empathy}.
\newblock \bibinfo{journal}{\emph{International journal of social robotics}}  \bibinfo{volume}{5} (\bibinfo{year}{2013}), \bibinfo{pages}{171--191}.
\newblock


\bibitem[Ofuji and Ogasawara(2018)]%
        {ofuji2018verbal}
\bibfield{author}{\bibinfo{person}{Kenta Ofuji} {and} \bibinfo{person}{Naomi Ogasawara}.} \bibinfo{year}{2018}\natexlab{}.
\newblock \showarticletitle{Verbal disaster warnings and perceived intelligibility, reliability, and urgency: The effects of voice gender, fundamental frequency, and speaking rate}.
\newblock \bibinfo{journal}{\emph{Acoustical Science and Technology}} \bibinfo{volume}{39}, \bibinfo{number}{2} (\bibinfo{year}{2018}), \bibinfo{pages}{56--65}.
\newblock


\bibitem[Ouyang and Kaiser(2015)]%
        {ouyang2015prosody}
\bibfield{author}{\bibinfo{person}{Iris~Chuoying Ouyang} {and} \bibinfo{person}{Elsi Kaiser}.} \bibinfo{year}{2015}\natexlab{}.
\newblock \showarticletitle{Prosody and information structure in a tone language: an investigation of Mandarin Chinese}.
\newblock \bibinfo{journal}{\emph{Language, Cognition and Neuroscience}} \bibinfo{volume}{30}, \bibinfo{number}{1-2} (\bibinfo{year}{2015}), \bibinfo{pages}{57--72}.
\newblock


\bibitem[Owens et~al\mbox{.}(2022)]%
        {owens2022exploring}
\bibfield{author}{\bibinfo{person}{Kentrell Owens}, \bibinfo{person}{Johanna Gunawan}, \bibinfo{person}{David Choffnes}, \bibinfo{person}{Pardis Emami-Naeini}, \bibinfo{person}{Tadayoshi Kohno}, {and} \bibinfo{person}{Franziska Roesner}.} \bibinfo{year}{2022}\natexlab{}.
\newblock \showarticletitle{Exploring deceptive design patterns in voice interfaces}. In \bibinfo{booktitle}{\emph{Proceedings of the 2022 European Symposium on Usable Security}}. \bibinfo{pages}{64--78}.
\newblock


\bibitem[Pal et~al\mbox{.}(2022)]%
        {pal2022medmcqa}
\bibfield{author}{\bibinfo{person}{Ankit Pal}, \bibinfo{person}{Logesh~Kumar Umapathi}, {and} \bibinfo{person}{Malaikannan Sankarasubbu}.} \bibinfo{year}{2022}\natexlab{}.
\newblock \showarticletitle{MedMCQA: A Large-scale Multi-Subject Multi-Choice Dataset for Medical domain Question Answering}. In \bibinfo{booktitle}{\emph{Proceedings of the Conference on Health, Inference, and Learning}} \emph{(\bibinfo{series}{Proceedings of Machine Learning Research}, Vol.~\bibinfo{volume}{174})}, \bibfield{editor}{\bibinfo{person}{Gerardo Flores}, \bibinfo{person}{George~H Chen}, \bibinfo{person}{Tom Pollard}, \bibinfo{person}{Joyce~C Ho}, {and} \bibinfo{person}{Tristan Naumann}} (Eds.). \bibinfo{publisher}{PMLR}, \bibinfo{pages}{248--260}.
\newblock
\urldef\tempurl%
\url{https://proceedings.mlr.press/v174/pal22a.html}
\showURL{%
\tempurl}


\bibitem[Pal et~al\mbox{.}(2023)]%
        {pal2023affects}
\bibfield{author}{\bibinfo{person}{Debajyoti Pal}, \bibinfo{person}{Vajirasak Vanijja}, \bibinfo{person}{Himanshu Thapliyal}, {and} \bibinfo{person}{Xiangmin Zhang}.} \bibinfo{year}{2023}\natexlab{}.
\newblock \showarticletitle{What affects the usage of artificial conversational agents? An agent personality and love theory perspective}.
\newblock \bibinfo{journal}{\emph{Computers in Human Behavior}}  \bibinfo{volume}{145} (\bibinfo{year}{2023}), \bibinfo{pages}{107788}.
\newblock


\bibitem[Park et~al\mbox{.}(2024)]%
        {park2024effects}
\bibfield{author}{\bibinfo{person}{Donggun Park}, \bibinfo{person}{Yushin Lee}, {and} \bibinfo{person}{Yong~Min Kim}.} \bibinfo{year}{2024}\natexlab{}.
\newblock \showarticletitle{Effects of autonomous driving context and anthropomorphism of in-vehicle voice agents on intimacy, trust, and intention to use}.
\newblock \bibinfo{journal}{\emph{International Journal of Human--Computer Interaction}} \bibinfo{volume}{40}, \bibinfo{number}{22} (\bibinfo{year}{2024}), \bibinfo{pages}{7179--7192}.
\newblock


\bibitem[Pias et~al\mbox{.}(2024)]%
        {pias2024impact}
\bibfield{author}{\bibinfo{person}{Sabid Bin~Habib Pias}, \bibinfo{person}{Ran Huang}, \bibinfo{person}{Donald~S Williamson}, \bibinfo{person}{Minjeong Kim}, {and} \bibinfo{person}{Apu Kapadia}.} \bibinfo{year}{2024}\natexlab{}.
\newblock \showarticletitle{The Impact of Perceived Tone, Age, and Gender on Voice Assistant Persuasiveness in the Context of Product Recommendations}. In \bibinfo{booktitle}{\emph{Proceedings of the 6th ACM Conference on Conversational User Interfaces}}. \bibinfo{pages}{1--15}.
\newblock


\bibitem[Pisanski and Reby(2021)]%
        {pisanski2021efficacy}
\bibfield{author}{\bibinfo{person}{Katarzyna Pisanski} {and} \bibinfo{person}{David Reby}.} \bibinfo{year}{2021}\natexlab{}.
\newblock \showarticletitle{Efficacy in deceptive vocal exaggeration of human body size}.
\newblock \bibinfo{journal}{\emph{Nature communications}} \bibinfo{volume}{12}, \bibinfo{number}{1} (\bibinfo{year}{2021}), \bibinfo{pages}{968}.
\newblock


\bibitem[Poyatos(1993)]%
        {poyatos1993paralanguage}
\bibfield{author}{\bibinfo{person}{Fernando Poyatos}.} \bibinfo{year}{1993}\natexlab{}.
\newblock \bibinfo{booktitle}{\emph{Paralanguage: A Linguistic and Interdisciplinary Approach to Interactive Speech and Sounds}}.
\newblock \bibinfo{publisher}{John Benjamins Publishing Company}, \bibinfo{address}{Amsterdam/Philadelphia}.
\newblock


\bibitem[Qian et~al\mbox{.}(2025)]%
        {qian2025exploring}
\bibfield{author}{\bibinfo{person}{Zhigu Qian}, \bibinfo{person}{Jiaojiao Fu}, {and} \bibinfo{person}{Yangfan Zhou}.} \bibinfo{year}{2025}\natexlab{}.
\newblock \showarticletitle{Exploring Cultural and Intergenerational Dynamics in Voice Assistant Design for Chinese Older Adults}.
\newblock \bibinfo{journal}{\emph{Proceedings of the ACM on Interactive, Mobile, Wearable and Ubiquitous Technologies}} \bibinfo{volume}{9}, \bibinfo{number}{1} (\bibinfo{year}{2025}), \bibinfo{pages}{1--18}.
\newblock


\bibitem[Rhee and Choi(2020)]%
        {rhee2020effects}
\bibfield{author}{\bibinfo{person}{Chong~Eun Rhee} {and} \bibinfo{person}{Junho Choi}.} \bibinfo{year}{2020}\natexlab{}.
\newblock \showarticletitle{Effects of personalization and social role in voice shopping: An experimental study on product recommendation by a conversational voice agent}.
\newblock \bibinfo{journal}{\emph{Computers in Human Behavior}}  \bibinfo{volume}{109} (\bibinfo{year}{2020}), \bibinfo{pages}{106359}.
\newblock


\bibitem[Romportl(2014)]%
        {romportl2014speech}
\bibfield{author}{\bibinfo{person}{Jan Romportl}.} \bibinfo{year}{2014}\natexlab{}.
\newblock \showarticletitle{Speech synthesis and uncanny valley}. In \bibinfo{booktitle}{\emph{International conference on text, speech, and dialogue}}. Springer, \bibinfo{pages}{595--602}.
\newblock


\bibitem[Russell and Chi(2014)]%
        {russell2014looking}
\bibfield{author}{\bibinfo{person}{Daniel~M Russell} {and} \bibinfo{person}{Ed~H Chi}.} \bibinfo{year}{2014}\natexlab{}.
\newblock \showarticletitle{Looking back: Retrospective study methods for HCI}.
\newblock In \bibinfo{booktitle}{\emph{Ways of Knowing in HCI}}. \bibinfo{publisher}{Springer}, \bibinfo{pages}{373--393}.
\newblock


\bibitem[Rzepka et~al\mbox{.}(2023)]%
        {rzepka2023voice}
\bibfield{author}{\bibinfo{person}{Christine Rzepka}, \bibinfo{person}{Benedikt Berger}, \bibinfo{person}{Anton Koslow}, {and} \bibinfo{person}{Thomas Hess}.} \bibinfo{year}{2023}\natexlab{}.
\newblock \showarticletitle{Voice Assistant, Buy Coffee Capsules!: Understanding the Determinants of Consumers' Intention to Use Voice Commerce}.
\newblock \bibinfo{journal}{\emph{ACM SIGMIS Database: the DATABASE for Advances in Information Systems}} \bibinfo{volume}{54}, \bibinfo{number}{3} (\bibinfo{year}{2023}), \bibinfo{pages}{137--159}.
\newblock


\bibitem[Sapienza(1997)]%
        {sapienza1997aerodynamic}
\bibfield{author}{\bibinfo{person}{Christine~M Sapienza}.} \bibinfo{year}{1997}\natexlab{}.
\newblock \showarticletitle{Aerodynamic and acoustic characteristics of the adult AfricanAmerican voice}.
\newblock \bibinfo{journal}{\emph{Journal of Voice}} \bibinfo{volume}{11}, \bibinfo{number}{4} (\bibinfo{year}{1997}), \bibinfo{pages}{410--416}.
\newblock


\bibitem[{Sathvik Prasad and Bradley Reaves}(2023)]%
        {robocallDatasetTechReport}
\bibfield{author}{\bibinfo{person}{{Sathvik Prasad and Bradley Reaves}}.} \bibinfo{year}{2023}\natexlab{}.
\newblock \bibinfo{booktitle}{\emph{{Robocall Audio from the FTC's Project Point of No Entry}}}.
\newblock \bibinfo{type}{{T}echnical {R}eport} TR-2023-1. \bibinfo{institution}{{North Carolina State University}}.
\newblock


\bibitem[Schirmer et~al\mbox{.}(2019)]%
        {schirmer2019angry}
\bibfield{author}{\bibinfo{person}{Annett Schirmer}, \bibinfo{person}{Yenju Feng}, \bibinfo{person}{Antarika Sen}, {and} \bibinfo{person}{Trevor~B Penney}.} \bibinfo{year}{2019}\natexlab{}.
\newblock \showarticletitle{Angry, old, male--and trustworthy? How expressive and person voice characteristics shape listener trust}.
\newblock \bibinfo{journal}{\emph{PloS one}} \bibinfo{volume}{14}, \bibinfo{number}{1} (\bibinfo{year}{2019}), \bibinfo{pages}{e0210555}.
\newblock


\bibitem[Seaborn et~al\mbox{.}(2024)]%
        {seaborn2024cross}
\bibfield{author}{\bibinfo{person}{Katie Seaborn}, \bibinfo{person}{Iona Gessinger}, \bibinfo{person}{Suzuka Yoshida}, \bibinfo{person}{Benjamin~R Cowan}, {and} \bibinfo{person}{Philip~R Doyle}.} \bibinfo{year}{2024}\natexlab{}.
\newblock \showarticletitle{Cross-cultural validation of partner models for voice user interfaces}. In \bibinfo{booktitle}{\emph{Proceedings of the 6th ACM Conference on Conversational User Interfaces}}. \bibinfo{pages}{1--10}.
\newblock


\bibitem[Seaborn et~al\mbox{.}(2021a)]%
        {seaborn2021voice}
\bibfield{author}{\bibinfo{person}{Katie Seaborn}, \bibinfo{person}{Norihisa~P Miyake}, \bibinfo{person}{Peter Pennefather}, {and} \bibinfo{person}{Mihoko Otake-Matsuura}.} \bibinfo{year}{2021}\natexlab{a}.
\newblock \showarticletitle{Voice in human--agent interaction: A survey}.
\newblock \bibinfo{journal}{\emph{ACM Computing Surveys (CSUR)}} \bibinfo{volume}{54}, \bibinfo{number}{4} (\bibinfo{year}{2021}), \bibinfo{pages}{1--43}.
\newblock


\bibitem[Seaborn et~al\mbox{.}(2021b)]%
        {10.1145/3386867}
\bibfield{author}{\bibinfo{person}{Katie Seaborn}, \bibinfo{person}{Norihisa~P. Miyake}, \bibinfo{person}{Peter Pennefather}, {and} \bibinfo{person}{Mihoko Otake-Matsuura}.} \bibinfo{year}{2021}\natexlab{b}.
\newblock \showarticletitle{Voice in Human–Agent Interaction: A Survey}.
\newblock \bibinfo{journal}{\emph{ACM Comput. Surv.}} \bibinfo{volume}{54}, \bibinfo{number}{4}, Article \bibinfo{articleno}{81} (\bibinfo{date}{may} \bibinfo{year}{2021}), \bibinfo{numpages}{43}~pages.
\newblock
\showISSN{0360-0300}
\urldef\tempurl%
\url{https://doi.org/10.1145/3386867}
\showDOI{\tempurl}


\bibitem[Seaborn and Urakami(2021)]%
        {seaborn2021measuring}
\bibfield{author}{\bibinfo{person}{Katie Seaborn} {and} \bibinfo{person}{Jacqueline Urakami}.} \bibinfo{year}{2021}\natexlab{}.
\newblock \showarticletitle{Measuring voice UX quantitatively: A rapid review}. In \bibinfo{booktitle}{\emph{Extended abstracts of the 2021 CHI conference on human factors in computing systems}}. \bibinfo{pages}{1--8}.
\newblock


\bibitem[Sun et~al\mbox{.}(2018)]%
        {sun2018lip}
\bibfield{author}{\bibinfo{person}{Ke Sun}, \bibinfo{person}{Chun Yu}, \bibinfo{person}{Weinan Shi}, \bibinfo{person}{Lan Liu}, {and} \bibinfo{person}{Yuanchun Shi}.} \bibinfo{year}{2018}\natexlab{}.
\newblock \showarticletitle{Lip-interact: Improving mobile device interaction with silent speech commands}. In \bibinfo{booktitle}{\emph{Proceedings of the 31st Annual ACM Symposium on User Interface Software and Technology}}. \bibinfo{pages}{581--593}.
\newblock


\bibitem[Sweller(2011)]%
        {sweller2011cognitive}
\bibfield{author}{\bibinfo{person}{John Sweller}.} \bibinfo{year}{2011}\natexlab{}.
\newblock \bibinfo{title}{Cognitive load theory}.
\newblock
\newblock


\bibitem[Van~den Steen(2009)]%
        {van2009authority}
\bibfield{author}{\bibinfo{person}{Eric Van~den Steen}.} \bibinfo{year}{2009}\natexlab{}.
\newblock \showarticletitle{Authority versus persuasion}.
\newblock \bibinfo{journal}{\emph{American Economic Review}} \bibinfo{volume}{99}, \bibinfo{number}{2} (\bibinfo{year}{2009}), \bibinfo{pages}{448--453}.
\newblock


\bibitem[Van~Zant and Berger(2020)]%
        {van2020voice}
\bibfield{author}{\bibinfo{person}{Alex~B Van~Zant} {and} \bibinfo{person}{Jonah Berger}.} \bibinfo{year}{2020}\natexlab{}.
\newblock \showarticletitle{How the voice persuades.}
\newblock \bibinfo{journal}{\emph{Journal of personality and social psychology}} \bibinfo{volume}{118}, \bibinfo{number}{4} (\bibinfo{year}{2020}), \bibinfo{pages}{661}.
\newblock


\bibitem[Viswanathan and Viswanathan(2005)]%
        {viswanathan2005measuring}
\bibfield{author}{\bibinfo{person}{Mahesh Viswanathan} {and} \bibinfo{person}{Madhubalan Viswanathan}.} \bibinfo{year}{2005}\natexlab{}.
\newblock \showarticletitle{Measuring speech quality for text-to-speech systems: development and assessment of a modified mean opinion score (MOS) scale}.
\newblock \bibinfo{journal}{\emph{Computer speech \& language}} \bibinfo{volume}{19}, \bibinfo{number}{1} (\bibinfo{year}{2005}), \bibinfo{pages}{55--83}.
\newblock


\bibitem[Wang et~al\mbox{.}(2021)]%
        {wang2021demystifying}
\bibfield{author}{\bibinfo{person}{Dawei Wang}, \bibinfo{person}{Kai Chen}, {and} \bibinfo{person}{Wei Wang}.} \bibinfo{year}{2021}\natexlab{}.
\newblock \showarticletitle{Demystifying the vetting process of voice-controlled skills on markets}.
\newblock \bibinfo{journal}{\emph{Proceedings of the ACM on Interactive, Mobile, Wearable and Ubiquitous Technologies}} \bibinfo{volume}{5}, \bibinfo{number}{3} (\bibinfo{year}{2021}), \bibinfo{pages}{1--28}.
\newblock


\bibitem[Wang(2017)]%
        {wang2017liar}
\bibfield{author}{\bibinfo{person}{William~Yang Wang}.} \bibinfo{year}{2017}\natexlab{}.
\newblock \showarticletitle{“Liar, Liar Pants on Fire”: A New Benchmark Dataset for Fake News Detection}. In \bibinfo{booktitle}{\emph{Proceedings of the 55th Annual Meeting of the Association for Computational Linguistics (Volume 2: Short Papers)}}. \bibinfo{pages}{422--426}.
\newblock


\bibitem[Wang et~al\mbox{.}(2023)]%
        {wang2023dark}
\bibfield{author}{\bibinfo{person}{Xian Wang}, \bibinfo{person}{Lik-Hang Lee}, \bibinfo{person}{Carlos Bermejo~Fernandez}, {and} \bibinfo{person}{Pan Hui}.} \bibinfo{year}{2023}\natexlab{}.
\newblock \showarticletitle{The dark side of augmented reality: Exploring manipulative designs in AR}.
\newblock \bibinfo{journal}{\emph{International Journal of Human--Computer Interaction}} (\bibinfo{year}{2023}), \bibinfo{pages}{1--16}.
\newblock


\bibitem[Widdicks et~al\mbox{.}(2020)]%
        {widdicks2020backfiring}
\bibfield{author}{\bibinfo{person}{Kelly Widdicks}, \bibinfo{person}{Daniel Pargman}, {and} \bibinfo{person}{Staffan Bjork}.} \bibinfo{year}{2020}\natexlab{}.
\newblock \showarticletitle{Backfiring and favouring: How design processes in HCI lead to anti-patterns and repentant designers}. In \bibinfo{booktitle}{\emph{Proceedings of the 11th Nordic Conference on Human-Computer Interaction: Shaping Experiences, Shaping Society}}. \bibinfo{pages}{1--12}.
\newblock


\bibitem[Williams and Moser(2019)]%
        {williams2019art}
\bibfield{author}{\bibinfo{person}{Michael Williams} {and} \bibinfo{person}{Tami Moser}.} \bibinfo{year}{2019}\natexlab{}.
\newblock \showarticletitle{The art of coding and thematic exploration in qualitative research}.
\newblock \bibinfo{journal}{\emph{International management review}} \bibinfo{volume}{15}, \bibinfo{number}{1} (\bibinfo{year}{2019}), \bibinfo{pages}{45--55}.
\newblock


\bibitem[Yarosh et~al\mbox{.}(2018)]%
        {yarosh2018children}
\bibfield{author}{\bibinfo{person}{Svetlana Yarosh}, \bibinfo{person}{Stryker Thompson}, \bibinfo{person}{Kathleen Watson}, \bibinfo{person}{Alice Chase}, \bibinfo{person}{Ashwin Senthilkumar}, \bibinfo{person}{Ye Yuan}, {and} \bibinfo{person}{AJ~Bernheim Brush}.} \bibinfo{year}{2018}\natexlab{}.
\newblock \showarticletitle{Children asking questions: speech interface reformulations and personification preferences}. In \bibinfo{booktitle}{\emph{Proceedings of the 17th ACM conference on interaction design and children}}. \bibinfo{pages}{300--312}.
\newblock


\bibitem[Yilmazyildiz et~al\mbox{.}(2015)]%
        {yilmazyildiz2015gibberish}
\bibfield{author}{\bibinfo{person}{Selma Yilmazyildiz}, \bibinfo{person}{Werner Verhelst}, {and} \bibinfo{person}{Hichem Sahli}.} \bibinfo{year}{2015}\natexlab{}.
\newblock \showarticletitle{Gibberish speech as a tool for the study of affective expressiveness for robotic agents}.
\newblock \bibinfo{journal}{\emph{Multimedia Tools and Applications}}  \bibinfo{volume}{74} (\bibinfo{year}{2015}), \bibinfo{pages}{9959--9982}.
\newblock


\bibitem[Zaga et~al\mbox{.}(2016)]%
        {zaga2016help}
\bibfield{author}{\bibinfo{person}{Cristina Zaga}, \bibinfo{person}{Roelof~AJ De~Vries}, \bibinfo{person}{Sem~J Spenkelink}, \bibinfo{person}{Khiet~P Truong}, {and} \bibinfo{person}{Vanessa Evers}.} \bibinfo{year}{2016}\natexlab{}.
\newblock \showarticletitle{Help-giving robot behaviors in child-robot games: Exploring semantic free utterances}. In \bibinfo{booktitle}{\emph{2016 11th ACM/IEEE International Conference on Human-Robot Interaction (HRI)}}. IEEE, \bibinfo{pages}{541--542}.
\newblock


\bibitem[Zagal et~al\mbox{.}(2013)]%
        {zagal2013dark}
\bibfield{author}{\bibinfo{person}{Jos{\'e}~P Zagal}, \bibinfo{person}{Staffan Bj{\"o}rk}, {and} \bibinfo{person}{Chris Lewis}.} \bibinfo{year}{2013}\natexlab{}.
\newblock \showarticletitle{Dark patterns in the design of games}. In \bibinfo{booktitle}{\emph{Foundations of Digital Games 2013}}.
\newblock


\bibitem[Zhan et~al\mbox{.}(2023)]%
        {zhan2023deceptive}
\bibfield{author}{\bibinfo{person}{Xiao Zhan}, \bibinfo{person}{Yifan Xu}, {and} \bibinfo{person}{Stefan Sarkadi}.} \bibinfo{year}{2023}\natexlab{}.
\newblock \showarticletitle{Deceptive AI ecosystems: The case of ChatGPT}. In \bibinfo{booktitle}{\emph{Proceedings of the 5th International Conference on Conversational User Interfaces}}. \bibinfo{pages}{1--6}.
\newblock


\bibitem[Zhang et~al\mbox{.}(2011)]%
        {zhang2011effects}
\bibfield{author}{\bibinfo{person}{Lu Zhang}, \bibinfo{person}{Lee~B Erickson}, {and} \bibinfo{person}{Heidi~C Webb}.} \bibinfo{year}{2011}\natexlab{}.
\newblock \showarticletitle{Effects of “emotional text” on online customer service chat}.
\newblock  (\bibinfo{year}{2011}).
\newblock


\bibitem[Zhang et~al\mbox{.}(2018)]%
        {zhang2018modeling}
\bibfield{author}{\bibinfo{person}{Zhuosheng Zhang}, \bibinfo{person}{Jiangtong Li}, \bibinfo{person}{Pengfei Zhu}, \bibinfo{person}{Hai Zhao}, {and} \bibinfo{person}{Gongshen Liu}.} \bibinfo{year}{2018}\natexlab{}.
\newblock \showarticletitle{Modeling multi-turn conversation with deep utterance aggregation}.
\newblock \bibinfo{journal}{\emph{arXiv preprint arXiv:1806.09102}} (\bibinfo{year}{2018}).
\newblock


\bibitem[Zhong and Ma(2022)]%
        {zhong2022effects}
\bibfield{author}{\bibinfo{person}{Runting Zhong} {and} \bibinfo{person}{Mengyao Ma}.} \bibinfo{year}{2022}\natexlab{}.
\newblock \showarticletitle{Effects of communication style, anthropomorphic setting and individual differences on older adults using voice assistants in a health context}.
\newblock \bibinfo{journal}{\emph{BMC geriatrics}} \bibinfo{volume}{22}, \bibinfo{number}{1} (\bibinfo{year}{2022}), \bibinfo{pages}{751}.
\newblock


\end{thebibliography}


\appendix

\section{Ethical Considerations}

We acknowledge the ethical concerns in our studies and have taken measures to ensure compliance. Guided by the Menlo Report~\cite{bailey2012menlo} and the Belmont Report~\cite{beauchamp2008belmont}. The study design was reviewed with faculty and approved by our university's ethics committee. Due to the study's deceptive elements, we included debriefing sessions to clarify the true aims, inform participants of their right to delete data, and offer additional compensation if they were dissatisfied, although no participants expressed dissatisfaction. Participants were compensated properly and could withdraw from the study at any time. Behavioral data, recordings, and interview transcripts were stored temporarily on a local encrypted device, anonymized, and deleted after analysis.

\section{Themes Emerged During the Main Study}

Table~\ref{tbl:appen_theme} showed the themes merged during the main study.

\begin{table}[ht]
\centering
\begin{tabular}{c|c|p{13cm}}
\toprule
\rowcolor{gray!30} \textbf{Scenario} & \textbf{Aim} & \textbf{Description} \\ \hline
\multirow{3}{*}{shop} & emp & \colorbox{green!10}{Lively, warm, and engaging tone (9)}, \colorbox{green!10}{Faster speed to intuitively highlight product features (4)}, \colorbox{green!10}{``charming girl'' voice (2)}, \colorbox{red!10}{Avoid speaking too fast (4)}, \colorbox{green!10}{Greater impact of voice and tone (14)}, \colorbox{green!10}{Clear articulation and clarity (5)}, \colorbox{green!10}{Pleasant  tone and pitch (3)}, \colorbox{green!10}{Attractive and creates a desire to purchase (4)}, \colorbox{red!10}{Avoid strange pauses (1)}, \colorbox{red!10}{Avoid being too serious or mechanical (5)}, \colorbox{green!10}{Concise (1)} \\ \cline{2-3}
& per & \colorbox{green!10}{Tone has an impact (22)}, \colorbox{green!10} {Prefers enthusiasm/passion/vitality (8)}, \colorbox{green!10} {Attractiveness of words (6)}, \colorbox{green!10}{Tone has a major impact (9)}, \colorbox{green!10} {Natural and comfortable (3)}, \colorbox{red!10}{Avoid offensive/urgent tone (8)}, \colorbox{red!10}{Avoid being overly mechanical or stiff (4)}, \colorbox{green!10}{Voice of people of the same age (1)} \\ \cline{2-3}
& weak & \colorbox{green!10}{Tone has an impact (18)}, \colorbox{green!10}{Solution-oriented content (9)}, \colorbox{green!10}{Tone dominates (5)}, \colorbox{green!10}{Enthusiasm/sincerity/emotional value (12)}, \colorbox{red!10}{ Avoid friendly/neutral/monotone (6)}, \colorbox{red!10}{Avoid being too lively (2)}, \colorbox{green!10}{Comfortable mood (2)}, \colorbox{red!10}{Avoid authoritative/robotic/offensive tone (2)}, \colorbox{green!10}{Natural (1)}\\ \hline
qa & emp & \colorbox{green!10}{Tone has an impact (18)}, \colorbox{green!10}{Tone dominates (6)}, \colorbox{green!10}{Enthusiastic/natural (7)}, \colorbox{green!10}{Friendly (1)}, \colorbox{red!10}{Dislikes neutral/stiff/human-machine interaction (5)}, \colorbox{red!10}{Avoid being too serious (2)}, \colorbox{red!10}{Source is not very useful (1)}, \colorbox{green!10}{Directness (1)}, \colorbox{red!10}{Avoid adding emotional tone, as it would resemble short videos (1)}, \colorbox{green!10}{Prefers authoritative tone (5)}, \colorbox{green!10}{More credible sources (2)}, \colorbox{green!10}{Correspondence with text content (1)}, \colorbox{red!10}{Avoid speaking too fast (2)}, \colorbox{red!10}{Avoid urgency (1)}, \colorbox{green!10}{Pleasant tone (1)}, \colorbox{green!10}{Pauses (1)}, \colorbox{red!10}{Description of drama (1)}\\ \cline{2-3}
& per & \colorbox{green!10}{Tone has an impact (21)}, \colorbox{green!10}{Tone dominates (4)}, \colorbox{green!10}{Neutral/authoritative/mature (11)}, \colorbox{red!10}{Tone sounds somewhat like sales or short video (7)}, \colorbox{green!10}{Affinity (1)}, \colorbox{green!10}{"Charming girl" voice (3)}, \colorbox{green!10}{Energetic/passionate/willing to listen (5)}, \colorbox{green!10}{Friendly (2)}, \colorbox{red!10}{Avoid speaking too slowly (2)}, \colorbox{red!10}{Avoid stuttering (4)} \colorbox{red!10}{Avoid robotic/stiff tone (1)}, \colorbox{green!10}{Natural (1)}, \colorbox{red!10}{Avoid speaking too quickly (3)}, \colorbox{green!10}{Young (1)}, \colorbox{green!10}{Content (1)}\\ \cline{2-3}
& weak & \colorbox{green!10}{Tone has an impact (17)}, \colorbox{green!10}{Tone dominates (4)}, \colorbox{green!10}{Contrast (1)}, \colorbox{green!10}{Enthusiastic (1)}, \colorbox{red!10}{Avoid slow speech (1)}, \colorbox{red!10}{Avoid disinterest in text content/text content (8)}, \colorbox{red!10}{Avoid being too enthusiastic (1)}, \colorbox{green!10}{Be sincere (1)}, \colorbox{red!10}{Neutral tone lacks professionalism (1)}, \colorbox{red!10}{Avoid speaking too quickly (3)}, \colorbox{green!10}{Be natural and normal (1)}, \colorbox{red!10}{Avoid mechanical tone (1)}, \colorbox{green!10}{Calm and mature (1)}, \colorbox{green!10}{Authoritative (2)}, \colorbox{green!10}{Enjoyable and uplifting (1)}, \colorbox{red!10}{Avoid pauses (1)}\\ \hline
\end{tabular}
\caption{The themes affecting users' decision. The phrases highlighted \colorbox{green!10}{in green} denotes positive descriptions, whereas those \colorbox{red!10}{in red} denotes negative descriptions.}
\label{tbl:appen_theme}
\end{table}

\begin{table}[htbp]
\centering
\caption{Classification of Dark Patterns - Part 1}
\label{tbl:dark_patterns_classification_part1}
\resizebox{\textwidth}{!}{
\begin{tabular}{p{2cm}|p{2.5cm}|ccccc ccccc ccccc ccccc c}
\toprule
\rowcolor{gray!20} 
\textbf{Category} & \textbf{Type} & \raisebox{-0.5ex}{\cite{brignull2018dark}} & \raisebox{-0.5ex}{\cite{gray2018dark}} & \raisebox{-0.5ex}{\cite{bosch2016tales}} & \raisebox{-0.5ex}{\cite{conti2010malicious}} & \raisebox{-0.5ex}{\cite{greenberg2014dark}} & \raisebox{-0.5ex}{\cite{10.1145/3357236.3395486}} & \raisebox{-0.5ex}{\cite{mathur2021makes}} & \raisebox{-0.5ex}{\cite{mathur2019dark}} & \raisebox{-0.5ex}{\cite{10.1145/3411764.3445467}} & \raisebox{-0.5ex}{\cite{burr2018analysis}} & \raisebox{-0.5ex}{\cite{widdicks2020backfiring}} & \raisebox{-0.5ex}{\cite{bongard2021definitely}} & \raisebox{-0.5ex}{\cite{10.1145/3532106.3533562}} & \raisebox{-0.5ex}{\cite{zagal2013dark}} & \raisebox{-0.5ex}{\cite{maier2019dark}} & \raisebox{-0.5ex}{\cite{10.1145/3479600}} & \raisebox{-0.5ex}{\cite{luguri2021shining}} & \raisebox{-0.5ex}{\cite{chatellier2019shaping}} & \raisebox{-0.5ex}{\cite{di2020ui}} & \raisebox{-0.5ex}{\cite{10.1145/3313831.3376672}} & \raisebox{-0.5ex}{\cite{10.1145/3491102.3501899}} \\
\midrule

\multirow{9}{*}\textbf{Emphasizing}
& Interface Interference & $\bullet$ & $\bullet$ & $\circ$ & $\bullet$ & $\circ$ & $\bullet$ & $\bullet$ & $\bullet$ & $\bullet$ & $\circ$ & $\circ$ & $\bullet$ & $\bullet$ &  & $\bullet$ & $\bullet$ & $\bullet$ & $\bullet$ & $\bullet$ & $\bullet$ & $\bullet$ \\
& Visual Interference & $\circ$ & $\bullet$ & $\circ$ & $\bullet$ & $\circ$ &  & $\bullet$ & $\bullet$ & $\bullet$ & $\circ$ &  & $\bullet$ &  &  & $\bullet$ &  & $\bullet$ & $\bullet$ & $\bullet$ & $\bullet$ & $\bullet$ \\
& False Hierarchy &  & $\bullet$ & $\circ$ & $\bullet$ &  & $\circ$ & $\bullet$ & $\circ$ & $\bullet$ & $\circ$ &  & $\bullet$ &  &  & $\bullet$ &  & $\bullet$ &  & $\bullet$ & $\circ$ & $\circ$ \\
& Aesthetic Manipulation &  & $\bullet$ & $\circ$ & $\bullet$ & $\bullet$ &  & $\bullet$ & $\circ$ & $\bullet$ & $\bullet$ &  & $\circ$ &  &  & $\bullet$ & $\circ$ & $\bullet$ & $\circ$ & $\bullet$ & $\circ$ & $\bullet$ \\
& Manipulating Navigation &  &  &  & $\bullet$ & $\circ$ &  &  & $\circ$ & $\bullet$ & $\circ$ &  &  &  &  & $\bullet$ & $\bullet$ & $\circ$ & $\bullet$ &  & $\bullet$ & $\bullet$ \\
& Bias Grind &  &  &  &  &  &  &  &  &  &  &  &  & $\bullet$ &  &  & $\circ$ &  &  &  &  &  \\
& Habitual Feature Tour &  &  &  &  &  &  &  &  &  &  &  &  &  &  &  & $\bullet$ &  &  &  & $\circ$ & $\circ$ \\
& Trick Wording & $\bullet$ & $\bullet$ & $\bullet$ & $\bullet$ &  &  & $\bullet$ & $\bullet$ &  & $\circ$ &  & $\bullet$ &  &  & $\bullet$ &  & $\bullet$ & $\bullet$ & $\bullet$ &  &  \\
& Bad Defaults &  &  & $\bullet$ & $\circ$ &  &  & $\circ$ & $\circ$ &  & $\bullet$ &  &  &  &  &  &  & $\bullet$ & $\circ$ & $\bullet$ & $\bullet$ & $\bullet$ \\
\hline

\multirow{16}{*}\textbf{Persuading}
& Confirmshaming & $\bullet$ & $\circ$ & $\circ$ & $\bullet$ &  & $\bullet$ & $\bullet$ & $\bullet$ &  & $\circ$ &  & $\bullet$ &  &  & $\bullet$ &  & $\bullet$ &  & $\circ$ &  & $\circ$ \\
& Fake Social Proof & $\bullet$ & $\circ$ & $\bullet$ & $\bullet$ & $\circ$ &  & $\bullet$ & $\bullet$ &  &  & $\circ$ &  &  &  &  &  & $\bullet$ &  & $\bullet$ & $\circ$ & $\circ$ \\
& Fake Urgency & $\bullet$ & $\circ$ & $\circ$ & $\circ$ &  &  & $\bullet$ & $\bullet$ & $\circ$ &  &  & $\bullet$ &  &  &  &  & $\bullet$ &  & $\bullet$ &  & $\circ$ \\
& Fake Scarcity & $\bullet$ & $\circ$ & $\circ$ & $\circ$ &  &  & $\bullet$ & $\bullet$ &  &  &  & $\bullet$ &  &  &  &  & $\bullet$ &  & $\bullet$ &  &  \\
& Toying with Emotion &  & $\bullet$ & $\bullet$ & $\bullet$ &  &  & $\bullet$ & $\circ$ &  & $\bullet$ &  &  &  &  & $\bullet$ &  & $\bullet$ & $\bullet$ & $\bullet$ & $\circ$ & $\bullet$ \\
& Trading &  &  &  &  &  &  &  &  &  & $\bullet$ &  &  &  &  &  &  &  &  &  &  &  \\
& Nudging &  &  &  &  &  &  &  &  &  & $\bullet$ &  &  &  &  &  &  & $\circ$ & $\bullet$ &  & $\circ$ &  \\
& Loss-gain framing &  &  &  &  &  &  &  &  &  &  &  & $\bullet$ &  &  &  &  & $\bullet$ & $\bullet$ &  & $\circ$ &  \\
& Cuteness &  &  &  &  &  &  &  &  &  &  &  &  &  &  &  &  & $\bullet$ &  &  &  &  \\
& Captive Audience &  &  &  & $\circ$ & $\bullet$ & $\circ$ &  &  & $\circ$ &  &  &  &  &  &  &  & $\circ$ & $\circ$ &  & $\bullet$ & $\bullet$ \\
& Milk Factor &  &  &  & $\circ$ & $\bullet$ & $\circ$ &  &  &  &  &  &  &  &  &  &  &  &  &  &  &  \\
& Playing by Appointment &  &  &  &  &  &  &  &  &  &  &  &  &  & $\bullet$ &  &  &  &  & $\bullet$ &  &  \\
& Cross-Device Switching &  &  &  &  &  &  &  &  &  &  &  &  &  &  &  &  &  &  &  & $\bullet$ &  \\
& Emotional Exploitation &  &  &  &  &  &  &  &  &  &  &  &  &  &  &  &  &  &  &  & $\bullet$ & $\bullet$ \\
& Zone States &  &  &  &  &  &  &  &  &  &  &  &  &  &  &  &  &  &  &  & $\circ$ & $\bullet$ \\
& Personalization &  &  &  &  &  &  &  &  &  &  &  &  &  &  &  &  & $\circ$ & $\circ$ & $\bullet$ & $\circ$ &  \\
\hline

\multirow{17}{*}\textbf{Weakening} 
& Obstruction & $\bullet$ & $\bullet$ & $\bullet$ & $\bullet$ & $\bullet$ & $\bullet$ & $\bullet$ & $\bullet$ & $\circ$ & $\bullet$ & $\circ$ & $\bullet$ & $\bullet$ & $\bullet$ & $\bullet$ & $\circ$ & $\bullet$ & $\bullet$ & $\circ$ & $\circ$ & $\circ$ \\
& Hard to Cancel & $\bullet$ & $\bullet$ & $\bullet$ & $\bullet$ & $\circ$ &  & $\bullet$ & $\bullet$ &  & $\circ$ & $\circ$ & $\circ$ &  &  & $\bullet$ &  & $\bullet$ &  & $\bullet$ & $\circ$ & $\bullet$ \\
& Comparison Prevention & $\bullet$ & $\circ$ & $\circ$ & $\bullet$ &  &  &  & $\circ$ &  &  &  &  &  &  &  &  & $\circ$ &  & $\bullet$ &  &  \\
& Switchoff Delay &  &  &  &  &  &  &  &  &  &  &  &  & $\bullet$ &  &  &  &  &  &  &  &  \\
& Favouring Anti-Patterns &  &  &  &  &  &  &  &  &  &  & $\bullet$ &  &  &  &  &  &  &  &  &  &  \\
& Pay to Skip &  &  &  &  &  &  &  &  &  &  &  &  &  & $\bullet$ &  &  &  &  &  &  &  \\
& Grinding &  &  &  &  &  &  &  &  &  &  &  &  &  & $\bullet$ &  &  &  &  &  &  &  \\
& Infinite Scroll &  &  &  &  &  &  &  &  &  &  & $\circ$ &  &  &  &  & $\bullet$ & $\circ$ &  &  & $\bullet$ & $\bullet$ \\
& Roach Motel &  &  &  &  &  &  &  &  &  &  &  &  &  &  &  &  & $\bullet$ &  & $\bullet$ & $\circ$ &  \\
& Distraction &  &  &  & $\bullet$ & $\bullet$ &  &  & $\circ$ &  &  & $\bullet$ &  &  &  &  & $\bullet$ & $\circ$ & $\circ$ &  & $\bullet$ & $\bullet$ \\
& Attention Grabber &  &  &  & $\circ$ & $\bullet$ & $\circ$ &  & $\circ$ &  &  &  & $\circ$ &  &  &  & $\bullet$ & $\circ$ & $\bullet$ &  & $\bullet$ & $\bullet$ \\
& Interruption &  & $\circ$ &  & $\bullet$ & $\bullet$ &  &  &  &  & $\circ$ & $\circ$ & $\bullet$ &  &  &  & $\circ$ & $\circ$ & $\bullet$ & $\bullet$ & $\bullet$ & $\bullet$ \\
& Attention Quicksand &  &  &  &  &  &  &  &  &  &  &  &  & $\bullet$ &  &  & $\bullet$ &  & $\circ$ &  & $\bullet$ & $\bullet$ \\
& Goal Displacement &  &  &  &  &  &  &  &  &  &  &  &  &  &  &  &  &  &  &  & $\bullet$ & $\bullet$ \\
& Normative Dissociation &  &  &  &  &  &  &  &  &  &  &  &  &  &  &  &  &  &  &  & $\circ$ & $\bullet$ \\
& Immortal Accounts &  & $\circ$ & $\bullet$ & $\bullet$ & $\circ$ &  &  & $\circ$ &  &  &  &  &  &  &  &  &  &  & $\bullet$ & $\circ$ &  \\
& Intermediate Currency &  & $\bullet$ &  &  &  &  &  &  &  &  &  &  &  &  &  &  &  &  & $\bullet$ &  &  \\
\bottomrule
\end{tabular}}

\vspace{1ex}
\begin{tabular}{lll}
\textbf{Legend:} & $\bullet$ = Directly mentioned/fully covered & $\circ$ = Similar concept/partially covered \\
\end{tabular}
\end{table}

\begin{table}[htbp]
\centering
\caption{Classification of Dark Patterns- Part 2}
\label{tbl:dark_patterns_classification_part2}
\resizebox{\textwidth}{!}{
\begin{tabular}{p{2cm}|p{2.5cm}|ccccc ccccc ccccc ccccc c}
\toprule
\rowcolor{gray!20} 
\textbf{Category} & \textbf{Type} & \raisebox{-0.5ex}{\cite{brignull2018dark}} & \raisebox{-0.5ex}{\cite{gray2018dark}} & \raisebox{-0.5ex}{\cite{bosch2016tales}} & \raisebox{-0.5ex}{\cite{conti2010malicious}} & \raisebox{-0.5ex}{\cite{greenberg2014dark}} & \raisebox{-0.5ex}{\cite{10.1145/3357236.3395486}} & \raisebox{-0.5ex}{\cite{mathur2021makes}} & \raisebox{-0.5ex}{\cite{mathur2019dark}} & \raisebox{-0.5ex}{\cite{10.1145/3411764.3445467}} & \raisebox{-0.5ex}{\cite{burr2018analysis}} & \raisebox{-0.5ex}{\cite{widdicks2020backfiring}} & \raisebox{-0.5ex}{\cite{bongard2021definitely}} & \raisebox{-0.5ex}{\cite{10.1145/3532106.3533562}} & \raisebox{-0.5ex}{\cite{zagal2013dark}} & \raisebox{-0.5ex}{\cite{maier2019dark}} & \raisebox{-0.5ex}{\cite{10.1145/3479600}} & \raisebox{-0.5ex}{\cite{luguri2021shining}} & \raisebox{-0.5ex}{\cite{chatellier2019shaping}} & \raisebox{-0.5ex}{\cite{di2020ui}} & \raisebox{-0.5ex}{\cite{10.1145/3313831.3376672}} & \raisebox{-0.5ex}{\cite{10.1145/3491102.3501899}} \\
\midrule

\multirow{21}{*}\textbf{Malicious} 
& Sneaking & $\bullet$ & $\bullet$ & $\bullet$ & $\bullet$ & $\bullet$ & $\bullet$ & $\bullet$ & $\bullet$ & $\circ$ & $\bullet$ & $\circ$ & $\bullet$ & $\bullet$ & $\bullet$ & $\bullet$ & $\bullet$ & $\bullet$ & $\circ$ & $\bullet$ & $\bullet$ & $\bullet$ \\
& Hidden Costs & $\bullet$ & $\bullet$ &  & $\bullet$ &  &  & $\bullet$ & $\bullet$ &  & $\circ$ &  &  &  &  & $\bullet$ &  & $\bullet$ &  &  & $\circ$ & $\bullet$ \\
& Hidden Subscription & $\bullet$ & $\bullet$ &  & $\circ$ &  &  & $\bullet$ & $\bullet$ &  & $\circ$ &  &  &  &  & $\bullet$ &  & $\bullet$ &  &  &  &  \\
& Hidden Information &  & $\bullet$ & $\bullet$ & $\bullet$ & $\circ$ &  & $\bullet$ & $\circ$ &  & $\circ$ &  & $\bullet$ &  &  & $\bullet$ &  & $\bullet$ & $\bullet$ & $\bullet$ & $\circ$ & $\circ$ \\
& Hidden Legalese Stipulations &  & $\circ$ & $\bullet$ & $\bullet$ &  &  &  & $\circ$ &  &  &  &  &  &  &  &  & $\circ$ &  & $\bullet$ &  &  \\
& Feature Fog &  &  &  &  &  &  &  &  &  &  &  &  & $\bullet$ &  &  &  &  &  &  &  &  \\
& Pre-delivered Content &  &  &  &  &  &  &  &  &  &  &  &  &  & $\bullet$ &  &  &  &  &  &  &  \\
& Backfiring Anti-Patterns &  &  &  &  &  &  &  &  &  &  & $\bullet$ &  &  &  &  &  &  &  &  &  &  \\
& Sneak into Basket &  &  &  &  &  &  &  &  &  &  &  &  &  &  &  &  & $\bullet$ &  & $\bullet$ &  &  \\
& Disguised Ads & $\bullet$ & $\bullet$ &  & $\bullet$ & $\circ$ & $\bullet$ &  &  &  &  &  &  &  &  &  &  & $\circ$ & $\circ$ & $\bullet$ &  &  \\
& Bait and Switch & $\bullet$ & $\bullet$ &  & $\circ$ & $\bullet$ & $\circ$ &  &  &  &  &  &  &  &  &  &  & $\bullet$ & $\bullet$ & $\bullet$ & $\bullet$ & $\bullet$ \\
& Disguised Data Collection &  &  & $\circ$ & $\circ$ & $\bullet$ & $\circ$ &  &  &  &  &  &  &  &  &  &  & $\circ$ & $\bullet$ &  &  &  \\
& Value Alignment &  &  &  &  &  &  &  &  &  & $\bullet$ &  &  &  &  &  &  &  &  &  &  &  \\
& False or Misleading Messages &  &  &  &  &  &  &  &  &  &  &  &  &  &  &  &  & $\bullet$ & $\circ$ &  &  &  \\
& Reference Pricing &  &  &  &  &  &  &  &  &  &  &  &  &  &  &  &  & $\circ$ &  & $\bullet$ & $\circ$ &  \\
& Second-Order Effects &  &  &  &  &  &  &  &  &  & $\bullet$ &  &  &  &  &  & $\circ$ &  & $\circ$ &  &  &  \\
& Belief Change &  &  &  &  &  &  &  &  &  & $\bullet$ &  &  &  &  &  &  &  &  &  &  &  \\
& Behavioral Addiction &  &  &  &  &  &  &  &  &  & $\bullet$ & $\bullet$ &  &  &  &  & $\bullet$ &  & $\bullet$ &  & $\bullet$ & $\bullet$ \\
& Feedback Loops &  &  &  &  &  &  &  &  &  & $\bullet$ &  &  &  &  &  & $\bullet$ &  & $\bullet$ &  &  &  \\
& Preselection &  &  &  &  &  &  &  &  &  &  &  &  &  &  &  &  & $\bullet$ &  & $\bullet$ & $\circ$ & $\circ$ \\
& Habit Formation &  &  &  &  &  &  &  &  &  &  &  &  &  &  &  &  &  &  &  & $\bullet$ & $\bullet$ \\
\hline

\multirow{14}{*}\textbf{Abuse} 
& Forced Action & $\bullet$ & $\bullet$ & $\bullet$ & $\bullet$ & $\bullet$ & $\bullet$ & $\bullet$ & $\bullet$ & $\circ$ & $\bullet$ & $\circ$ & $\bullet$ & $\bullet$ & $\bullet$ & $\bullet$ & $\bullet$ & $\bullet$ & $\bullet$ & $\bullet$ & $\circ$ & $\circ$ \\
& Forced Registration &  & $\circ$ & $\bullet$ & $\bullet$ &  &  & $\bullet$ & $\bullet$ &  & $\bullet$ &  &  &  &  & $\bullet$ &  & $\circ$ & $\bullet$ & $\circ$ &  &  \\
& Privacy Zuckering & $\circ$ & $\bullet$ & $\bullet$ & $\circ$ & $\circ$ &  & $\circ$ &  &  & $\circ$ & $\circ$ &  &  &  &  &  & $\circ$ & $\circ$ & $\bullet$ &  &  \\
& Address Book Leeching &  &  & $\bullet$ & $\circ$ &  &  &  &  &  &  &  &  &  &  &  &  &  &  &  &  &  \\
& Shadow User Profiles &  &  & $\bullet$ & $\circ$ & $\circ$ &  &  &  &  &  &  &  &  &  &  &  &  &  &  &  &  \\
& Extreme Countdown &  &  &  &  &  &  &  &  &  &  &  &  & $\bullet$ &  &  &  &  &  & $\bullet$ &  &  \\
& Bundled consent &  &  &  &  &  &  &  &  &  &  &  & $\bullet$ &  &  &  &  & $\bullet$ & $\circ$ &  &  &  \\
& Impersonation &  &  &  &  &  &  &  &  &  &  &  &  &  & $\bullet$ &  &  &  &  &  &  &  \\
& Coercion &  &  &  & $\bullet$ & $\circ$ &  &  & $\circ$ &  & $\bullet$ & $\circ$ & $\bullet$ &  &  &  &  & $\bullet$ & $\bullet$ &  &  &  \\
& Nagging & $\bullet$ & $\bullet$ & $\circ$ & $\bullet$ & $\bullet$ & $\bullet$ & $\bullet$ & $\circ$ & $\bullet$ & $\circ$ & $\circ$ & $\bullet$ & $\bullet$ & $\circ$ & $\bullet$ & $\bullet$ & $\bullet$ &  & $\bullet$ & $\circ$ & $\circ$ \\
& Forced Continuity &  &  &  &  &  &  &  &  &  &  &  &  &  &  &  &  & $\bullet$ &  &  &  &  \\
& Gamification &  & $\bullet$ & $\circ$ &  &  &  &  &  &  & $\circ$ & $\circ$ &  &  &  &  & $\bullet$ & $\circ$ & $\bullet$ & $\circ$ & $\circ$ & $\circ$ \\
& Auto-play &  &  &  &  &  &  &  &  &  &  & $\circ$ & $\bullet$ & $\bullet$ &  &  & $\bullet$ & $\circ$ & $\circ$ &  & $\circ$ & $\circ$ \\
& Social Pyramid &  &  &  &  &  &  &  &  &  &  &  &  &  &  &  &  &  &  & $\bullet$ &  &  \\
\bottomrule
\end{tabular}}

\vspace{1ex}
\begin{tabular}{lll}
\textbf{Legend:} & $\bullet$ = Directly mentioned/fully covered & $\circ$ = Similar concept/partially covered \\
\end{tabular}
\end{table}

\section{Voice Characteristics' Statistics}\label{sec:audio_feature}

Table~\ref{tab:core_audio_features} reported the voice characteristics' statistics. Specifically, the reported dimensions are defined as follows:

$\bullet$ \textbf{Category:} The label indicating the experimental condition or type associated with the audio sample, typically derived from the subdirectory name.

$\bullet$ \textbf{Duration (s):} The total length of the audio recording in seconds.

$\bullet$ \textbf{Mean F0 (Hz):} The average fundamental frequency (perceived as pitch) across the voiced segments of the speech signal, measured in Hertz. This reflects the overall pitch level of the voice. 

$\bullet$ \textbf{Std Dev F0 (Hz):} The standard deviation of the fundamental frequency across voiced segments, measured in Hertz. This quantifies the amount of pitch variation or intonation used by the speaker.

$\bullet$ \textbf{Mean Int. (dB):} The average acoustic intensity (related to perceived loudness) of the speech signal, measured in decibels (dB).

$\bullet$ \textbf{Speech Rate (Onsets/s):} An estimation of the speech articulation rate, calculated based on the number of detected acoustic onsets (approximating syllables or stressed events) per second.

$\bullet$ \textbf{Pause Ratio (\%):} The percentage of the total audio duration that consists of pauses or silence (identified based on fundamental frequency absence and a minimum duration threshold). This reflects the fluency or hesitancy of the speech.

$\bullet$ \textbf{Jitter (\%):} A measure of the cycle-to-cycle variability of the fundamental frequency period, typically reported as a percentage (e.g., Jitter local). It relates to perceived pitch instability or vocal roughness.

$\bullet$ \textbf{Shimmer (\%):} A measure of the cycle-to-cycle variability of the speech signal's amplitude, typically reported as a percentage (e.g., Shimmer local). It relates to perceived amplitude instability.

$\bullet$ \textbf{Mean HNR (dB):} The average Harmonics-to-Noise Ratio, measured in decibels. It represents the ratio of harmonic (periodic) energy to noise (aperiodic) energy in the voice signal, often associated with voice clarity versus hoarseness.

$\bullet$ \textbf{Spec Centroid (Hz):} The Spectral Centroid indicates the spectrum's average frequency (``center of mass'') in Hz, often related to perceived brightness.

$\bullet$ \textbf{Spec Band (Hz):} Spectral Bandwidth measures the frequency spread around this centroid, showing if the sound energy is concentrated (purer) or widely distributed (richer/noisier).

\end{document}